\documentclass[a4paper,11pt]{article}
\pdfoutput=1 

\usepackage{jheppub} 
\usepackage{amssymb}
\usepackage{mathtools}
\usepackage{slashed}
\usepackage{cleveref}

\newcommand{\be}{\begin{equation}}
\newcommand{\ee}{\end{equation}}
\newcommand{\bea}{\begin{eqnarray}}
\newcommand{\eea}{\end{eqnarray}}

\newcommand{\epm}{e^+e^-}

\newcommand{\ftsoft}[1]{\widetilde{\mathbb{S}}_{\mbox{\small {#1}-h}}}

\allowdisplaybreaks

\title{\boldmath Factorization of $e^+e^- \to H \, X$ cross section, differential in $z_h$, $P_T$
 and thrust, 
 in the $2$-jet limit}

\author{M. Boglione,}
\author{and A. Simonelli}
 
\affiliation{Dipartimento di Fisica, Universit\`a di Torino,\\
                Via P.~Giuria 1, I-10125 Torino, Italy}
\affiliation{INFN, Sezione di Torino, Via P.~Giuria 1, I-10125 Torino, Italy}

\emailAdd{boglione@to.infn.it}
\emailAdd{andrea.simonelli@unito.it}

 \abstract{
 Factorizing the cross section for single hadron production in $e^+e^-$ annihilations is a highly non trivial task when the transverse momentum of the outgoing hadron with respect to the thrust axis is taken into account. We work in a scheme that allows to factorize the $e^+e^- \to HX$ cross section as a convolution of a calculable hard coefficient and a Transverse Momentum Dependent (TMD) fragmentation function. 
 The result, differential in $z_h$, $P_T$ and thrust, will be given to all orders in perturbation theory and explicitly
 computed to Next to Leading Order (NLO) and Next to Leading Log (NLL) accuracy.
 The predictions obtained from our computation, applying the simplest and most natural ansatz to model the non-perturbative part of the TMD, are in exceptional agreement with the experimental measurements of the BELLE Collaboration. 
 The factorization scheme we propose relates the TMD parton densities defined in 1-hadron and 2-hadron processes, restoring the possibility to perform global phenomenological studies of TMD physics including experimental data from semi-inclusive deep inelastic scattering, Drell-Yan processes, $e^+e^- \to H_1 H_2 X$ and $e^+e^- \to HX$ annihilations.
}

\begin{document} 

\maketitle
 
%
\section{Introduction\label{sec:intro}}

\bigskip

Recently, the BELLE collaboration has measured $\epm \to H\,X$ cross section, as functions of $P_T$, the transverse momentum of the detected hadron with respect to the thrust axis~\cite{Seidl:2019jei} at a c.m. energy of about 10 GeV. This can be considered a break-through measurement in the investigation of the 3D structure of hadrons, as it offers a direct glance to the transverse motion of the fragmenting partons.

Unfortunately the factorization mechanism, which was first devised by Collins, Soper and Sterman (CSS) for Drell-Yan processes and later proven to work for Semi-Inclusive Deep Inelastic Scattering (SIDIS)~\cite{Collins:1984kg,Collins:1989bt,Collins:1989gx,Collins:1992kk,Collins:1993kq} and $\epm \to H_1\,H_2\,X$~Ref.~\cite{Collins:2011zzd}, cannot be directly applied to $\epm \to H\,X$ processes. In fact, having only one hadron detected in the final state makes it impossible 
to cast the cross section in a form that allows to define the Transverse Momentum Dependent parton densities (TMDs) in the conventional way, i.e. by including part of the soft radiation generated in the process inside the TMD itself.
This is a serious impediment which endangers the possibility to exploit the valuable 
information encoded in the BELLE experimental data in the phenomenological study of the mechanisms which lead to hadronization and, ultimately, to the structure of hadrons in terms of their elementary constituents, quarks and gluons.

In the attempt to overcome this obstacle, in Ref.~\cite{Boglione:2020cwn} we introduced a factorization scheme, closely following the CSS methodology of Ref.~\cite{Collins:2011zzd}, in which the definition of the TMD is slightly modified in such a way to leave out all its  
soft content. 
As a result, the final cross section is written as the convolution of a hard coefficient, which represents the partonic core of the process and is calculable in pQCD at any desired order, with a Transverse Momentum Dependent Fragmentation Function (TMD FF), which carries non-perturbative information and provides a direct probe of the 3D-dynamics of the hadronization mechanism.
The advantage of this scheme is that it allows to relate the TMD fragmentation function extracted in $e^+e^- \to H\,X$, which is a 1-hadron class process, to the TMD parton densities extracted from SIDIS, Drell-Yan and $e^+e^- \to H_1\,H_2\,X$, i.e. in any 2-hadron class process.
This is the main point of strength of the proposed scheme: the immense effort already dedicated in the extraction of TMD distribution and fragmentation functions in the past two decades does  not go wasted, as the relation to those function is uniquely determined and preserved. This point will be discussed in more detail in Section~\ref{sec:final_xs} where we will present a basic phenomenological application of our final result.

These issues are also treated in a 
in the context of Soft and Collinear Effective Field Theories (SCET) 
, see for instance Refs.~\cite{Neill:2016vbi,Kang:2017glf,Kang:2017mda}. 
Very recently two papers, Refs.~\cite{Kang:2020yqw} and~\cite{Makris:2020ltr}, have been dedicated to the study of $e^+e^- \to H\,X$ cross sections as derived in a SCET framework.

This article is structured as follows. Section~\ref{sec:partonic_xs} shows the computation of the hard part of the process, i.e. the partonic cross section: two separate subsections are dedicated to the cases in which the fragmenting parton is either a gluon or a fermion, respectively. 
The computation at Next Lowest Order (NLO) is performed explicitly.
In Section~\ref{sec:nll_tmd} we revise the structure of the 
TMD fragmentation function and its general perturbative and non perturbative content, giving the explicit expressions of all relevant quantities to Next to Leading Log (NLL) accuracy. 
In Section~\ref{sec:final_xs} we build the final expression of the $e^+e^- \to HX$ cross section, we discuss its properties and comment on issues related to its resummation. In Section~\ref{sec:pheno} we show how our final formula can be applied to reproduce the experimental cross section as measured by the BELLE Collaboration. Even without a proper fit the agreement with data turns out to be excellent, confirming the validity and strength of the formalism we propose. In Section~\ref{sec:concl} we draw our conclusions.

\bigskip

\subsection{Kinematics of the \texorpdfstring{$\epm \to H\,X$}{e+e- --> HX}process and cross section}\label{subsec:epm_1h}

\bigskip

In the process measured by the BELLE collaboration an electron and a positron, with momenta $l_1$ and $l_2$ respectively, annihilate in a virtual boson with momentum $q$
(since the c.m. energy is around $10$ GeV, it can only be a photon).
The observed hadron, of momentum $P$, belongs to a jet initiated by a parton produced in the $e^+e^-$ annihilation.
The BELLE cross section is sensitive to the fractional energy $z_h$ of the detected hadron $H$ with respect to the total energy $Q = \sqrt{q^2}$ available in the c.m.:
\begin{align}
z_h = 2 \frac{P \cdot q}{q^2}.
\label{eq:zh_def}
\end{align}
The \textbf{parton frame} is defined by fixing the $z$-axis of the c.m. frame to coincide with the axis of the jet, which is also the direction of the fragmenting parton.
Then, the momentum $P$ of the detected hadron can be written as:
\begin{align}
P = 
\left( P^+,\,P_p^-,\,\vec{P}_{p,\,T} \right) \quad 
\mbox{with } P^+ = z_h \, q^+, 
\label{eq:P_pframe}
\end{align}
and the measurement of $\vec{P}_{p,\,T}$ probes the transverse motion of the partons inside the detected hadron.
However, the determination of the jet axis is not easy. 
In what follows, we will identify it with the reconstructed thrust axis $\vec{n}$, which is the direction that maximizes thrust, $T$, defined as:
\begin{align}
T = \mbox{Max}\,\frac{\sum_i |\vec{p}_i \cdot \vec{n}|}{\sum_i |\vec{p}_i|},
\label{eq:thrust_def}
\end{align}
where the sum is over all the final state particles $i$.
The variable $T$ describes the topology of the final state and it ranges from $0.5$ to $1$, where the lower limit corresponds to a spherical distribution of final state particles, while the upper limit realizes a pencil-like event.
Therefore, in this paper we will interpret the transverse momentum measured by BELLE as $P_T \equiv |\vec{P}_{p,\,T}|$ and we will provide a cross section differential in the three variables: $z_h$, $P_T$ and $T$.
Following the scheme presented in Ref.~\cite{Boglione:2020cwn}, the final cross section will be written schematically as:
\begin{align}
\frac{d \sigma}{dz_h \, dT \, dP_T^2} = 
\pi \sum_j \, \int_{z_h}^1 \,
\frac{d z}{z} \, 
\frac{d \widehat{\sigma}}{d {z_h}/{z} \, dT} \,
D_{1,\,H/j}(z,\,P_T)
\left[
1+\mathcal{O}\left( \frac{P_T^2}{Q^2}\right) 
\right]
\,
\left[
1+\mathcal{O}\left( \frac{M_H^2}{Q^2}\right)
\right],
\label{eq:xs_final_general}
\end{align}
where $M_H$ is the mass of the detected hadron and the integration variable $z$ is the light-cone momentum fraction of the detected hadron with respect to its parent (fragmenting) parton. Kinematics constrain $z$ to lie in the range $z_h \leq z \leq 1$.
We will focus on spinless hadrons, hence only the unpolarized TMD FFs $D_1$ appears in the cross section.
The TMD FFs are defined in the Fourier conjugate space of the transverse momentum of the fragmenting parton with respect to the direction of the detected hadron.
A proper definition of this transverse momentum is obtained by boosting (or equivalently by rotating) the parton frame defined in Eq.~\eqref{eq:P_pframe} in order to re-define the $z$-axis so that it coincides to the direction of the detected hadron. 
According to Ref.~\cite{Collins:2011zzd}, we will refer to the resulting frame as the \textbf{hadron frame}, where the momentum $k$ of the fragmenting parton can be written as:
\begin{align}
k = 
\left( k^+,\,k_h^-,\,\vec{k}_{h,\,T} \right) \quad 
\mbox{with } k^+ = {P^+}/{z}.
\label{eq:k_hframe}
\end{align}
The transverse momenta $\vec{k}_{h,\,T}$ and $\vec{P}_{p,\,T}$ are related by the following expression:
\begin{align}
\vec{k}_{h,\,T} = -\frac{1}{z} \; \vec{P}_{p,\,T}\left[1 + 
\mathcal{O}( \frac{P_{p,\,T}^2}{Q^2} )\right].
\label{eq:h_p_frame}
\end{align}
Therefore, in Eq.~\eqref{eq:xs_final_general} the unpolarized TMD FFs 
have to be intended as:
\begin{align}
&D_{1,\,H/j}(z,\,P_T)
\left[
1+\mathcal{O}\left( \frac{P_T^2}{Q^2}\right) 
\right]=
\int \, \frac{d^2\vec{b}_T}{(2\pi)^2} \,
e^{i \frac{\vec{P}_T}{z} \cdot \vec{b}_T} \,
\widetilde{D}_{1,\,H/j}(z,\,b_T).
\label{eq:TMD_FT_PT}
\end{align}
Notice that the Fourier transform acts like an analytic continuation of the TMD in momentum space and consequently it is totally inadequate when used in the range of large values of $P_T$, where TMDs lose their physical meaning.

\bigskip

The cross section in Eq.~\eqref{eq:xs_final_general} can also be written as the contraction of two tensors: the leptonic tensor $L_{\mu\nu}$ and the hadronic tensor $W_H^{\mu\nu}$, which encode the initial and final state information, respectively.
The leptonic tensor is defined as the lowest order of the electromagnetic vertex $\epm \to \gamma^\star$ and it is given by:
\begin{align}
L^{\mu\nu}(\theta) = 
l_1^\mu \, l_2^\nu +
l_2^\mu \, l_1^\nu -
g^{\mu \nu} \, l_1 \cdot l_2,
\label{eq:lept_tens}
\end{align}
where $l_1$ and $l_2$ are the momenta of the electron and the positron, respectively, and $\theta$ is the zenith angle of the electron with respect to the thrust axis $\vec{n}$: 
\begin{align}
&l_1 = \frac{Q}{2} \left(
1,\,\vec{u}
\right); \\
&l_2 = \frac{Q}{2} \left(
1,\,-\vec{u}
\right),
\label{eq:epm_momenta}
\end{align}
with $\vec{u} \cdot \vec{n} = \cos{\theta}$.
The final cross section is insensitive to the value of $\theta$, hence we will integrate over the whole solid angle specified by the leptons.
On the other hand, the hadronic tensor is defined by:
\begin{align}
&W^{\mu \, \nu}_H(z_h,\,T,\,P_T) = 
4 \pi^3 \, \sum_X \delta^{(4)} 
\left( p_X +P - q \right) \,
\times 
\notag \\
&\quad \times \,
\langle 0 | \,
j^\mu(0) | \,
P,\,X,\,\mbox{out} \,
\rangle_T \,
{}_T \langle P,\,X,\,\mbox{out} |
j^\nu(0) | \,
0 \rangle = 
\notag \\
&\quad= 
\frac{1}{4\pi} \, \sum_X \, \int d^4 z\,
e^{i q \cdot z}
\langle 0 | \,
j^\mu\left(z/2\right) | \,
P,\,X,\,\mbox{out} \,
\rangle_T \,
{}_T \langle P,\,X,\,\mbox{out} |
j^\nu\left(-z/2\right) | \,
0 \rangle,
\label{eq:current_def_W}
\end{align}
where the factor $1/(4\pi)$ coincides with the normalization factor used in Ref.~\cite{Collins:2011zzd}, while the final states vectors are labeled by ``T" as a reminder of the event topology, which has to be selected according to the value of thrust. 
Finally, the cross section is given by:
\begin{align}
&\frac{d \sigma}{dz_h \, dT \, dP_T^2} = 
z_h \, \frac{\alpha^2}{4 Q^4} \,
\int_0^{2\pi} d\phi \, 
\int_0^\pi d\theta \,
L_{\mu\nu}(\theta) \,
\frac{d W^{\mu \, \nu}_H(z_h,\,T,\,P_T)}
{d P_T^2},
\label{eq:xs_tensors_general}
\end{align}
where $\phi$ is the azimuthal angle of the electron with respect to the $x$-axis of the c.m. frame. Since we are considering unpolarized leptons, the cross section does not depend on this variable.
For practical purposes, the hadronic tensor can be written by using the structure functions obtained by projecting onto its relevant Lorentz structures:
\begin{align} 
W^{\mu \, \nu}_H &= 
\left(-g^{\mu \, \nu} + 
\frac{q^\mu q^\nu}{q^2} \right) 
F_{1,\,H} \, +
\frac{\left( P^\mu - q^\mu \frac{P \cdot q}{q^2} \right) 
\left( P^\nu - q^\nu \frac{P \cdot q}{q^2} \right)}
{P \cdot q} \,
F_{2,\,H}.
\label{eq:had_tens_structfun}
\end{align}
We can easily compute the projections: 
\begin{align}
&-g_{\mu\,\nu}W^{\mu \, \nu}_H = 
3 F_{1,\,H} + \frac{z_h}{2} \, F_{2,\,H} + 
\mathcal{O} \left( \frac{M^2}{Q^2} \right); 
\label{eq:W_proj1} \\
&\frac{P_\mu P_\nu}{Q^2} W^{\mu \, \nu}_H = 
\left( \frac{z_h}{2} \right)^2 \, \left[ 
F_{1,\,H} + 
\frac{z_h}{2} \, F_{2,\,H}\right] + 
\mathcal{O} \left( \frac{M^2}{Q^2} \right).
\label{eq:W_proj2}
\end{align}
Such decomposition allows to define the transverse (T) and the longitudinal (L) component of the cross section as:
\begin{align}
&\frac{1}{\sigma_B} \, 
\frac{d \sigma_T}{dz_h \, dT \, dP_T^2} = 
z_h \, \frac{dF_{1,\,H}(z_h,\,T,\,P_T)}
{dP_T^2}; 
\label{eq:xs_T}\\
&\frac{1}{\sigma_B} \, 
\frac{d \sigma_L}{dz_h \, dT \, dP_T^2} = 
\frac{z_h}{2} \, \left(
\frac{dF_{1,\,H}(z_h,\,T,\,P_T)}{dP_T^2} +
 \frac{z_h}{2} \, \frac{dF_{2,\,H}(z_h,\,T,\,P_T)}{dP_T^2}
 \right),
\label{eq:xs_L}
\end{align}
where we used the Born cross section:
\begin{align}
\sigma_B = \frac{4 \pi \alpha^2}{3 Q^2}.
\label{eq:born_xs}
\end{align}
This remarkably simplifies the integration over $\theta$, in fact we have:
\begin{align}
&\frac{d \sigma}
{dz_h \, d\cos{\theta} \, dT \, dP_T^2} =
\frac{3}{8} \, (1+\cos^2{\theta}) \,
\frac{d \sigma_T}{dz_h \, dT \, dP_T^2}+
 \frac{3}{4} \, \sin^2{\theta} \,
\frac{d \sigma_L}{dz_h \, dT \, dP_T^2}.
\label{eq:xs_diff_theta}
\end{align}
Therefore it follows straightforwardly:
\begin{align}
&\frac{d \sigma}{dz_h \, dT \, dP_T^2} = 
\frac{d \sigma_T}{dz_h \, dT \, dP_T^2} +
\frac{d \sigma_L}{dz_h \, dT \, dP_T^2} =
\notag \\
&\quad=
\frac{4 \pi \alpha^2}{3 Q^2} \,
\left[
\frac{3}{2} \, z_h \,
\frac{dF_{1,\,H}(z_h,\,T,\,P_T)}{dP_T^2}+ 
\frac{z_h^2}{4} \,
\frac{dF_{2,\,H}(z_h,\,T,\,P_T)}{dP_T^2}
\right]
\label{eq:xs_general_structfun}
\end{align}
Another simplification occurs when the projection with respect to $P_\mu \, P_\nu$ (see Eq.~\eqref{eq:W_proj2}) is zero (or can be neglected), making the longitudinal cross section to vanish.
In this case, the two structure functions are not independent anymore and in fact $F_{2,\,H} = -\frac{2}{z_h} \, F_{1,\,H} $.
As a consequence, the hadronic tensor can be written as:
\begin{align}
W^{\mu \, \nu}_{H,\,T}(z_h,\,T,\,P_T) =
H_T^{\mu\nu} \,
F_{1,\,H}(z_h,\,T,\,P_T).
\label{eq:k2prog_null_W}
\end{align}
where we defined the transverse tensor:
\begin{align}
H_T^{\mu\nu} = 
\left[ 
-g^{\mu \nu} + 
\frac{P^\mu q^\nu + P^\nu q^\mu}
{P \cdot q} -
q^2 \, \frac{P^\mu P^\nu}
{\left(P \cdot q\right)^2}
\right].
\label{eq:HT_tensor}
\end{align}

\bigskip

%
\section{Partonic Cross Section 
\label{sec:partonic_xs}}

\bigskip

In this section, we will compute the partonic cross section of $\epm \to H\,X$  
in the $2$-jet limit, providing its explicit expression to 1-loop (NLO) precision and also its all-order, resummed, formulation.
According to the scheme presented in Ref.~\cite{Boglione:2020cwn}, the partonic cross section is the hard part of the factorized full cross section and encodes its short-distance contributions.
Therefore, it is fully computable in perturbative QCD as it represents the partonic analogue of the whole process. 
In other words, the partonic cross section describes the process $\epm \to j \, X$, where $j$ indicates the parton of type $j$ that fragments into the detected hadron $H$;
$j$ can either be a gluon or a fermion (quark/antiquark) of flavor $f$.
However, in the $2$-jet limit, the contribution of the fragmenting gluon is strongly suppressed by kinematics (we  will present the explicit computation in Sect.~\ref{sec:frag_gluon}) so that the only relevant contribution is given by the fragmenting fermions.
The topology corresponding to a quasi $2$-jet configuration is obtained by requiring the thrust $T$ to be almost $1$ or, analogously, $\tau = 1-T \sim 0$.
Since the lowest order is an exact $2$-jet configuration (pencil-like), the first genuinely non trivial term is generated at 1-loop, and is due to the emission of a real parton crossing the final state cut.
Practically speaking, we will introduce a sharp cut-off $\tau_{\mbox{\tiny MAX}}$ that restricts the range of $\tau$ values to $ 0 \leq \tau \leq \tau_{\mbox{\tiny MAX}}$.
Then, the $2$-jet-like topology for the final state will correspond to the limit $\tau_{\mbox{\tiny MAX}} \to 0$.

\bigskip

The reference frame for the computation is the partonic analogue of the hadron frame, see Eq.~\eqref{eq:k_hframe}, where the momentum $k_1$ of the outgoing parton lies along the plus direction:
\begin{align}
k_1 = \left( k_1^+,\,0,\,\vec{0}_T \right)
\quad 
\mbox{with } k_1^+ = z \, q^+.
\label{eq:k1_mom}
\end{align}
All the formulas valid for the whole cross section (see Section~\ref{sec:intro}) naturally extend to their partonic version.
Clearly the partonic cross section cannot merely be obtained by summing over all the relevant Feynman graphs in the small-$\tau$ limit. In fact, in the final result, the overlapping between the hard and the collinear momentum regions has to be appropriately subtracted, while renormalization is achieved through an Ultra-Violet (UV) counterterm.
According to Ref.~\cite{Boglione:2020cwn}, the subtracted and renormalized final state tensor of the partonic cross section is given, order by order in perturbation theory, by:
\begin{align}
&\widehat{W}_j^{\mu \nu,\,{[n]}}{}^{\mbox{\small , sub}}
(z,\,\tau,\,\tau_{\mbox{\tiny MAX}} ) = 
\widehat{W}_f^{\mu \nu,\,{[n]}}
{}^{\mbox{\small , unsub}} (\epsilon; \,z, \,\tau,\,\tau_{\mbox{\tiny MAX}}) - 
\notag \\
&-
\sum_j \, 
\sum_{m = 1}^n \;
\int_z^1 \frac{d\widehat{z}}{\widehat{z}}\;
\widehat{W}_{j,\;R}^{\mu \nu,\,{[n-m]}}
{}^{\mbox{\small , sub}} 
({z}/{\widehat{z}},\,\tau,\,\tau_{\mbox{\tiny MAX}})
\,
\left[ 
\widehat{z} \, 
\widetilde{D}_{j,\,f}^{[m]} 
(\epsilon; \, \widehat{z},\,\tau_{\mbox{\tiny MAX}})
\right],
\label{eq:sub_mech_W}
\end{align}
here $\epsilon$ is the dimensional regularization parameter, where the dimension of space-time is $D = 4-2\epsilon$.
In the following, we will separately compute the contributions of a fragmenting gluon and a fragmenting fermion, in Sections~\ref{sec:frag_gluon} and~\ref{sec:frag_ferm} respectively.

\bigskip

%
\subsection{Fragmenting Gluon \label{sec:frag_gluon}}

\bigskip

\begin{figure}[t]
\centering
\includegraphics[width=3.5cm]{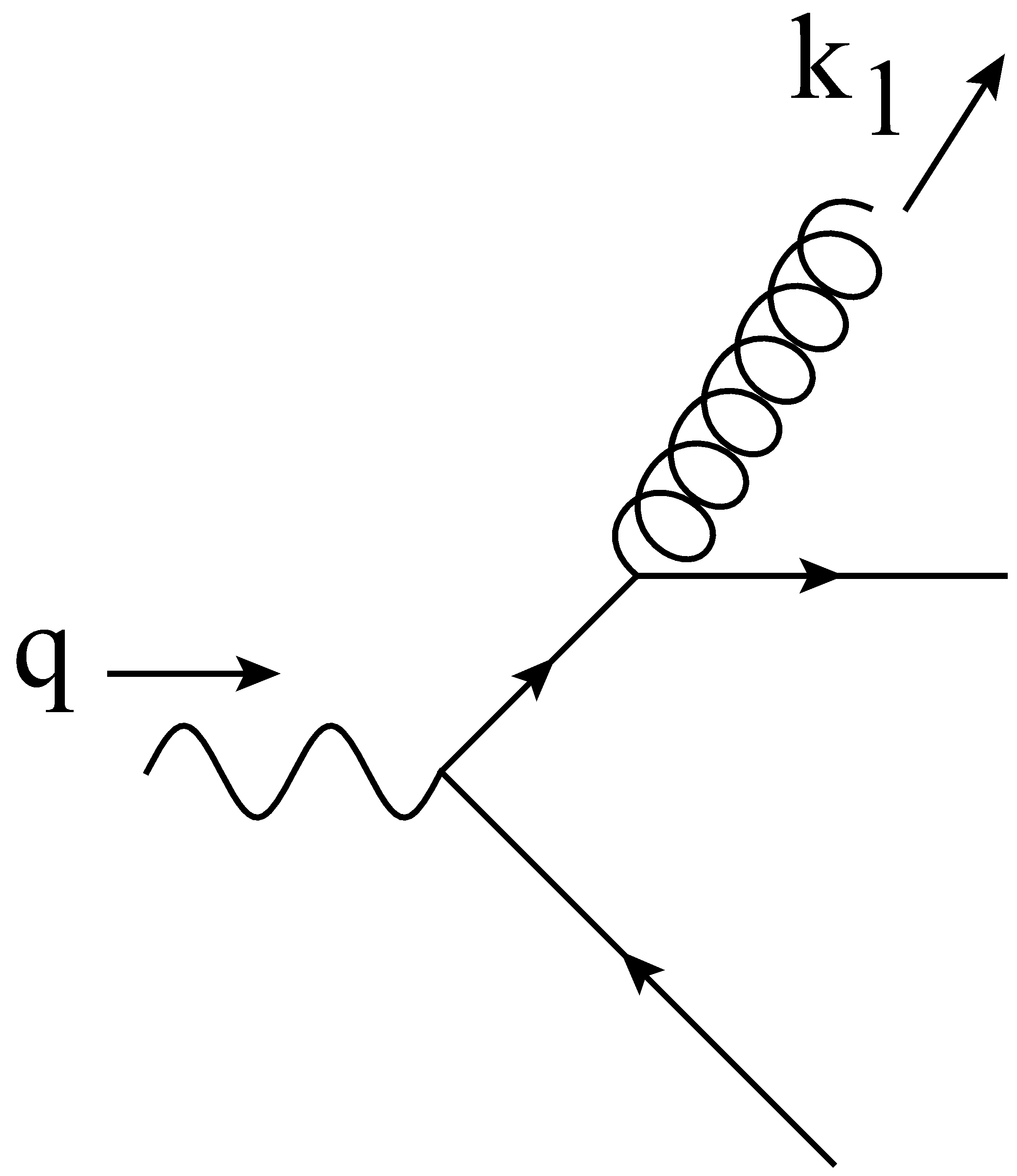}
\caption{The only 1-loop Feynman graph contributing to $\widehat{W}_g^{\mu \, \nu}$, when the gluon is emitted by the quark.
The emission from the antiquark line is analogous.}
\label{fig:frag_gluon}
\end{figure}

In a 1-loop computation, the configuration in which the detected hadron is produced by the fragmentation of a gluon can only occur through the emission of a real gluon from either the quark or the antiquark line.
Then, to compute the final state tensor $\widehat{W}_g^{\mu \, \nu}$ one simply considers the Feynman graph in Fig.~\ref{fig:frag_gluon}.
%
%
If the $2$-jet limit is applied to the final state topology, then $\widehat{W}_g^{\mu \, \nu}$ is suppressed.
In fact, after the emission of the gluon, the fermion cannot deviate drastically from its original direction (otherwise it would generate a third jet) and hence it can only proceed  almost collinearly to the gluon.
In principle, a $2$-jet configuration may be achieved also if the fermion turns soft or if it reflects backwards after the emission of the gluon. 
However, such configurations are suppressed by power counting (see Chapter 5 of Ref.~\cite{Collins:2011zzd}).
As a consequence, the only relevant kinematic configuration in the $2$-jet limit is given by the emitting fermion being collinear to the fragmenting gluon.
However, this is exactly the same configuration that has to be \emph{subtracted out} in the final result, in order to avoid double counting due to the overlapping with the collinear momentum region.
This simple argument is enough to conclude that the subtracted final state tensor associated to the fragmenting gluon has to be suppressed in a quasi $2$-jet configuration.
Nevertheless, we will perform the explicit computation of $\widehat{W}_g^{\mu \, \nu}$ anyway, for pedagogical reasons: it is very simple compared to the more important case of the fragmenting fermion (only one Feynman diagram, no rapidity cut-offs, etc ...) and it will serve as a benchmark to introduce the main features associated with the computation of the final state tensor.
Section~\ref{subsubsec:frag_gluon_unsub} will present the explicit computation of the partonic cross section, while in Section~\ref{subsubsec:frag_gluon_sub} we will describe the mechanism of subtraction.

\bigskip

%
\subsubsection{Unsubtracted Final State Tensor \label{subsubsec:frag_gluon_unsub}}

\bigskip

The unsubtracted final state tensor is obtained by applying the hard approximation ($T_H$ in  Ref.~\cite{Collins:2011zzd}) to the Feynman graphs contributing to the desired order in perturbation theory. This implies considering their massless limit and replacing every soft-collinear divergent quantity with its lowest order approximation.
In momentum space, the lowest order partonic TMD is simply a delta function, both in $z$ and in the transverse momentum of the fragmenting parton $\vec{k}_T$, therefore in the Fourier conjugate space (where we will work) the  unsubtracted final state tensor does not depend on $\vec{b}_T$  (see Section 6 of Ref.~\cite{Boglione:2020cwn}).

In the following, we will assume that the fragmenting gluon is emitted by the quark, as in Fig.~\ref{fig:frag_gluon}.
The case of emission by an antiquark is perfectly analogous.
The final state hosts three particles: the fragmenting gluon, the quark and the antiquark.
The squared amplitude and the phase space will depend on all the possible combination of scalar products among their momenta.
Then, by labeling $k_2$ the momentum of the antiquark and $k_3$ that of the quark, it is extremely useful to express all the quantities in terms of the following variables:
\begin{align}
&y_1 = \frac{2}{Q^2} \, k_2 \cdot k_3 \,;
\quad
y_3 = \frac{2}{Q^2} \, k_1 \cdot k_2 \,;
\quad
y_2 = \frac{2}{Q^2} \, k_3 \cdot k_1 \,,
\label{eq:y123_def}
\end{align}
subjected to the constraint $\sum_i y_i = 1$, due to the momentum conservation $q = k_1 + k_2 + k_3$.
We will use dimensional regularization, hence the dimension of spacetime is $D = 4 - 2 \epsilon$.

\bigskip

According to standard conventions (Ref.~\cite{Collins:2011zzd}) the polarization vector $e(k_1, \,\lambda)$ of the on-shell fragmenting gluon is defined to have zero plus and minus components\footnote{In general, the only requirement on $e(k_1, \,\lambda)$ is $k_1 \cdot e(k_1, \,\lambda) = 0$ and $e(k_1, \,\lambda) \cdot e(k_1, \,\lambda)^\star = 1$.}. 
Then, the squared amplitude associated to the Feynman graph in Fig.~\ref{fig:frag_gluon}, summed over all the polarization values $\lambda$, is given by:
\begin{align}
&M_g^{\mu\,\nu}{}^{\;[1]}(\epsilon; \,\mu, \,\{ y_i \}) = 
\begin{gathered}
\includegraphics[width=5.5cm]{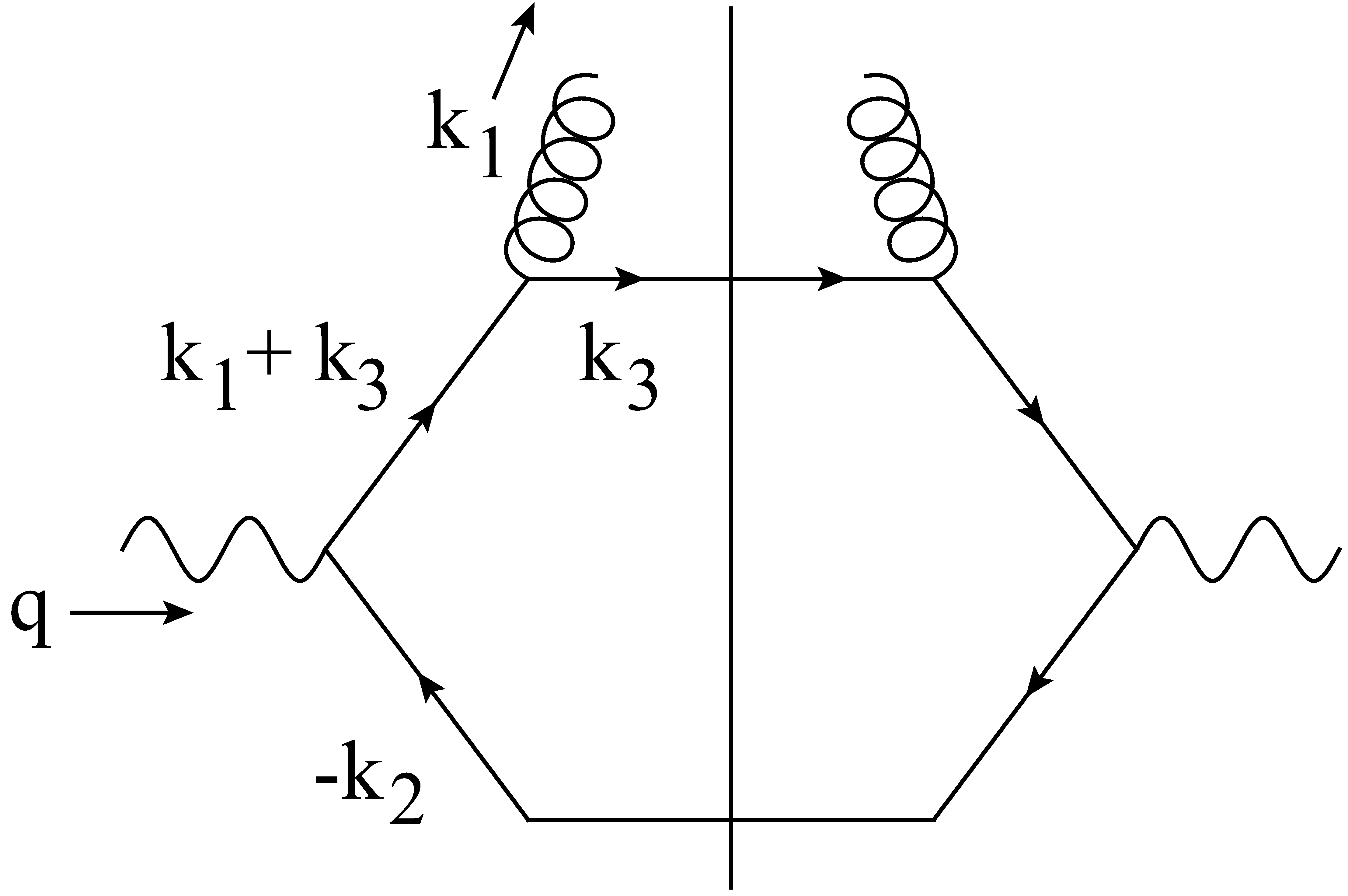}
\end{gathered}
= 
\notag \\
&\quad= 
\sum_f \, e_f^2 \, 
\sum_\lambda \, (- i g_0 \slashed{e}_\lambda t^a) \,
\frac{i (\slashed{k}_1 + \slashed{k}_3)}{(k_1 + k_3)^2 + i 0} \,
\gamma^\mu \, \slashed{k}_2 \, \gamma^\nu \, 
\frac{-i (\slashed{k}_1 + \slashed{k}_3)}{(k_1 + k_3)^2 - i 0} \,
(i g_0 \slashed{e}_\lambda^\star t_a) \, \slashed{k}_3
= 
\notag \\
&\quad= 
\sum_f \, e_f^2 \, 
g^2 \, \mu^{2\epsilon} \, C_F \, N_C \, \frac{1}{\left( (k_1 + k_3)^2 \right)^2} \,
\mbox{Tr} \left\{
(\slashed{k}_1 + \slashed{k}_3) \,
\gamma^\mu \, \slashed{k}_2 \, \gamma^\nu \, 
(\slashed{k}_1 + \slashed{k}_3) \, 
\gamma^j \, \slashed{k}_3 \, \gamma_j
\right\}
= 
\notag \\
&\quad= 
\sum_f \, e_f^2 \, 
g^2 \, \mu^{2\epsilon} \, 2 \, C_F \, N_C \, 
\frac{1}{\left( (k_1 + k_3)^2 \right)^2} \,
\Bigg[
(1-\epsilon) \, \mbox{Tr} \left\{
(\slashed{k}_1 + \slashed{k}_3) \,
\gamma^\mu \, \slashed{k}_2 \, \gamma^\nu \, 
(\slashed{k}_1 + \slashed{k}_3) \, 
\slashed{k}_3
\right\} + 
\notag \\
&\quad\hspace{5.75cm}+
k_3^j \, \, \mbox{Tr} \left\{
(\slashed{k}_1 + \slashed{k}_3) \,
\gamma^\mu \, \slashed{k}_2 \, \gamma^\nu \, 
(\slashed{k}_1 + \slashed{k}_3) \, 
\gamma_j
\right\}
\Bigg],
\label{eq:M_gluon}
\end{align}
where $e_f$ is the charge fraction of the (anti)quark generated by the virtual photon. The third line of the previous equation has been obtained by using $\sum_\lambda e^\mu(k_1, \,\lambda) e^\nu(k_1, \,\lambda) = - g^{\mu \, \nu}$.
The denominator can easily be written in terms of the variables $y_i$ defined in Eq.~\eqref{eq:y123_def}:
\begin{align}
\left( (k_1 + k_3)^2 \right)^2 = \left( 2 k_1 \cdot k_3 \right)^2 = 
Q^4 \, y_2^2 .
\label{eq:den_Mg}
\end{align}
On the other hand, we can compute the projections of the numerator onto its relevant Lorentz structures as in Eqs.~\eqref{eq:W_proj1} and~\eqref{eq:W_proj2}:
\begin{align}
&-g_{\mu \nu} \, \mbox{Tr} \left\{
(\slashed{k}_1 + \slashed{k}_3) \,
\gamma^\mu \, \slashed{k}_2 \, \gamma^\nu \, 
(\slashed{k}_1 + \slashed{k}_3) \, 
\slashed{k}_3
\right\} = 
\notag \\
&\quad= 
8 (1-\epsilon) \, 
\left\{
2 \, (k_1 + k_3) \cdot k_2 \; (k_1 + k_3) \cdot k_3 -
k_2 \cdot k_3 \; (k_1 + k_3)^2
\right\} = 
\notag \\
&\quad= 
Q^4 \, 4 \, (1-\epsilon) \, y_2 \, y_3 ; 
\label{eq:num_g_Mg_1} \\
&-g_{\mu \nu} \, \mbox{Tr} \left\{
(\slashed{k}_1 + \slashed{k}_3) \,
\gamma^\mu \, \slashed{k}_2 \, \gamma^\nu \, 
(\slashed{k}_1 + \slashed{k}_3) \, 
\gamma_j
\right\} =
\notag \\
&\quad= 
8 (1-\epsilon) \, 
\left\{
2 \, (k_1 + k_3) \cdot k_2 \; (k_1 + k_3)_j -
k_{2,\,j} \; (k_1 + k_3)^2
\right\} =
\notag \\
&\quad= 
Q^2 \, 8 \, (1-\epsilon) \, k_{3,\,j},
\label{eq:num_g_Mg_2}
\end{align}
where the last line of Eq.~\eqref{eq:num_g_Mg_2} has been obtained by using transverse momentum conservation, $k_{2,\,j} = - k_{3,\,j}$. Furthermore:
\begin{align}
&\frac{k_{1,\,\mu} \, k_{1,\,\nu}}{Q^2} \, \mbox{Tr} \left\{
(\slashed{k}_1 + \slashed{k}_3) \,
\gamma^\mu \, \slashed{k}_2 \, \gamma^\nu \, 
(\slashed{k}_1 + \slashed{k}_3) \, 
\slashed{k}_3
\right\} = 
\notag \\
&\quad= 
\frac{1}{Q^2} \, 16 \, k_1 \cdot k_2 \left( k_1 \cdot (k_1 + k_3)\right)^2 =
\notag \\
&\quad= 
Q^4 \, 2 \, y_2^2 \, y_3;
\label{eq:num_k2_Mg_1} \\
&\frac{k_{1,\,\mu} \, k_{1,\,\nu}}{Q^2} \, \mbox{Tr} \left\{
(\slashed{k}_1 + \slashed{k}_3) \,
\gamma^\mu \, \slashed{k}_2 \, \gamma^\nu \, 
(\slashed{k}_1 + \slashed{k}_3) \, 
\gamma_j
\right\} = 
\notag \\
&\quad= 
\frac{1}{Q^2} \, 16 \, k_1 \cdot k_2 \; k_1 \cdot (k_1 + k_3) \; 
(k_1 + k_3)_j = 
\notag \\
&\quad= 
Q^2 \, 4 \, y_2 \, y_3 \, k_{3,\,j} .
\label{eq:num_k2_Mg_2}
\end{align}
Therefore, the projection of the full squared amplitude onto the metric tensor gives:
\begin{align}
&-g_{\mu \nu} 
M_g^{\mu\,\nu}{}^{\;[1]}(\epsilon; \,\mu, \,\{ y_i \}) =
\notag \\
&\quad= 
\sum_f \, 
\left[
e_f^2 \, 2 \, N_C \, (1-\epsilon)
\right]
g^2 \, \mu^{2\epsilon} \, 4 \, C_F \, 
\left[ 
(1-\epsilon) \, \frac{y_3}{y_2} +
2 \, \frac{1}{y_2^2} \, \frac{k_{3,\,T}^2}{Q^2}
\right] = 
\notag \\
&\quad= 
H_0 \,
g^2 \, \mu^{2\epsilon} \, 4 \, C_F \, 
\left[ 
1 + 2 \, \frac{y_1}{(1-y_1)^2} -\epsilon
\right] \, \frac{y_3}{y_2} .
\label{eq:g_Mg}
\end{align}
The last line has been obtained by expressing the ratio ${k_{3,\,T}^2}/{Q^2}$ in terms of the $y_i$ variables and by defining:
\begin{align}
&H_{0,\,f} = e_f^2 \, 2 \, N_C \, (1-\epsilon);
\label{eq:H0f_def} \\
&H_0 = \sum_f \, H_{0,\,f},
\label{eq:H0_def}
\end{align}
where $H_{0,\,f}$ is the constant factor appearing in front of the lowest order final state tensor, see Eq.~\eqref{eq:g_W_lo}.
Notice that the expression in Eq.~\eqref{eq:g_Mg} only diverges if $y_2 = 0$, which corresponds to the configuration in which the quark is collinear to the fragmenting gluon. 
On the other hand, in $y_1 = 0$ the function is regular, in agreement with the power counting prediction about the soft and the collinear-to-antiquark configurations.
The other projection gives:
\begin{align}
&\frac{k_{1,\,\mu} \, k_{1,\,\nu}}{Q^2} M_g^{\mu\,\nu}{}^{\;[1]}(\epsilon; \,\mu, \,\{ y_i \}) =
\notag \\
&\quad= 
H_0 \,
g^2 \, \mu^{2\epsilon} \, 2 \, C_F \,
\frac{1}{1-\epsilon}
\Bigg[
(1-\epsilon) \, y_3 + 
2 \frac{y_3}{y_2} \frac{k_{3,\,T}^2}{Q^2}
\Bigg] =
\notag \\
&\quad= 
H_0 \,
g^2 \, \mu^{2\epsilon} \, 2 \, C_F \,
\frac{1}{1-\epsilon} 
\Bigg[
1 + 2 \, \frac{y_1}{(1-y_1)^2} \, y_3 -\epsilon
\Bigg] \, y_3 .
\label{eq:k2_Mg}
\end{align}
Notice that this expression is regular in both $y_1 = 0$ and $y_2 = 0$, consequently it is  suppressed in the $2$-jet limit.

\bigskip

\begin{figure}[t]
\centering
\includegraphics[width=9cm]{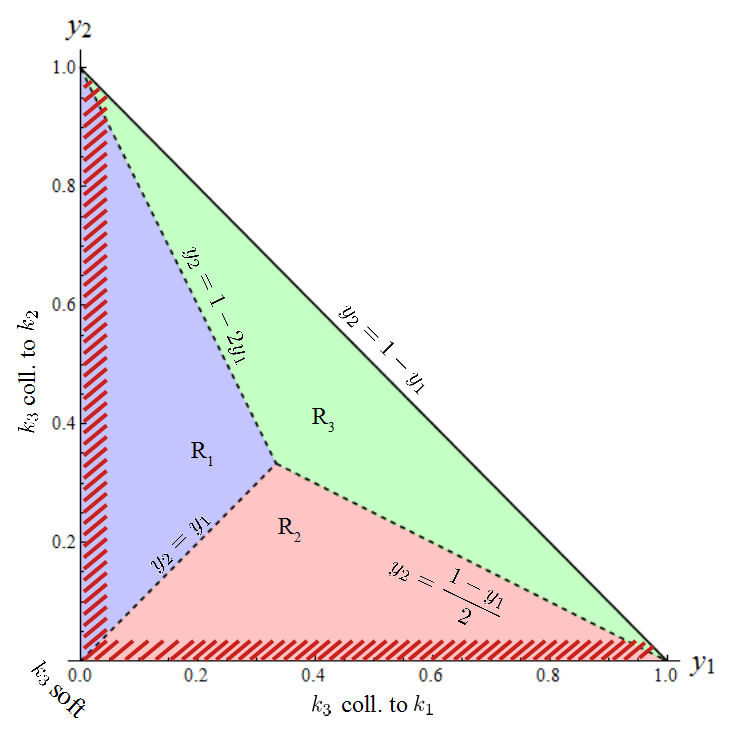}
\caption{The phase space available for the three final state massless particles $g$, $q$ and $\Bar{q}$, with momenta $k_1$, $k_3$ and $k_2$, respectively. The dashed (red) bands represent the quasi $2$-jet configurations, where $y_1$ and/or $y_2$ are zero. The sub-regions $R_i$ correspond to a value of thrust given by $\tau = y_i$.}
\label{fig:phase_space_real}
\end{figure}

The phase space available for the three final state massless particles can be represented by the triangle in Fig.~\ref{fig:phase_space_real}.
The phase space cannot extend beyond the edge given by $y_2 = 1 - y_1$ (or $y_3 = 0$), due to momentum conservation.
%
%
Since the fragmenting gluon does not cross the final state cut, the phase space integral only  involves $k_2$ and $k_3$, and it reduces to a single integration over $y_2$ after applying the momentum conservation delta. It is given by:
\begin{align}
&\int \,
\frac{d^{3 - 2 \epsilon} \vec{k}_2}
{(2 \pi)^{3 - 2 \epsilon} \, 2 |\vec{k}_2|} \,
\frac{d^{3 - 2 \epsilon} \vec{k}_3}
{(2 \pi)^{3 - 2 \epsilon} \, 2 |\vec{k}_3|} \,
(2 \pi)^{4 - 2 \epsilon} \,
\delta( q - k_1 + k_2 + k_3 ) = 
\notag \\
&\quad=
(4\pi) \, \frac{1}{2} \, \frac{1}{(4 \pi)^2} \, S_\epsilon \, 
Q^{-2 \epsilon} \, (1-y_1)^{-1+2\epsilon} \, 
\int_0^{1-y_1} \, d y_2 \, (y_1 \, y_2 \, y_3)^{-\epsilon},
\label{eq:phsp_int_noT}
\end{align}
where $S_\epsilon$ is a shorthand notation for ${(4 \pi)^\epsilon}/{\Gamma(1-\epsilon)}$, as in Ref.~\cite{Collins:2011zzd}.
The $(4\pi)$ factor in front of the phase space integral simplifies with the normalization chosen for the hadronic tensor, see Eq.~\eqref{eq:current_def_W}.
Since the phase space integration involves only $y_2$, it can be very useful to change variables and write all the quantities in terms of the following:
\begin{align}
&y_1 = 1-z, \quad y_2 = \alpha \, z, \quad y_3 = z \, (1-\alpha). 
\label{eq:change_var}
\end{align}
Notice that here we are not interested to the whole phase space, but only to the (very narrow) region corresponding to a quasi $2$-jet configuration,  where $y_1$ and/or $y_2$  tend to zero. This region is indicated by a dashed red band in Fig.~\ref{fig:phase_space_real}.
The integration over these regions is well defined once we appropriately limit the range of thrust, $T = 1 - \tau$, within  the interval $0 \leq \tau \leq \tau_{\mbox{\tiny MAX}}$.
At the partonic level, $\tau$ can be computed explicitly and its value changes inside the triangle of Fig.~\ref{fig:phase_space_real}.
In particular, $\tau$ always coincides with the minimum among $y_1$, $y_2$ and $y_3$, allowing the definition of three sub-regions inside the triangle, $R_1$, $R_2$ and $R_3$, according to the value assumed by $\tau$.
Then, the $2$-jet limit of each region is obtained by considering the limit of vanishing $\tau_{\mbox{\tiny MAX}}$ and keeping only the leading divergence in $\tau = 0$.
Notice that a pure $3$-jet configuration is the point of intersection of all three sub-regions, where $y_1 = y_2 = 1/3$.

\bigskip

The integration over $R_1$ can be derived from Eq.~\eqref{eq:phsp_int_noT} by letting $y_2$  vary between $y_1$ and $1 - 2 y_1$, and by imposing $y_1 < 1/3$. 
Furthermore, in this case, $\tau = 1- y_1 \leq \tau_{\mbox{\tiny MAX}}$.
Notice that in the $2$-jet limit the sub-region $R_1$ approaches the configurations where the quark is either soft or collinear to the antiquark.
Then, its contribution must be suppressed by some power of $\tau$.
In the following we will verify this power counting prediction.
Let us consider the $R_1$-contribution to the projections of the unsubtracted final state tensor in the $2$-jet limit.
In terms of the set of variables defined in Eq.~\eqref{eq:change_var} we have:
\begin{align}
&\frac{\alpha_S}{4 \pi}\left(-g_{\mu \nu}
\widehat{W}_{g,\,R_1}^{\mu\,\nu}{}^{\;[1]}\right) =
H_0 \,
\frac{\alpha_S}{4 \pi} \, 
2 \, C_F \, S_\epsilon \, 
\left( \frac{\mu}{Q} \right)^{2\epsilon} \, 
\delta\left( \tau - (1-z) \right) \,
\theta \left( \tau_{\mbox{\tiny MAX}} - \tau \right) 
\, \times 
\notag \\
&\quad \times \, 
\theta\left(z - \frac{2}{3}\right) \, 
\frac{1 + (1-z)^2 - \epsilon z^2}{z^2} \,
(1-z)^{-\epsilon} \,
\int_{\frac{1-z}{z}}^{2-\frac{1}{z}} \, 
d \alpha \, \alpha^{-1-\epsilon} \, (1-\alpha)^{1-\epsilon}  =
\notag \\
&\quad=
H_0 \,
\frac{\alpha_S}{4 \pi} \, 
2 \, C_F \, S_\epsilon \, 
\left( \frac{\mu}{Q} \right)^{2\epsilon} \, 
\delta\left( \tau - (1-z) \right) \,
\theta \left( \frac{1}{3} - \tau \right) 
\theta \left( \tau_{\mbox{\tiny MAX}} - \tau \right) \,\times 
\notag \\
&\quad \times \, 
\frac{1 + \tau^2 - \epsilon (1-\tau)^2}{(1-\tau)^2} \,
\tau^{-\epsilon} \, I_{1,\,-1}(\epsilon, \,\tau)
\label{eq:gluon_gW_R1_unapprox}
\end{align}
having defined the integrals:
\begin{align}
&I_{a,\,b}(\epsilon, \,\tau) = 
\int_{\frac{\tau}{1-\tau}}^{2 - \frac{1}{1-\tau}} \,
d \alpha \, \alpha^{-a-\epsilon} \, (1-\alpha)^{-b-\epsilon}, 
\quad \mbox{with } \mbox{Re~}{\epsilon}<\mbox{min}(-a+1,-b+1) .
\label{eq:R1_int_def}
\end{align}
Since $\tau_{\mbox{\tiny MAX}}$ is the maximum value that $\tau$ can reach, beyond which the final state goes into a $3$-jet configuration, we surely can consider $\tau_{\mbox{\tiny MAX}} < 1/3$ and drop the first theta function in Eq.~\eqref{eq:gluon_gW_R1_unapprox}.
Furthermore, since $\tau$ is limited to be small, in the interval $[0, \tau_{\mbox{\tiny MAX}}]$, we can take the following approximation, valid when $\tau_{\mbox{\tiny MAX}} \to 0$:
\begin{align}
&I_{a,\,b}(\epsilon, \,\tau) 
\stackrel{\tau_{\mbox{\tiny MAX}} \to 0}{\sim}
\frac{\tau^{-a+1 -\epsilon}}{a-1+\epsilon} +
B(1-a-\epsilon,\,1-b-\epsilon) +
\frac{\tau^{-b+1 -\epsilon}}{b-1+\epsilon},
\notag \\
&\quad \mbox{with } \mbox{Re~}{\epsilon}<\mbox{min}(-a+1,-b+1) .
\label{eq:R1_int_def_smallt}
\end{align}
which in our case means:
\begin{align}
&I_{1,\,-1}(\epsilon, \,\tau)
\stackrel{\tau_{\mbox{\tiny MAX}} \to 0}{\sim}
\frac{\tau^{-\epsilon}}{\epsilon} +
B(-\epsilon,\,2-\epsilon) +
\frac{\tau^{2 -\epsilon}}{2+\epsilon},
\quad \mbox{with } \mbox{Re~}{\epsilon}<0 .
\label{eq:R1_int_def_smallt_1}
\end{align}
Finally, the upper limit on $\tau$ translates into a lower limit on $z$, due to the Dirac delta. Hence we have $1-\tau_{\mbox{\tiny MAX}} \leq z \leq 1$. In the limit of vanishing cut-off, the contribution of the sub-region $R_1$ has to be proportional to $\delta(1-z)$. In fact, the integration with a test function $T(z)$ gives:
\begin{align}
\int_0^1 \, dz \, T(z) \, 
\delta\left( \tau - (1-z) \right) \,
\theta 
\left( \tau_{\mbox{\tiny MAX}} - \tau \right) = 
T(1)\, \theta 
\left( \tau_{\mbox{\tiny MAX}} - \tau \right) +
\mathcal{O}\left( \tau \right),
\label{eq:delta_R1_test}
\end{align}
therefore:
\begin{align}
\delta\left( \tau - (1-z) \right) \,
\theta \left( \tau_{\mbox{\tiny MAX}} - \tau \right)
\stackrel{\tau_{\mbox{\tiny MAX}} \to 0}{\sim}
\delta(1-z) \,
\theta \left( \tau_{\mbox{\tiny MAX}} - \tau \right)
\label{eq:delta_R1}
\end{align}
Finally, we have:
\begin{align}
&\frac{\alpha_S}{4 \pi}\left(-g_{\mu \nu} 
\widehat{W}_{g,\,R_1}^{\mu\,\nu}{}^{\;[1]} \right)
\stackrel{\tau_{\mbox{\tiny MAX}} \to 0}{\sim}
H_0 \,
\frac{\alpha_S}{4 \pi} \, 
2 \, C_F \, S_\epsilon \, 
\left( \frac{\mu}{Q} \right)^{2\epsilon} \, 
\delta\left( 1-z \right) \, 
\theta 
\left(  \tau_{\mbox{\tiny MAX}} - \tau \right) \, \times
\notag \\
&\quad \times \, 
\frac{1 + \tau^2 - \epsilon (1-\tau)^2}{(1-\tau)^2} \,
\tau^{-\epsilon} \, 
\left( \frac{\tau^{-\epsilon}}{\epsilon} +
B(-\epsilon,\,2-\epsilon) +
\frac{\tau^{2 -\epsilon}}{2+\epsilon}
\right),
\quad \mbox{with } \mbox{Re~}{\epsilon}<0 .
\label{eq:gluon_gW_R1_approx}
\end{align}
The above equation shows that the whole $2$-jet contribution of the sub-region $R_1$ is of order $\mathcal{O}(\tau^{-\epsilon})$. 
Since the limit $\tau \to 0$ has to be taken \emph{before} expanding in powers of $\epsilon$ and since $\mbox{Re~}{\epsilon}<0$, then the whole expression of Eq.~\ref{eq:gluon_gW_R1_approx} is suppressed in the $2$-jet approximation, as expected from power counting.

The other projection of the final state tensor gives an analogous result:
\begin{align}
&\frac{\alpha_S}{4 \pi}\left(\frac{k_{1\,\mu} k_{1,\,\nu}}{Q^2}
\widehat{W}_{g,\,R_1}^{\mu\,\nu}{}^{\;[1]}\right)=
H_0 \,
\frac{\alpha_S}{4 \pi} \, 
2 \, C_F \, S_\epsilon \, 
\left( \frac{\mu}{Q} \right)^{2\epsilon} \, 
\delta\left( \tau - (1-z) \right) \,
\theta \left( \tau_{\mbox{\tiny MAX}} - \tau \right) \,\times 
\notag \\
&\quad \times \, 
\theta\left(z - \frac{2}{3}\right) \, 
\frac{(1-z)^{-\epsilon}}{1-\epsilon} \,
\int_{\frac{1-z}{z}}^{2-\frac{1}{z}} \, 
d \alpha \, \alpha^{-\epsilon} \, (1-\alpha)^{1-\epsilon} \,
\left[ 
(1- \epsilon) \, z + 2 \,(1-z) \, (1-\alpha)
\right] = 
\notag \\
&\quad=
H_0 \,
\frac{\alpha_S}{4 \pi} \, 
2 \, C_F \, S_\epsilon \, 
\left( \frac{\mu}{Q} \right)^{2\epsilon} \, 
\delta\left( \tau - (1-z) \right) \,
\theta\left(\frac{1}{3} - \tau \right) \,
\theta \left( \tau_{\mbox{\tiny MAX}} - \tau \right) \, \times 
\notag \\
&\quad \times \,
\frac{\tau^{-\epsilon}}{1-\epsilon} \,
\left[
(1-\tau) \, (1-\epsilon) \, I_{0,\,-1}(\epsilon,\,\tau) +
2 \tau \, I_{0,\,-2}(\epsilon,\,\tau)
\right].
\label{eq:gluon_k2W_R1_unapprox}
\end{align}
Notice that the previous expression is regular for any $\epsilon$ and vanishes if $\mbox{Re~}{\epsilon} < 0$ when $\tau$ goes to zero.
In conclusion, $R_1$ does not contribute to the unsubtracted final state tensor in the $2$-jet limit.

Let's now consider the contribution of $R_2$.
In the $2$-jet limit, this region approaches the configuration where the quark is either soft or  collinear to the fragmenting gluon.
While the soft configuration has to be suppressed by some power of $\tau$, the contribution of the quark collinear to the gluon has to become larger and larger as $\tau$ goes to zero.
Starting from Eq.~\eqref{eq:phsp_int_noT}, if $y_1 > 1/3$, then $0 < y_2 < {(1-y_1)}/{2}$, otherwise if $y_1 < 1/3$, then $0 < y_2 < y_1$. In terms of the variables defined in Eq.~\eqref{eq:change_var} we have:
\begin{align}
&\frac{\alpha_S}{4 \pi}\left(-g_{\mu \nu} 
\widehat{W}_{g,\,R_2}^{\mu\,\nu}{}^{\;[1]}\right) =
H_0 \,
\frac{\alpha_S}{4 \pi} \, 
2 \, C_F \, S_\epsilon \, 
\left( \frac{\mu}{Q} \right)^{2\epsilon} \,  
\frac{(1-z)^{-\epsilon}}{z} \,
\frac{1 + (1-z)^2 - \epsilon z^2}{z^2} \, 
\theta \left( \tau_{\mbox{\tiny MAX}} - \tau \right) \,
\times
\notag \\
&\quad \times \,
\left[
\theta\left( \frac{2}{3} - z \right) \, \int_0^{\frac{1}{2}} + \,
\theta\left( z - \frac{2}{3} \right) \, \int_0^{\frac{1-z}{z}}
\right] \,
d \alpha \, \alpha^{-1-\epsilon} \, (1-\alpha)^{1-\epsilon} \,
\delta \left( \alpha - \frac{\tau}{z} \right) =
\notag \\
&\quad=
H_0 \,
\frac{\alpha_S}{4 \pi} \, 
2 \, C_F \, S_\epsilon \, 
\left( \frac{\mu}{Q} \right)^{2\epsilon} \,  
\frac{1 + (1-z)^2 - \epsilon z^2}{z^2} \,
\frac{(1-z)^{-\epsilon}}{z} \,
\theta \left( \tau_{\mbox{\tiny MAX}} - \tau \right)
\, \times
\notag \\
&\quad \times
\left[
\theta\left( \frac{2}{3} - z \right) \, 
\theta\left( z - 2\tau \right) + \,
\theta\left( z - \frac{2}{3} \right) \, 
\theta\left( z - \tau \right) - \,
\theta\left( z - \frac{2}{3} \right) \, 
\theta\left( z - (1-\tau) \right)
\right] \,\times
\notag \\
&\quad \times
\left(\frac{\tau}{z}\right)^{-1-\epsilon} \,
\left(1-\frac{\tau}{z}\right)^{1-\epsilon}.
\label{eq:gluon_gW_R2_unapprox}
\end{align}
Since $z>0$, we can expand $\left(1-{\tau}/{z}\right)^{1-\epsilon}$ in powers of $\tau$. Furthermore, the theta functions can be approximated as:
\begin{align}
&\theta \left( \tau_{\mbox{\tiny MAX}} - \tau \right) \,
\left[
\theta\left( \frac{2}{3} - z \right) \, 
\theta\left( z - 2\tau \right) + \,
\theta\left( z - \frac{2}{3} \right) \, 
\theta\left( z - \tau \right) 
\right] 
\stackrel{\tau_{\mbox{\tiny MAX}} \to 0}{\sim}
\notag \\
&\quad
\stackrel{\tau_{\mbox{\tiny MAX}} \to 0}{\sim}
\theta(1-z) \,
\theta \left( \tau_{\mbox{\tiny MAX}} - \tau \right) ;
\label{eq:R2_theta_appr_1}
\\
&\theta \left( \tau_{\mbox{\tiny MAX}} - \tau \right) \,
\theta\left( z - \frac{2}{3} \right) \, 
\theta\left( z - (1-\tau) \right) 
\stackrel{\tau_{\mbox{\tiny MAX}} \to 0}{\sim}
\notag \\
&\quad
\stackrel{\tau_{\mbox{\tiny MAX}} \to 0}{\sim}
\theta(1-z) \, \theta\left(z-(1-\tau)\right) \,
\theta \left( \tau_{\mbox{\tiny MAX}} - \tau \right).
\label{eq:R2_theta_appr_2}
\end{align}
Notice that the approximation in Eq.~\eqref{eq:R2_theta_appr_2} forces $z$ to be very close to $1$ as $\tau_{\mbox{\tiny MAX}}$ goes to zero, resulting in a configuration in which the quark is soft.
The factor $(1-z)^{-\epsilon}$ suppresses its contribution as $\mbox{Re~}{\epsilon} < 0$, confirming the power counting prediction.
Finally:
\begin{align}
&\frac{\alpha_S}{4 \pi}\left(-g_{\mu \nu} 
\widehat{W}_{g,\,R_2}^{\mu\,\nu}{}^{\;[1]}\right)
\stackrel{\tau_{\mbox{\tiny MAX}} \to 0}{\sim}
\notag \\
&\quad 
H_0 \,
\Bigg(
\frac{\alpha_S}{4 \pi} \, 2 \, C_F  \, S_\epsilon \, 
\left( \frac{\mu}{Q} \right)^{2\epsilon} \theta(1-z)\,
\left(\frac{1-z}{z}\right)^{-\epsilon} \,
\frac{1 + (1-z)^2 - \epsilon z^2}{z^2} \,
\tau^{-1-\epsilon} \Bigg) \,
\theta \left( \tau_{\mbox{\tiny MAX}} - \tau \right)
=
\notag \\
&\quad=
H_0 \,
\frac{\alpha_S}{4 \pi} \,
J_{g/q}^{\;[1]}\left( \epsilon;\,\tau,\,z \right) \, 
\theta \left( \tau_{\mbox{\tiny MAX}} - \tau \right)
, \quad \mbox{Re~}{\epsilon}<0,
\label{eq:gluon_gW_R2}
\end{align}
where $J_{g/q}^{\;[1]}$ is the 1-loop jet thrust function associated to the fragmenting gluon, divergent in $\tau = 0$ and defined in Eq.~\eqref{eq:jetgluon_thrust}.
The other projection gives a contribution of $\mathcal{O}(\tau^{-\epsilon})$ and hence it is suppressed in the $2$-jet limit as $\mbox{Re~}{\epsilon} < 0$.

The final contribution comes from the $R_3$ region.
Fig.~\ref{fig:phase_space_real} shows that this region reaches the 2-jet configuration (red dashed bands) only in the wedges very close to the axes, hence it is suppressed compared to  $R_1$ and $R_2$ in the $2$-jet limit.
The projections give:
\begin{align}
&\frac{\alpha_S}{4 \pi}\left(-g_{\mu \nu} 
\widehat{W}_{g,\,R_3}^{\mu\,\nu}{}^{\;[1]}\right) = 
H_0 \,
\frac{\alpha_S}{4 \pi} \, 
2 \, C_F \, S_\epsilon \, 
\left( \frac{\mu}{Q} \right)^{2\epsilon} \,
\frac{(1-z)^{-\epsilon}}{z} \,
\frac{1 + (1-z)^2 - \epsilon z^2}{z^2} \, 
\theta \left( \tau_{\mbox{\tiny MAX}} - \tau \right) \,
\times
\notag \\ 
&\quad \times
\left[
\theta\left( \frac{2}{3} - z \right) \, \int_{\frac{1}{2}}^1 + \,
\theta\left( z - \frac{2}{3} \right) \, \int_{2-\frac{1}{z}}^1
\right] \,
d \alpha \, \alpha^{-1-\epsilon} \,
(1-\alpha)^{1-\epsilon} \,
\delta \left( \alpha - (1-\frac{\tau}{z}) \right) .
\label{eq:gluon_gW_R3_unapprox}
\end{align}
Terms as $\left(1-{\tau}{z}\right)^{-n-\epsilon}$ can be expanded in powers of $\tau$, therefore the expression in the previous equation is suppressed in the limit of small $\tau$.
The same happens for the other projection.
Therefore, we can neglect the whole contribution of this region.

Finally, the only non zero contribution to the unsubtracted final state tensor associated to the fragmenting gluon in the $2$-jet limit is given by region $R_2$ in Eq.~\eqref{eq:gluon_gW_R2}:
\begin{align}
\widehat{W}_{g}^{\mu\,\nu}{}^{\;[1]}(\epsilon;\,z,\,\tau;\,\tau_{\mbox{\tiny MAX}})
\stackrel{\tau_{\mbox{\tiny MAX}} \to 0}{\sim}
H_T^{\mu\,\nu}(z) \,\widehat{F}_g^{[1]}(\epsilon;\,z,\,\tau;\,\tau_{\mbox{\tiny MAX}}),
\quad \mbox{Re~}{\epsilon}<0,
\label{eq:unsubW_gluon}
\end{align}
where the partonic structure function $\widehat{F}_g^{[1]}$ is:
\begin{align}
\widehat{F}_g^{[1]}(\epsilon;\,z,\,\tau;\,\tau_{\mbox{\tiny MAX}}) = 
\frac{H_0}{2} \,
J_{g/q}^{\;[1]}\left( \epsilon;\,\tau,\,z \right) \, 
\theta \left( \tau_{\mbox{\tiny MAX}} - \tau \right), \quad \mbox{Re~}{\epsilon}<0.
\label{eq:F1_gluon_1loop}
\end{align}
As expected, $\widehat{W}_{g}^{\mu\,\nu}$ ($\widehat{F}_g$) is proportional to the thrust jet function of the fragmenting gluon, equipped with a cut-off $\tau_{\mbox{\tiny MAX}}$ that constrains it to 
a $2$-jet-like final state topology. The expression in Eq.~\eqref{eq:unsubW_gluon} has been computed to 1-loop accuracy, but the computation could be repeated in a perfectly analogous way at any order in perturbation theory.

The effect of the combination of the sharp cut-off $\tau_{\mbox{\tiny MAX}}$ and the function $J_{g/q}$ can be obtained equivalently by allowing for a slight modification of the original definition of the thrust jet function of the fragmenting gluon given in  Eq.~\eqref{eq:jetgluon_thrust}.
Such definition involves a delta function that relates the value of $\tau$ to the transverse momentum $\vec{k}_T$ of the fermion (e.g. the quark) that flows collinearly to the fragmenting gluon. If this transverse momentum is small (compared to $Q$), $T$ is close to $1$ ensuring a  $2$-jet-like final state topology.
However, the original definition also involves an integration over $\vec{k}_T$ that stretches the jet function well beyond the region where it is defined\footnote{Actually, the integration does not extend to the whole spectrum of transverse momenta, since it must satisfy the implicit condition $\tau \leq 1$, i.e. $T>0$. The maximum value of $k_T$ is of order $Q$ and, to 1-loop, it is $k_T^{\mbox{\tiny MAX}\;[1]} = Q \, \sqrt{{(1-z)}/{z}}$.}.
Therefore, if $k_T$ is not allowed to become too large, then the function $J_{g/q}$ is strictly limited to describe 
the $2$-jet limit of the final state. 
This can be achieved by introducing 
a cut-off that constrains 
the integration range to the quasi $2$-jet configuration by 
setting an upper limit for the transverse momentum $k_T$.
Indeed, the new definition of $J_{g/q}$ will have to coincide to its original definition, Eq.~\eqref{eq:jetgluon_thrust}, at small values of $\tau$.
A proper cut-off for $k_T$ is the power counting energy scale $\lambda << Q$, used to size the collinear momenta.
With this choice, we have:
\begin{align}
&\frac{\alpha_S}{4 \pi} J_{g/q}^{[1],\,(\lambda)}(\epsilon;\,\tau, \, z) = 
\frac{\alpha_S}{4 \pi} \, 
2 \, C_F \, S_\epsilon \,
\theta(1-z)\,
\frac{1+(1-z)^2-\epsilon z^2}{z^2} \, 
\times
\notag \\
&\quad \times \,
\frac{\Gamma(1-\epsilon)}{\pi^{1-\epsilon}} \,
\mu^{2 \epsilon} \, 
\int d^{2-2\epsilon} \vec{k}_T \,
\frac{1}{k_T^2} \,
\delta\left(\tau - \frac{z}{1-z} \,\frac{k_T^2}{Q^2}\right) \,
\theta\left( \lambda^2 - k_T^2 \right)= 
\notag \\
&\quad=
\frac{\alpha_S}{4 \pi} \, 
2 \, C_F \, S_\epsilon \,
\left( \frac{\mu}{Q} \right)^{2 \epsilon} \, 
\times \notag \\
&\quad \times \,
\theta(1-z)\,
\left(\frac{z}{1-z}\right)^\epsilon \,
\frac{1+(1-z)^2-\epsilon z^2}{z^2} \,
\tau^{-1-\epsilon} \,
\theta \left( 
\frac{\lambda^2}{Q^2}-\frac{1-z}{z} \tau
\right) = 
\notag \\
&\quad=
\frac{\alpha_S}{4 \pi}
J_{g/q}^{[1]}(\epsilon;\,\tau, \, z) \,
\theta \left( 
\frac{\lambda^2}{Q^2}-\frac{1-z}{z} \tau
\right).
\label{eq:jetgluon_thrust_lambda}
\end{align}
We can relate this expression with that appearing in Eq.~\eqref{eq:unsubW_gluon}. In fact:
\begin{align}
\frac{\lambda^2}{Q^2}-\frac{1-z}{z} \tau > 0 
\Rightarrow
\begin{cases}
0 \leq z \leq \frac{1}{1+{\lambda^2}/{Q^2}}
\mbox{ and }
0 \leq \tau \leq \frac{z}{1-z} \, \frac{\lambda^2}{Q^2} \equiv 
\tau_{\mbox{\tiny MAX}}(\lambda)
\vspace{.7cm}
\\
\frac{1}{1+{\lambda^2}/{Q^2}} \leq z \leq 1
\mbox{ and }
0 \leq \tau \leq 1
\end{cases}
\label{eq:lambda_tmax}
\end{align}
%
%
\begin{figure}[t]
\centering
\includegraphics[width=9cm]{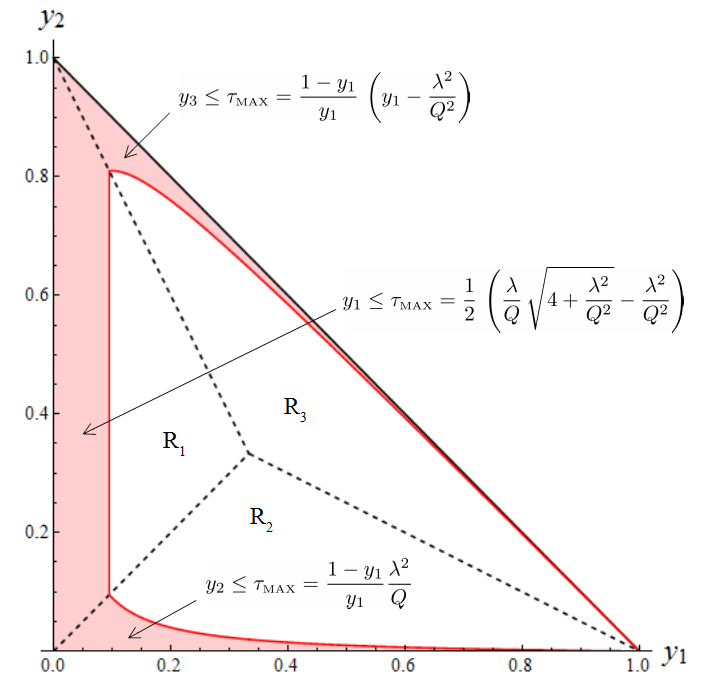}
\caption{Action of the cut-off $\tau_{\mbox{\tiny MAX}}(\lambda)$ on the phase space for the three final state particles, according to Eq.~\eqref{eq:lambda_tmax}. In this case $\lambda = 0.1 \, Q$. Notice that the red region shrinks as ${\lambda}/{Q} \to 0$.}
\label{fig:tmaxPlot}
\end{figure}
%
Notice that this choice sets $\tau_{\mbox{\tiny MAX}} = 1$ only in a thin slice of the phase space, where $z$ is very close to $1$. 
Furthermore, $\tau_{\mbox{\tiny MAX}} \to 0$
consistently implies ${\lambda^2}/{Q^2} \to 0$ (as showed in Fig.~\ref{fig:tmaxPlot}) and we can rewrite Eqs.~\eqref{eq:unsubW_gluon} and~\eqref{eq:F1_gluon_1loop} as follows:
\begin{align}
&\widehat{W}_{g}^{\mu\,\nu}{}^{\;[1]}(\epsilon;\,z,\,\tau;\,\tau_{\mbox{\tiny MAX}} \to 0) \equiv
\notag \\
&\quad \equiv
\widehat{W}_{g}^{\mu\,\nu}{}^{\;[1]}(\epsilon;\,z,\,\tau;\,{\lambda^2}/{Q^2} \to 0) = 
H_T^{\mu\,\nu}(z) \,
\widehat{F}_g^{[1]}(\epsilon;\,z,\,\tau;\,
{\lambda^2}/{Q^2} \to 0),
\quad \mbox{Re~}{\epsilon}<0.
\label{eq:unsubW_gluon_lambda}
\end{align}
and:
\begin{align}
\widehat{F}_g^{[1]}(\epsilon;\,z,\,\tau;\,
{\lambda^2}/{Q^2} \to 0) = 
\frac{H_0}{2} \,
J_{g/q}^{[1],\,(\lambda)}
\left(\epsilon;\,\tau, \, z\right),
\quad \mbox{Re~}{\epsilon}<0.
\label{eq:F1_gluon_1loop_lambda}
\end{align}
The $\epsilon$-expansion of the previous expressions is not straightforward.
The thrust jet function for the fragmenting gluon as defined in Section~\ref{app:jet_thrust_fun} can easily be expanded in powers of $\epsilon$ by using Eq.~\eqref{eq:thrust_distr_exp}, however in this case the presence of an explicit cut-off makes the computation a bit trickier. 
In fact, a direct application of Eq.~\eqref{eq:thrust_distr_exp} gives:
\begin{align}
\tau^{-1-\epsilon} \,
\theta \left( \frac{\lambda^2}{Q^2} - \frac{1-z}{z} \tau\right) = 
-\frac{1}{\epsilon} \, \delta(\tau) + 
\left( \frac{1}{\tau} \right)_+ \,
\theta \left( \frac{\lambda^2}{Q^2} - \frac{1-z}{z} \tau\right) +
\mathcal{O}\left(\epsilon\right).
\label{eq:thrust_distr_exp_lambda_1}
\end{align}
The interplay between the plus distribution and the theta function requires some extra care, especially in the limit of ${\lambda^2}/{Q^2} \to 0$. 
Being a distribution, it is best studied by considering its action on a test function $T(\tau)$.
In order to make the following expressions more readable, we define $r = {\lambda}/{Q}$.
Then, by definition of plus distribution and by using Eq.~\eqref{eq:lambda_tmax}, we have:
\begin{align}
&\int_0^1 \, d\tau \, T(\tau) \,
\left( \frac{1}{\tau} \right)_+ \,
\theta \left( r^2 - \frac{1-z}{z} \tau\right) = 
\notag \\
&\quad=
\theta \left( \frac{1}{1+r^2} - z \right) \,
\left[ 
\int_0^{\frac{z}{1-z} r^2} \, d\tau \, 
\frac{T(\tau) - T(0)}{\tau} + 
T(0) \, \log{\frac{z}{1-z} r^2}
\right] +
\, \notag \\
&\quad+ \,
\theta \left( z - \frac{1}{1+r^2} \right) \,
\int_0^1 \, d\tau \, 
\frac{T(\tau) - T(0)}{\tau} = 
\notag \\
&\quad=
T(0) \, 
\theta \left( \frac{1}{1+r^2} - z \right) \,
\log{\frac{z}{1-z} r^2} \, + 
\notag \\
&\quad +
\left[
\int_0^1 \, d\tau - 
\theta \left( \frac{1}{1+r^2} - z \right) \,
\int_{\frac{z}{1-z} r^2}^1 \, d\tau
\right] \, 
\frac{T(\tau) - T(0)}{\tau}.
\label{eq:plus_theta_int}
\end{align}
The last line (term in square brackets) of the previous expression is suppressed in the limit of $r \to 0$. In fact, both integrals are well defined thanks to the plus distribution prescription; this allows to perform an expansion in powers of $r^2$, where the first non-zero term appears at order $\mathcal{O}\left(r^2\right)$. Therefore it can be neglected, leading to the final result:
\begin{align}
&\tau^{-1-\epsilon} \,
\theta \left( \frac{\lambda^2}{Q^2} - \frac{1-z}{z} \tau\right) = 
\notag \\
&\quad=
\left[
-\frac{1}{\epsilon} \, + 
\theta \left( \frac{1}{1+{\lambda^2}/{Q^2}} - z \right) \,
\log{\frac{z}{1-z} \frac{\lambda^2}{Q^2}}
\right] \, \delta(\tau) +
\notag \\
&\quad+
\mathcal{O}\left(\frac{\lambda^2}{Q^2}\right) + 
\mathcal{O}\left(\epsilon\right),
\quad \mbox{Re~}{\epsilon}<0.
\label{eq:thrust_distr_exp_lambda_2}
\end{align}
Integration of the above expression over $\tau$ leads to the same result obtained by integrating the non-expanded initial function, apart from power suppressed corrections. In other words, by using Eq.~\eqref{eq:thrust_distr_exp_lambda_2} the operations of integration and the operation of $\epsilon$-expansion commute.
Then, we can write the thrust jet function of the fragmenting gluon, equipped with the cut-off as in Eq.~\eqref{eq:jetgluon_thrust_lambda}, as follows:
\begin{align}
&\frac{\alpha_S}{4 \pi}
J_{g/q}^{[1],\,(\lambda)}
(\epsilon;\,\tau, \, z) =
\notag \\
&\quad=
\frac{\alpha_S}{4 \pi} \, 
2 \, C_F \, S_\epsilon \,
\theta(1-z)\,
\frac{1+(1-z)^2-\epsilon z^2}{z^2} \,
\times
\notag \\
&\quad \times 
\left[
-\frac{1}{\epsilon} -
\log{\frac{\mu^2}{Q^2}} + 
\log{\frac{1-z}{z}} + 
\theta \left( 
\frac{1}{1+\frac{\lambda^2}{Q^2}} - z \right) \,
\log{\frac{z}{1-z} \frac{\lambda^2}{Q^2}}+
\mathcal{O}\left(\epsilon\right)
\right] \, \delta(\tau) +
\notag \\
&\quad +
\mathcal{O}\left(\frac{\lambda^2}{Q^2}\right) + 
\mathcal{O}\left(\epsilon\right),
\quad \mbox{Re~}{\epsilon}<0.
\label{eq:jetgluon_thrust_lambda_exp_1}
\end{align}
The previous expression can be further simplified, in fact the combination of the logarithmic  terms can be rewritten as:
\begin{align}
&\log{\frac{1-z}{z}} + 
\theta \left( 
\frac{1}{1+\frac{\lambda^2}{Q^2}} - z \right) \,
\log{\frac{z}{1-z} \frac{\lambda^2}{Q^2}} = 
\notag \\
&\quad = 
\theta \left( 
\frac{1}{1+\frac{\lambda^2}{Q^2}} - z \right) \,
\log{\frac{\lambda^2}{Q^2}} + 
\theta \left( 
z - \frac{1}{1+\frac{\lambda^2}{Q^2}} \right) \,
\log{\frac{1-z}{z}} = 
\log{\frac{\lambda^2}{Q^2}} + 
\mathcal{O}\left(\frac{\lambda^2}{Q^2}\right).
\label{eq:combo_logs_theta}
\end{align}
This result follows by considering the combination of logarithms in the first line of the previous expression as a distribution of $z$ and then integrating it with a test function $T(z)$. Since all the integrals are well defined, they can be expanded in powers of ${\lambda^2}/{Q^2}$ to give the final result of Eq.~\eqref{eq:combo_logs_theta}. Then, finally:
\begin{align}
&\frac{\alpha_S}{4 \pi}
J_{g/q}^{[1],\,(\lambda)}
(\epsilon;\,\tau, \, z) =
\frac{\alpha_S}{4 \pi} \, 
2 \, C_F \, S_\epsilon \,
\theta(1-z)\,
\frac{1+(1-z)^2-\epsilon z^2}{z^2} \,
\times
\notag \\
&\quad \times 
\left[
-\frac{1}{\epsilon} -
\log{\frac{\mu^2}{\lambda^2}} +
\mathcal{O}\left(\frac{\lambda^2}{Q^2}\right) + 
\mathcal{O}\left(\epsilon\right)
\right] \, \delta(\tau) +
\mathcal{O}\left(\frac{\lambda^2}{Q^2}\right) + 
\mathcal{O}\left(\epsilon\right),
\quad \mbox{Re~}{\epsilon}<0.
\label{eq:jetgluon_thrust_lambda_exp_2}
\end{align}

\bigskip

%
\subsubsection{Subtraction Mechanism \label{subsubsec:frag_gluon_sub}}

\bigskip

The unsubtracted final state tensor found in the previous section describes the fragmentation of a gluon resulting from a $\epm$ scattering at partonic level. However, this information is also encoded in the TMD FF of the gluon, which appears in the final cross section convoluted with the partonic cross section obtained in Eq.~\eqref{eq:unsubW_gluon}.
Therefore, $W_{g}^{\mu\,\nu}{}^{\;[1]}$ must be appropriately subtracted in order to remove all  contributions that overlap with the partonic version of the TMD FF of the gluon, in order to avoid double counting.
This subtraction will be performed in the $b_T$-space, where TMDs are defined explicitly in terms of operators, see e.g.  Refs.~\cite{Collins:2011zzd,Boglione:2020cwn}.

\bigskip

In momentum space, the partonic version of the 1-loop gluon-from-quark TMD FF is given by
(see e.g. Ref.~\cite{Collins:2011zzd}):
\begin{align}
\frac{\alpha_S}{4 \pi}
D_{g/q}^{[1]}(\epsilon;\,z,\,k_T) = 
\frac{\alpha_S}{4 \pi} \, 
2 \, C_F \, S_\epsilon \,
\frac{\Gamma(1-\epsilon)}{\pi^{1-\epsilon}} \, 
\mu^{2 \epsilon} \, \frac{1}{k_T^2} \,
\theta(1-z)\,
\frac{1+(1-z)^2-\epsilon z^2}{z^3}.
\label{eq:gluon_TMD_kT}
\end{align}
In principle, its Fourier transform involves an integral over the whole spectrum of $k_T$ transverse momenta, from zero to infinity, going far beyond the actual region of overlapping.
However, the presence of the cut-off $\lambda$ ensures that the Fourier transform of the gluon TMD FF, Eq.~\eqref{eq:gluon_TMD_kT}, will overlap with the final state tensor only
up to $\lambda$, matching the same range of $k_T$ that is involved in the thrust jet function of the fragmenting gluon equipped with the cut-off (see Eq.~\eqref{eq:jetgluon_thrust_lambda}). 
The incomplete Fourier transform of Eq.~\eqref{eq:gluon_TMD_kT} is computed with the help of the following expression:
\begin{align}
&\int d^{2-2\epsilon} \vec{k}_T \,
e^{i \vec{k}_T \cdot \vec{b}_T} \,
\mu^{2\epsilon} \, 
\frac{1}{k_T^2} \,
\theta \left( \lambda^2 - k_T^2 \right) = 
\notag \\
&\quad =
\frac{\pi^{-1-\epsilon}}{\Gamma(1-\epsilon)} \,
\left( \frac{\mu^2}{\lambda^2} \right)^\epsilon \,
\frac{\Gamma(-\epsilon)}{\Gamma(1-\epsilon)} \,
{}_1 F_2 \left( 
-\epsilon; \,1-\epsilon,\,1-\epsilon;\,
-\frac{\lambda^2 b_T^2}{4}
\right) =
\notag \\
&\quad=
\frac{\pi^{-1-\epsilon}}{\Gamma(1-\epsilon)} \,
\left[
-\frac{1}{\epsilon} - 
\log{\frac{\mu^2}{\lambda^2}} -
\frac{\lambda^2}{Q^2} \, 
\frac{\left(b_T Q\right)^2}{4} \,
{}_2 F_3 \left( 
1,\,1; 2,\,2,\,2;\,
-\frac{\lambda^2}{Q^2} \,
\frac{\left(b_T Q\right)^2}{4}
\right) + 
\mathcal{O}\left(\epsilon\right)
\right] = 
\notag \\
&\quad =
\frac{\pi^{-1-\epsilon}}{\Gamma(1-\epsilon)} \,
\left[
-\frac{1}{\epsilon} - 
\log{\frac{\mu^2}{\lambda^2}} +
\mathcal{O}\left(\frac{\lambda^2}{Q^2}\right)+
\mathcal{O}\left(\epsilon\right)
\right] , \quad \mbox{Re~}{\epsilon}<0.
\label{eq:FT_kT2_cutoff}
\end{align}
Then, we define the Fourier transform of the gluon-from-quark partonic TMD FF, equipped with the cut-off $\lambda$, as obtained from the incomplete Fourier transform. This gives:
\begin{align}
&\frac{\alpha_S}{4 \pi}
\widetilde{D}_{g/q}^{[1],\,(\lambda)}
(\epsilon;\,z) = 
\frac{\alpha_S}{4 \pi} \, 
2 \, C_F \, S_\epsilon \,
\frac{1+(1-z)^2-\epsilon z^2}{z^3} \, \times
\notag \\
&\quad \times
\left[
-\frac{1}{\epsilon} - 
\log{\frac{\mu^2}{\lambda^2}} +
\mathcal{O}\left(\frac{\lambda^2}{Q^2}\right)+
\mathcal{O}\left(\epsilon\right)
\right], \quad \mbox{Re~}{\epsilon}<0.
\label{eq:gluon_TMD_cutoff}
\end{align}
The order-by-order formula for the subtracted, renormalized partonic cross section is given in 
Eq.~\eqref{eq:sub_mech_W}.
Notice that the pole in Eq.~\eqref{eq:gluon_TMD_cutoff} embodies the collinear divergence associated to the TMD FF.
In fact, in the case of a fragmenting gluon, the TMD FF is not UV divergent and hence no UV counterterm has to be added to Eq.~\eqref{eq:gluon_TMD_cutoff} in order to obtain a renormalized quantity.
Therefore, the final expression for the partonic cross section has to be subtracted but not renormalized.
Its expression follows from the 1-loop version of Eq.~\eqref{eq:sub_mech_W}:
\begin{align}
&
\left.
\widehat{W}_g^{\mu \nu,\,[1]}
(z,\,\tau,\,\lambda )
\right \rvert_{\mbox{\small sub}}= 
\widehat{W}_{g}^{\mu\,\nu}{}^{\;[1]}(\epsilon;\,z,\,\tau;\,{\lambda^2}/{Q^2} \to 0) - 
\notag \\
&-
\sum_k \, 
\int_z^1 \frac{d\widehat{z}}{\widehat{z}}\;
\widehat{W}_k^{\mu\nu,\,[0]}
({z}/{\widehat{z}},\tau )
\,\left[ 
\widehat{z} \, 
\widetilde{D}_{g/q}^{[1],\,(\lambda)} 
(\epsilon; \, \widehat{z})
\right],
\label{eq:sub_mech_W_1loop}
\end{align}
in which Eqs.~\eqref{eq:unsubW_gluon_lambda},~\eqref{eq:W_lo_tensor} and~\eqref{eq:gluon_TMD_cutoff} have to be used.
Computations are easier by contracting both sides of Eq.~\eqref{eq:sub_mech_W_1loop} with the metric tensor $g_{\mu\nu}$. This gives:
\begin{align}
&
\left.
\widehat{F}_{1,\,g}^{[1]}
(z,\,\tau,\,\lambda )
\right \rvert_{\mbox{\small sub}}= 
\notag \\
&\quad=
\widehat{F}_{1,\,g}^{[1]}
(\epsilon;\,z,\,\tau;\,{\lambda^2}/{Q^2} \to 0) - 
\sum_k \, 
\int_z^1 \frac{d\widehat{z}}{\widehat{z}}\;
\widehat{F}_{1,\,g}^{[0]}
({z}/{\widehat{z}},\tau )
\,\left[ 
\widehat{z} \, 
\widetilde{D}_{g/q}^{[1],\,(\lambda)} 
(\epsilon; \, \widehat{z})
\right] =
\notag \\
&\quad=
\frac{H_0}{2} \,
\left[
J_{g/q}^{[1],\,(\lambda)}
\left(\epsilon;\,\tau, \, z\right)- 
z \,\delta(\tau) \, 
\widetilde{D}_{g/q}^{[1],\,(\lambda)} 
(\epsilon; \, z)
\right] =
\mathcal{O}\left(\frac{\lambda^2}{Q^2}\right).
\label{eq:F1_gluon_sub}
\end{align}
Then, ultimately the subtracted partonic cross section for the case of a fragmenting gluon is suppressed in the $2$-jet limit, i.e. when ${\lambda^2}/{Q^2} \to 0$:
\begin{align}
&\left.\frac{d \widehat{\sigma}^{[1]}_g}{dz \, dT} \right \rvert_{\mbox{\small sub}} =  
\sigma_B \, z \, 
\left.
\widehat{F}_{1,\,g}^{[1]}
(z,\,\tau,\,\lambda )
\right \rvert_{\mbox{\small sub}} = 
\mathcal{O}\left(\frac{\lambda^2}{Q^2}\right).
\label{eq:sub_xs_gluon}
\end{align}
Notice that, being $\lambda$ the IR energy scale that sizes the almost on-shell collinear momenta, 
our initial prediction about the power counting suppression of the fragmentation of a gluon is totally confirmed by Eq.~\eqref{eq:sub_xs_gluon}.
This also supports the intuition about 
a gluon-initiated jet in a $2$-jet-like final state: in such a topology, one jet is initiated by the quark, the other by the antiquark.
Finally, in Eq.~\eqref{eq:sub_xs_gluon} we can relate $\lambda$ directly to the measured value of thrust $\tau_{\mbox{\tiny meas.}}$. 
In fact, according to the approximation of thrust in the  $2$-jet limit, the measured thrust $\tau_{\mbox{\tiny meas.}}$ is well approximated by the sum of the invariant masses of the two jets (see Eq.~\eqref{eq:thrust_2jet}). According to power counting, such invariant masses are of order $\lambda^2$. Therefore, we simply have $\mathcal{O}(\tau_{\mbox{\tiny meas.}}) = \mathcal{O}\left({\lambda^2}/{Q^2}\right)$. 
This rather naive argument can be made more specific by exploting the definition of thrust (Eq.~\eqref{eq:thrust_def}) and the relation between the transverse momenta of the fragmenting parton and the detected hadron (Eq.~\eqref{eq:h_p_frame}). In fact, it is not difficult to prove that (see Ref.~\cite{Makris:2020ltr}):
\begin{align}
k_{h,\,T} \leq \frac{P_{p,\,T}}{z_h} \leq \sqrt{\tau_{\mbox{\tiny meas.}}} \, Q
\label{eq:kT_max_tmeas}
\end{align}
Therefore, the choice $\lambda^2 = \tau_{\mbox{\tiny meas.}} Q^2$ is supported by the kinematical bounds of the process in a total natural way.
Hence we can interpret the suppression of the partonic cross section in Eq.~\eqref{eq:sub_xs_gluon} as due to the topology of the final state of the $\epm$ scattering.

\bigskip

%
\subsection{Fragmenting Fermion \label{sec:frag_ferm}}

\bigskip

\begin{figure}[t]
 \centering 
\begin{tabular}{c@{\hspace*{3cm}}c}  
     \includegraphics[width=3.1cm]{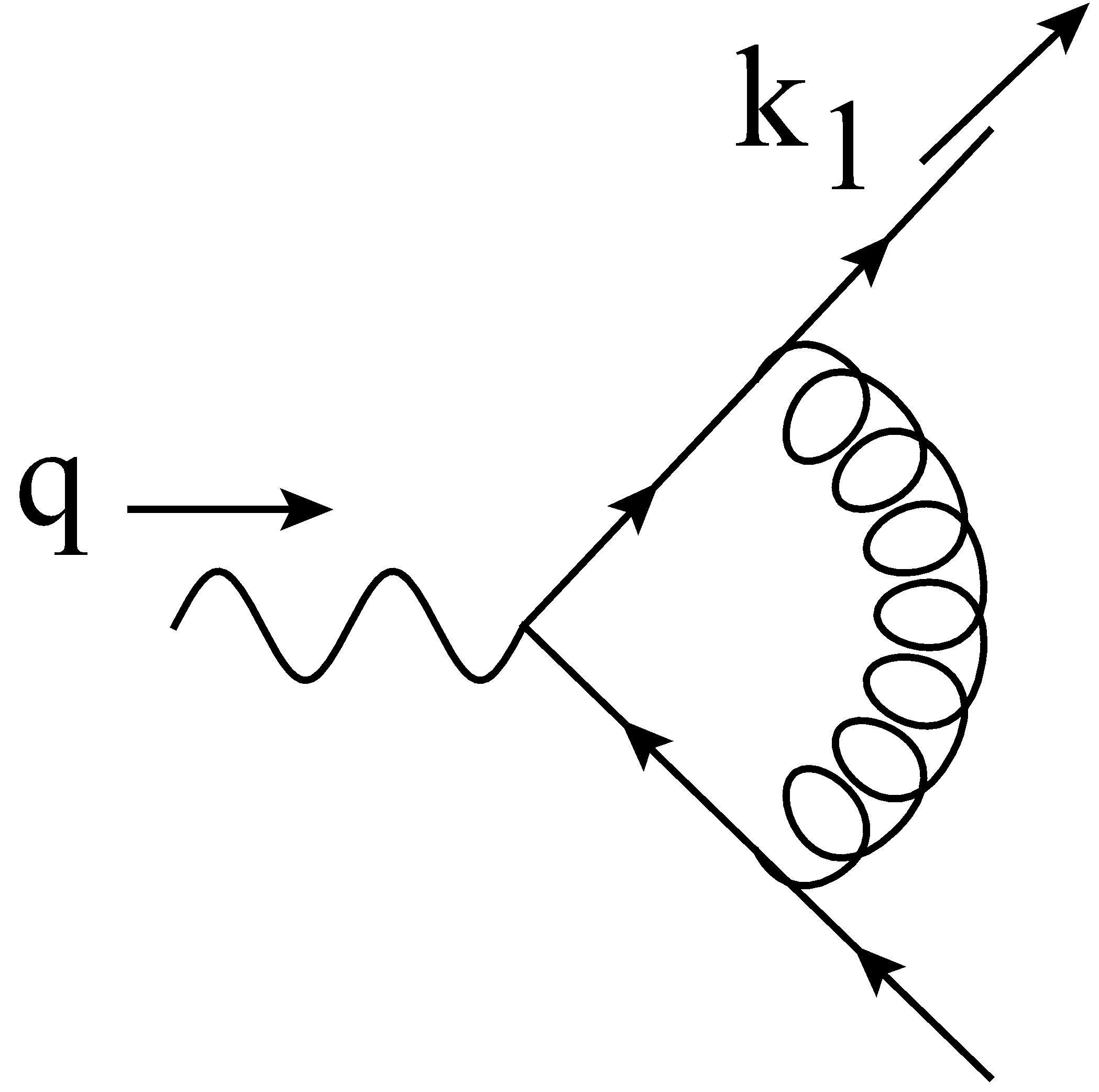}
&
     \includegraphics[width=3.5cm]{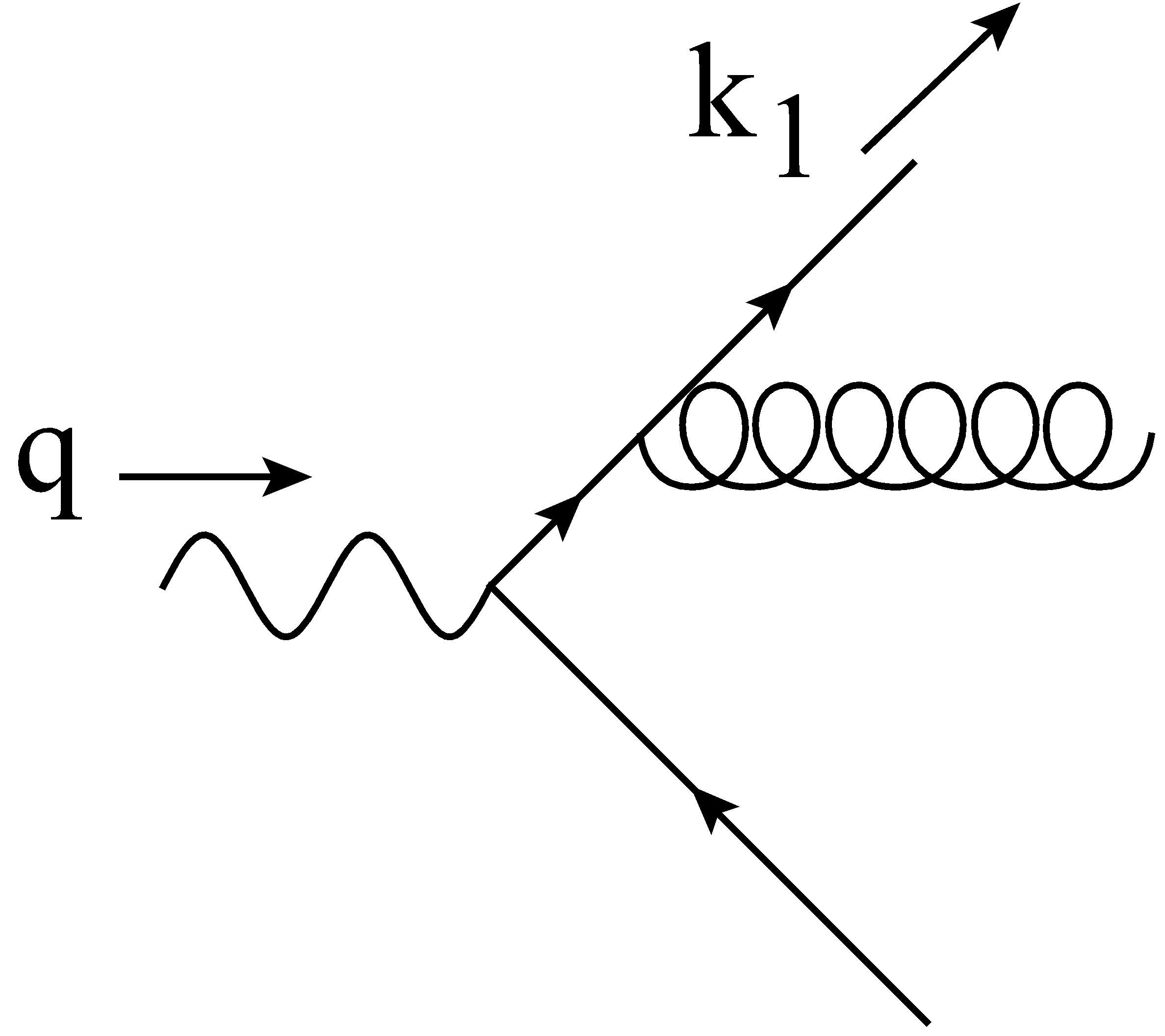}
 \\
 (a) & (b)
\end{tabular}
\caption{The 1-loop Feynman graphs contributing to $\widehat{W}^{\mu \nu}_f$, when the emitted gluon is virtual (a) and when it is real (b).}
\label{fig:frag_ferm}
\end{figure}

In this case, the detected hadron is produced by the fragmentation of a fermion of flavor $f$, which we will assume to be a quark.
The case of a fragmenting antiquark is totally analogous.
At 1-loop, one of the two fermionic legs emits a gluon, which can be either virtual or real, as represented by the Feynman graphs in Fig.~\ref{fig:frag_ferm}.
%
%
If the emitted gluon is virtual, then the final state topology is a perfect pencil-like event, as at LO (see Appendix~\ref{app:lo_partonic_xs}).
Instead, if the gluon is real, a $2$-jet-like configuration can be obtained only in the following three situations:
\begin{itemize}
\item The gluon is soft. 
Then the final state tensor is dominated by the soft thrust function $S$ as defined in Appendix~\ref{app:soft_thrust_fun}.
\item The gluon is collinear to the fermion which does not fragment (the antiquark in our case). Then we expect 
the final state tensor to be dominated by the backward thrust function $J_B$ as defined in Appendix~\ref{app:back_thrust_fun}.
\item The gluon is collinear to the fragmenting fermion. In this configuration the largest contribution to the final state tensor is given by the thrust jet function corresponding to the fragmenting quark $J_{q/q}$, as defined in Appendix~\ref{app:jet_thrust_fun}.
Notice, however, that this coincides to the configuration that has to be subtracted out in the final result, hence its computation has to be performed carefully.
\end{itemize}
All previous possibilities are allowed by power counting and all of them are expected to be dominant in the $2$-jet limit.
Note that the case of fragmenting fermion is way more complicated then the case of fragmenting gluon treated in Section.~\ref{sec:frag_gluon}, 
not only because it implies more squared matrix elements, but mostly due to the fact that the subtraction mechanism is made considerably more difficult by the presence of a rapidity cut-off in the partonic quark-from-quark TMD FF.
Moreover, the subtracted final state tensor will require the addition of a proper UV counterterm that renormalized its UV divergences.
We will deal with all those issues in the following sections.
In particular, in Section~\ref{subsubsec:frag_quark_unsub} we will present the explicit computation of the unsubtracted final state tensor $\widehat{W}^{\mu \nu}_f$, while in Section~\ref{subsubsec:frag_quark_sub} we will perform the subtractions that lead to the final expression for the NLO partonic cross section.
The resummed cross section will be presented in Section~\ref{subsubsec:evo_resumm}.

\bigskip

\subsubsection{Unsubtracted Final State Tensor \label{subsubsec:frag_quark_unsub}}

\bigskip

Similarly to the case of the fragmenting gluon, the unsubtracted final state tensor is obtained by applying the hard approximation ($T_H$ in Ref.~\cite{Collins:2011zzd}), which sets all the masses to zero and all the soft-collinear divergent quantities to their lowest order. According to the discussion at the beginning of Section~\ref{subsubsec:frag_gluon_unsub}, 
we will work in the Fourier conjugate space.
\paragraph{Virtual Emission.}
When the emitted gluon is virtual, the final state hosts two particles: the outgoing quark, of flavor $f$ and momentum $k_1$, and the antiquark crossing the final state cut, of momentum $k_2$.
Momentum conservation sets $q = k_1 + k_2$.
The 1-loop squared amplitude is given by:
\begin{align}
&M_{f,\,V}^{\mu\,\nu}{}^{\;[1]}(\epsilon;\,\mu,\,Q) = 
\begin{gathered}
\includegraphics[width=4.3cm]{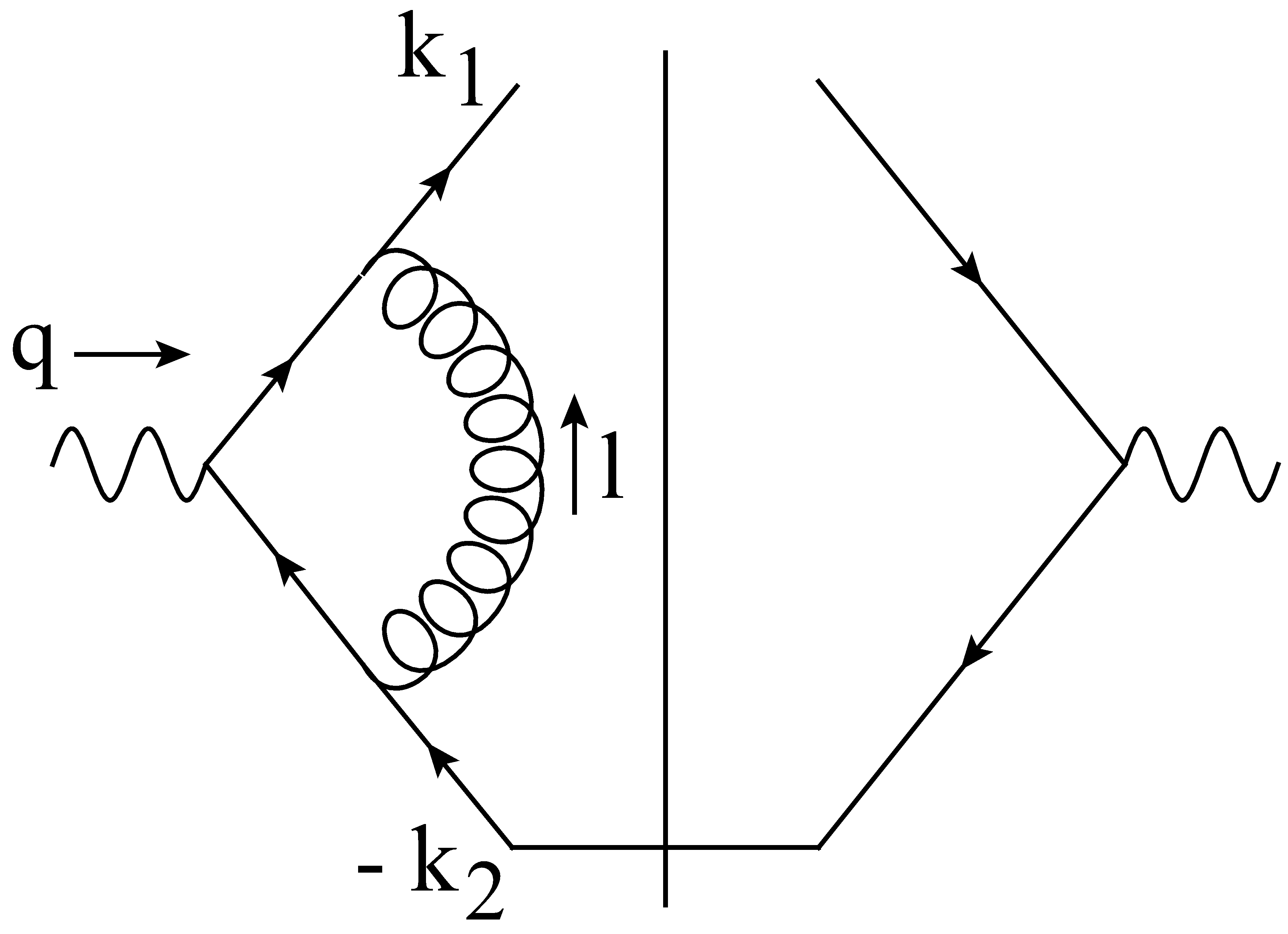}
\end{gathered}
+ h.c.
= 
\notag \\
&\quad= 
e_f^2 \, \int 
\frac{d^{4-2\epsilon} \, l}
{(2 \pi)^{4-2\epsilon}} \,
\Bar{u}(k_1) \, (-i g_0 \gamma^\alpha t^a) \, 
\frac{i (\slashed{k}_1 - \slashed{l})}{(k_1 - l)^2 + i \, 0} \,
\gamma^\mu \, 
\frac{i (-\slashed{k}_2 - \slashed{l})}{(k_2 + l)^2 + i \, 0} \,
\times \notag \\
&\quad \quad \times \,
(-i g_0 \gamma^\beta t^b) \, \slashed{k}_2 \,
\gamma^\nu \, u(k_1) \,
\frac{ -i \, g_{\alpha\,\beta} \, \delta_{a \, b}}{l^2 + i \, 0}
+ h.c.
= 
\notag \\
&\quad= 
i \, e_f^2 \, g^2 \, \mu^{2 \epsilon} \, C_F \, N_C \,
\int 
\frac{d^{4-2\epsilon} \, l}
{(2 \pi)^{4-2\epsilon}} \,
\frac{\mbox{Tr} \left \{ 
\slashed{k}_1 \, \gamma^\alpha \, (\slashed{k}_1 - \slashed{l}) \,
\gamma^\mu \, (\slashed{k}_2 + \slashed{l}) \, \gamma_\alpha \,
\slashed{k}_2 \, \gamma^\nu
\right \}}
{\left[ (k_1 - l)^2 + i \, 0 \right] \, 
\left[ (k_2 + l)^2 + i \, 0 \right] \,
\left[ l^2 + i \, 0 \right]} 
+ h.c.
\label{eq:M_V_1}
\end{align}
This expression can be properly simplified by decomposing the Dirac structure in its scalar, vector and tensor parts, by using momentum conservation and the Passarino-Veltman reduction formula~\cite{Passarino:1978jh}.
This leads to:
\begin{align}
&M_{f,\,V}^{\mu\,\nu}{}^{\;[1]}(\epsilon;\,\mu,\,Q) = M_f^{\mu\,\nu}{}^{\;[0]} \,
V^{[1]}(\epsilon;\,\mu,\,Q),
\label{eq:M_V_2}
\end{align}
which simply asserts that the 1-loop squared matrix element for the virtual emission of a gluon is the lowest order $M_f^{[0]}$, computed in Eq.~\eqref{eq:M_lo}, ``dressed" with the vertex factor $V$.
As a consequence, the corresponding contribution to the final state tensor will be simply proportional to the lowest order, computed in Eq.~\eqref{eq:g_W_lo}. Hence:
\begin{align}
W^{\mu \nu,\,[1]}_{f,\,V}
(\epsilon;\,z,\,\tau,\,\mu,\,Q) = 
H_T^{\mu \nu}(z) \,  \widehat{F}_{1,\,f}^{[0]}(z,\,\tau) \, 
V^{[1]}(\epsilon;\,\mu,\,Q).
\label{eq:W_ferm_virtual}
\end{align}
The 1-loop vertex factor is given by:
\begin{align}
&\frac{\alpha_S}{4 \pi}
V^{[1]}(\epsilon;\,\mu,\,Q) = 
i g^2 \, \mu^{2 \epsilon} \, C_F \,
\Bigg\{
-4(1-\epsilon)^2 \, \frac{I_0^{(3)}}{2(1-\epsilon)} + 
\notag \\
&\quad+
2 Q^2 \left[
I_0 + \frac{2}{Q^2} \, I_0^{(3)} - (1-\epsilon) 
\left(-\frac{1}{Q^2} \frac{I_0^{(3)}}{2(1-\epsilon)} \right)
\right]
\Bigg\} 
+ h.c.,
\label{eq:vertex_1loop_1}
\end{align}
where we introduced the integrals $I_0$ and $I_0^{(3)}$ defined as:
\begin{align}
&I_0 = \int 
\frac{d^{4-2\epsilon} \, l}
{(2 \pi)^{4-2\epsilon}} \, 
\frac{1}{\left[ (k_1 - l)^2 + i \, 0 \right] \, 
\left[ (k_2 + l)^2 + i \, 0 \right] \,
\left[ l^2 + i \, 0 \right]} = 
\notag \\
&\quad=  \frac{i}{\epsilon} \,
\frac{\Gamma(1+\epsilon)}{(4 \pi)^{2-\epsilon}} \,
(-Q^2)^{-1-\epsilon}\, B(-\epsilon,\,1-\epsilon);
\\
&I_0^{(3)} = 
\int 
\frac{d^{4-2\epsilon} \, l}
{(2 \pi)^{4-2\epsilon}} \, 
\frac{1}{\left[ (k_1 - l)^2 + i \, 0 \right] \, 
\left[ (k_2 + l)^2 + i \, 0 \right]} =
\notag \\
&\quad= 
i \frac{\Gamma(\epsilon)}{(4 \pi)^{2-\epsilon}} \,
(Q^2)^{-\epsilon} \, B(1-\epsilon,\,1-\epsilon).
\label{eq:integrals_PV}
\end{align}
Therefore:
\begin{align}
&\frac{\alpha_S}{4 \pi}
V^{[1]}(\epsilon;\,\mu,\,Q) = 
-\frac{\alpha_S}{4\pi} \, C_F \, S_\epsilon \,
\left( \frac{\mu}{Q} \right)^{2\epsilon} \,
(-1)^{-\epsilon} \,
\frac{\Gamma(1-\epsilon)^3\, \Gamma(1+\epsilon)}
{\Gamma(1-2\epsilon)} \,
\times
\notag \\
&\quad \times \,
\left(
-\frac{2\Gamma(-\epsilon)}
{\epsilon \, \Gamma(1-\epsilon)}+
\frac{(-1)^{-\epsilon}(3+2\epsilon)
\Gamma(\epsilon)}{(1-2\epsilon)\, \Gamma(1+\epsilon)}
\right)
+ h.c. = 
\notag \\
&\quad=
-\frac{\alpha_S}{4\pi} \,2\, C_F \, 
S_\epsilon \,
\left[ 
\frac{2}{\epsilon^2} +
\frac{2}{\epsilon} \,
\left( 
\frac{3}{2}+\log{\frac{\mu^2}{Q^2}}
\right)+
8 - \pi^2 + 3\log{\frac{\mu^2}{Q^2}} +
\left(\log{\frac{\mu^2}{Q^2}}\right)^2
\right]
\label{eq:vertex_1loop_2}
\end{align}
\paragraph{Real Emission.}
In this case, the emitted gluon is real and there are in total three particles in the final state: the fragmenting quark, the gluon and the antiquark. The squared amplitudes depend on all possible combinations of scalar products of the final state particle momenta, encoded in the variables $y_1$, $y_2$ and $y_3$ defined in Eq.~\eqref{eq:y123_def}. Momentum conservation ensures that $\sum_i y_i = 1$.
The squared amplitudes in which the gluon crosses diagonally the final state cut will be labeled by ``diag.". It is given by:
\begin{align}
&M_{f,\,\mbox{\small diag.}}^{\mu\nu,\,[1]}
(\epsilon;\,\mu,\,\{y_i\}) = 
\begin{gathered}
\includegraphics[width=4.5cm]{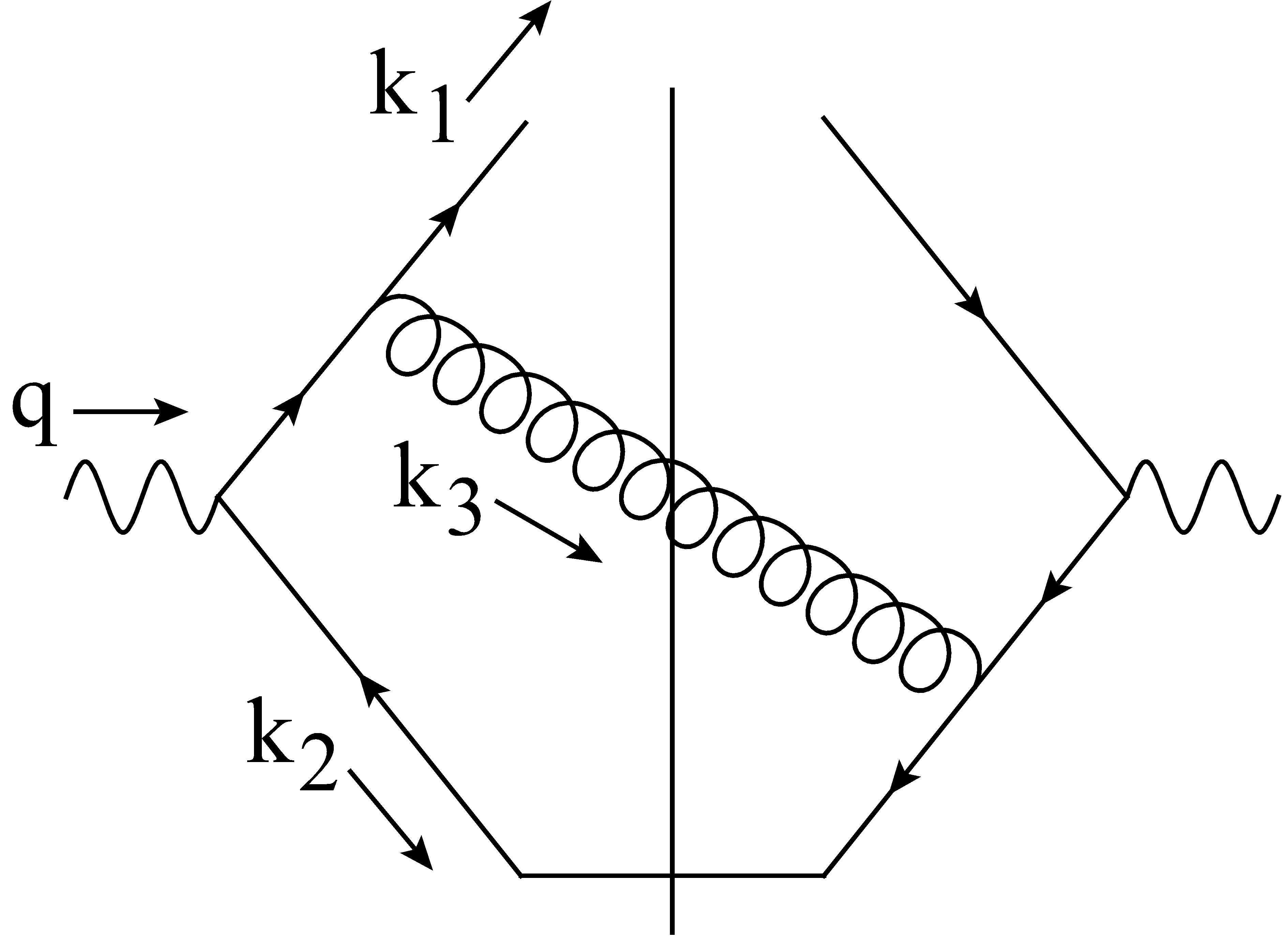}
\end{gathered}
+ h.c.
= 
\notag \\
&\quad=
e_f^2 \,
\Bar{u}(k_1) \, (-i g_0 \gamma^\alpha t^a) \,
\frac{i (\slashed{k}_1 + \slashed{k}_3) }{(k_1 + k_3)^2 + i 0} \, \gamma^\mu \,
\slashed{k}_2 \, (i g_0 \gamma^\beta t^b) \,
\times \notag \\
&\quad \quad \times
\frac{-i (-\slashed{k}_2 - \slashed{k}_3) }{(k_2 + k_3)^2 - i 0} \, \gamma^\nu \,
u(k_1) \, (-g_{\alpha\,\beta} \, \delta_{a\,b})
+ h.c. = 
\notag \\
&\quad=
e_f^2 \,g^2 \, \mu^{2\epsilon} \, C_F \, N_C \,
\frac{\mbox{Tr}_D \left\{ 
\slashed{k}_1 \, \gamma^\alpha \, (\slashed{k}_1 + \slashed{k}_3) \,
\gamma^\mu \, \slashed{k}_2 \, \gamma_\alpha \, 
(\slashed{k}_2 + \slashed{k}_3) \, \gamma^\nu
\right\}}
{(k_1 + k_3)^2 \, (k_2 + k_3)^2} 
+ h.c.
\label{eq:MX_R}
\end{align}
The projections give:
\begin{align}
&-g_{\mu\,\nu}
M_{f,\,\mbox{\small diag.}}^{\mu\nu,\,[1]}
(\epsilon;\,\mu,\,\{y_i\}) = 
H_{0,\,f} \,
g^2 \, \mu^{2\epsilon} \, C_F \, 8 \,
\left( 
\frac{y_3}{y_1 \, y_2} - \epsilon
\right); 
\label{eq:gMX_R} \\
&\frac{k_{1,\,\mu}k_{1,\,\nu}}{Q^2} M_{f,\,\mbox{\small diag.}}^{\mu\nu,\,[1]}
(\epsilon;\,\mu,\,\{y_i\}) 
= 0 ,
\label{eq:k2MX_R}
\end{align}
where $H_{0,\,f}$ has been defined in Eq.~\eqref{eq:H0f_def}.
Then we have to compute the squared amplitude in which the gluon attaches to both the upper fermionic lines. It will be labeled by ``up" and it gives:
\begin{align}
&M_{f,\,\mbox{\small up}}^{\mu\nu,\,[1]}
(\epsilon;\,\mu,\,\{y_i\})  = 
\begin{gathered}
\includegraphics[width=4.5cm]{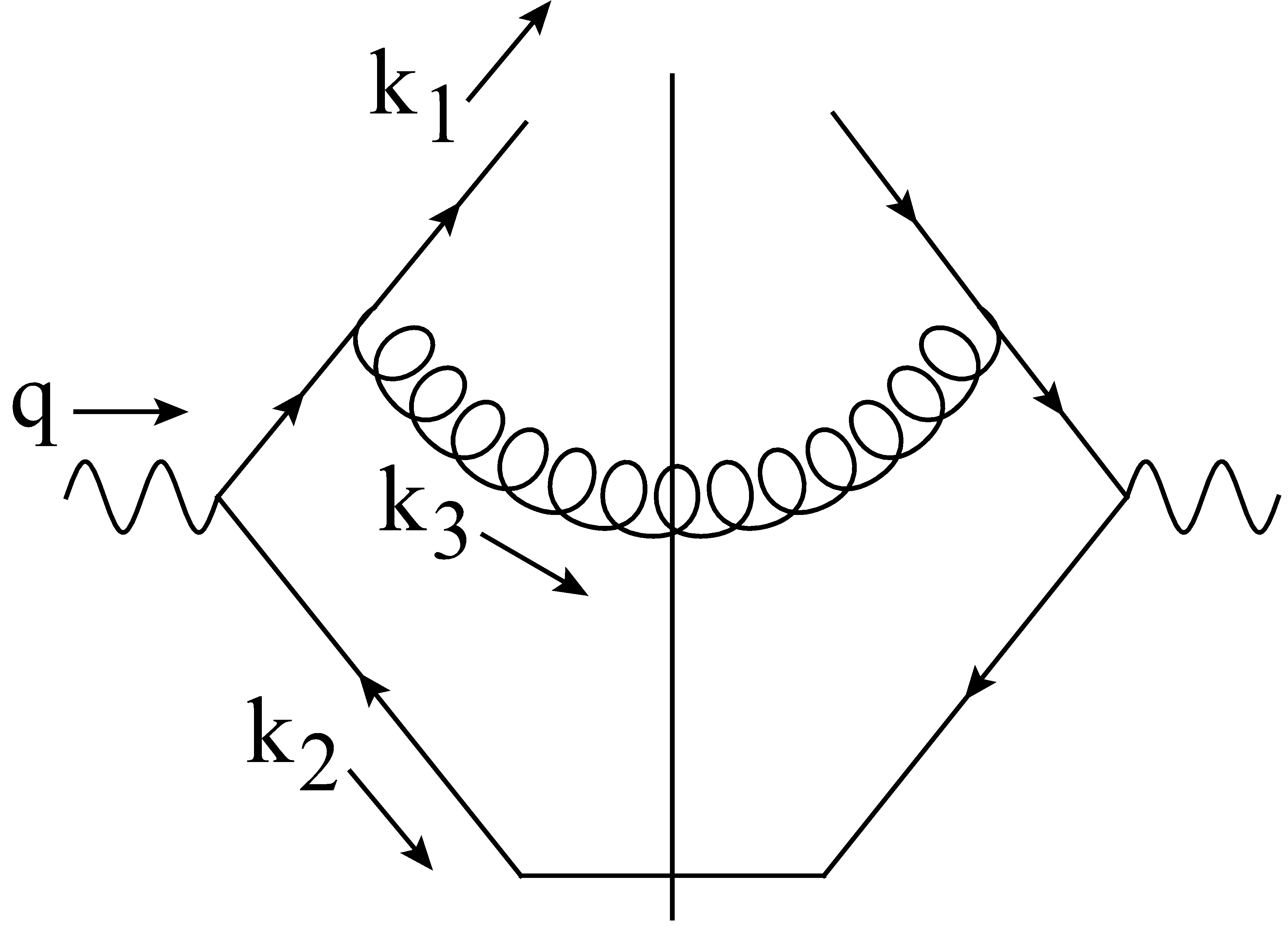}
\end{gathered}
= 
\notag \\
&\quad=
e_f^2 \,
\Bar{u}(k_1) \, (-i g_0 \gamma^\alpha t^a) \, 
\frac{i (\slashed{k}_1 + \slashed{k}_3) }{(k_1 + k_3)^2 + i 0} \, 
\gamma^\mu \, \slashed{k}_2 \, \gamma^\nu \, 
\frac{-i (\slashed{k}_1 + \slashed{k}_3) }{(k_1 + k_3)^2 + i 0} \,
\times 
\notag \\
&\quad\times \,
(i g_0 \gamma^\beta t^b) \, 
u(k_1) \, (-g_{\alpha\,\beta} \, \delta_{a\,b}) =
\notag \\
&\quad=
e_f^2 \,
g^2 \, \mu^{2\epsilon} \, C_F \, N_C \,
\frac{\mbox{Tr}_D \left\{ 
\slashed{k}_1 \, (\slashed{k}_1 + \slashed{k}_3) \, \gamma^\mu \,
 \slashed{k}_2 \, \gamma^\nu \, (\slashed{k}_1 + \slashed{k}_3)
\right\}}
{(k_1 + k_3)^4} \,2 \, (1-\epsilon) , 
\label{eq:MU_R}
\end{align}
Its projections are:
\begin{align}
&-g_{\mu\,\nu}
M_{f,\,\mbox{\small up}}^{\mu\nu,\,[1]}
(\epsilon;\,\mu,\,\{y_i\}) =
H_{0,\,f} \,
g^2 \, \mu^{2\epsilon} \, 
C_F \,4 \, (1-\epsilon) \, 
\frac{y_1}{y_2};
\label{eq:gMU_R} \\
&\frac{k_{1,\,\mu}k_{1,\,\nu}}{Q^2} M_{f,\,\mbox{\small up}}^{\mu\nu,\,[1]}
(\epsilon;\,\mu,\,\{y_i\})
= H_{0,\,f} \,
g^2 \, \mu^{2\epsilon} \, C_F \, 2 \, y_3 .
\label{eq:k2MU_R}
\end{align}
Notice that in this case there is a non vanishing contribution from the projection with $k_{1,\,\mu}k_{1,\,\nu}$. However, this is not divergent neither when $y_1 \to 0$, nor when $y_2 \to 0$ and hence we expect that it will be suppressed in a $2$-jet-like final state configuration.
The last squared amplitude involves the gluon attached to both the lower fermion lines and hence it will be labeled as ``down".
It is given by:
\begin{align}
&M_{f,\,\mbox{\small down}}^{\mu\nu,\,[1]}
(\epsilon;\,\mu,\,\{y_i\}) = 
\begin{gathered}
\includegraphics[width=4.5cm]{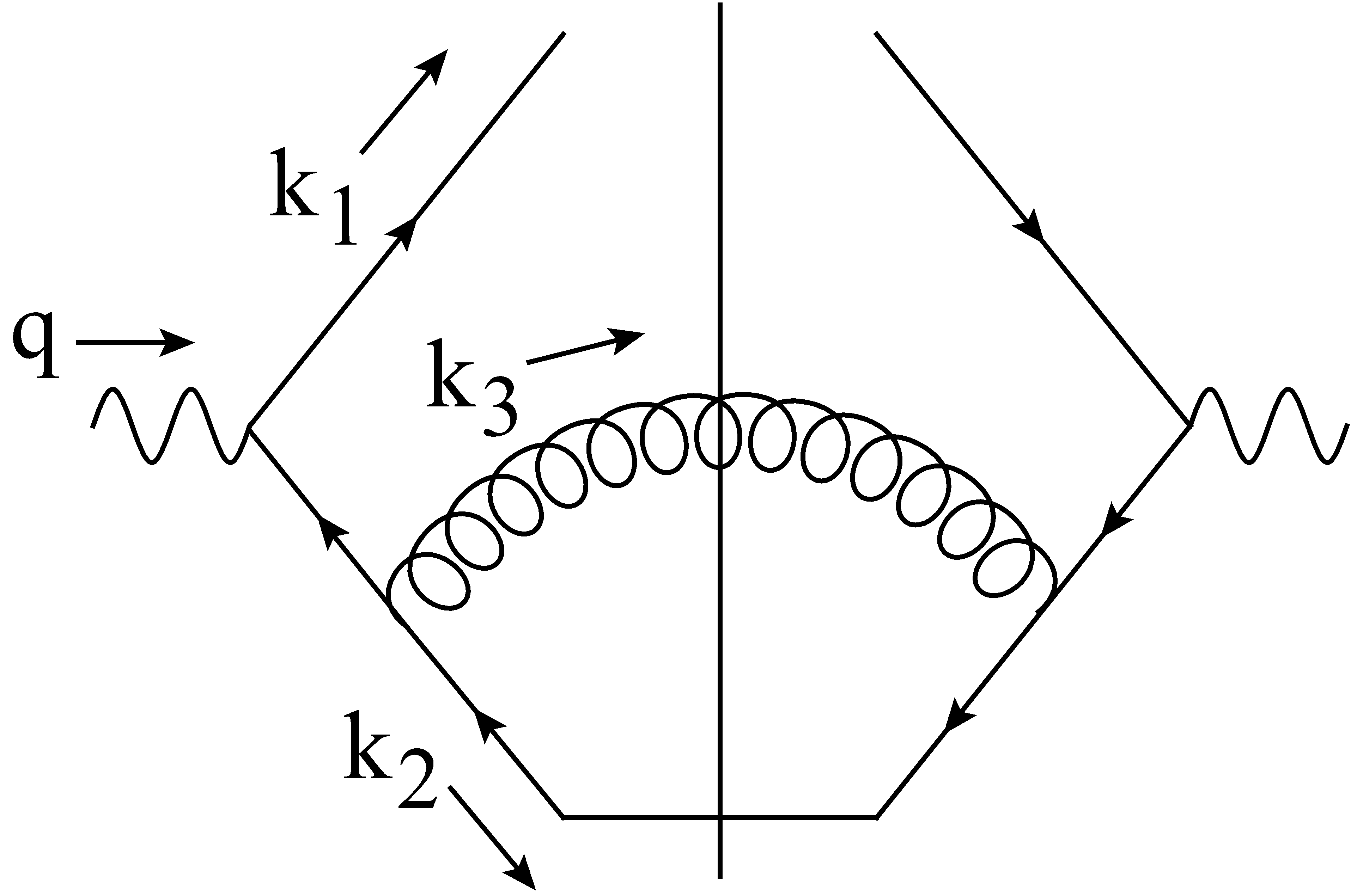}
\end{gathered}
= 
\notag \\
&\quad=
e_f^2 \,
\Bar{u}(k_1) \, \gamma^\mu \,
\frac{i (-\slashed{k}_2 - \slashed{k}_3) }{(k_2 + k_3)^2 + i 0} \,  (-i g_0 \gamma^\alpha t^a) \, \slashed{k}_2 \, 
(i g_0 \gamma^\beta t^b) \,
\times
\notag \\
&\quad \times \,
\frac{-i (-\slashed{k}_2 - \slashed{k}_3) }{(k_2 + k_3)^2 + i 0} \,
\gamma^\nu \, u(k_1) \, (-g_{\alpha\,\beta} \, \delta_{a\,b}) =
\notag \\
&\quad=
e_f^2 \, g^2 \, \mu^{2\epsilon} \, C_F \, N_C \,
\frac{\mbox{Tr}_D \left\{ 
\slashed{k}_1 \, \gamma^\mu \, (\slashed{k}_2 + \slashed{k}_3) \,
 \slashed{k}_2 \, (\slashed{k}_2 + \slashed{k}_3) \, \gamma^\nu
\right\}}
{(k_2 + k_3)^4} \,2 \, (1-\epsilon) , 
\label{eq:MUrev_R}
\end{align}
with projections:
\begin{align}
&-g_{\mu\,\nu}
M_{f,\,\mbox{\small down}}^{\mu\nu,\,[1]}
(\epsilon;\,\mu,\,\{y_i\}) = 
H_{0,\,f} \,
g^2 \, \mu^{2\epsilon} \, C_F \, 4 \, (1-\epsilon) \, 
\frac{y_2}{y_1};
\label{eq:gMUrev_R} \\
&\frac{k_{1,\,\mu}k_{1,\,\nu}}{Q^2} M_{f,\,\mbox{\small down}}^{\mu\nu,\,[1]}
(\epsilon;\,\mu,\,\{y_i\})
=  0.
\label{eq:k2MUrev_R}
\end{align}
Finally, we combine all the squared amplitudes in Eqs.~\eqref{eq:MX_R},~\eqref{eq:MU_R} and~\eqref{eq:MUrev_R} in the total contribution of the real emission $M_{f,\,R}^{\mu\nu,\,[1]}$, whose projections are:
\begin{align}
&-g_{\mu\,\nu}
M_{f,\,R}^{\mu\nu,\,[1]}
(\epsilon;\,\mu,\,\{y_i\}) =
H_{0,\,f} \,
g^2 \,8 C_F \, \mu^{2\epsilon}
\left[
\left(\frac{y_3}{y_1 \, y_2} - \epsilon \right) + \frac{(1-\epsilon)}{2} \left(\frac{y_1}{y_2 } + \frac{y_2}{y_1} \right)
\right] ;
\label{eq:gM_R} \\
&\frac{k_{1,\,\mu}k_{1,\,\nu}}{Q^2} M_{f,\,R}^{\mu\nu,\,[1]}
(\epsilon;\,\mu,\,\{y_i\})
= 
H_{0,\,f} \,
g^2 \,2 C_F  \, \mu^{2\epsilon} \, y_3 .
\label{eq:k2M_R}
\end{align}
As for the case of the fragmenting gluon, the available phase space for the three final state particles is represented by the dashed (red) vertical and horizontal boundaries of the triangle in Fig.~\ref{fig:phase_space_real}.
It is given by Eq.~\eqref{eq:phsp_int_noT}, where the delta function sets the values of thrust according to the sub-regions $R_1$, $R_2$ and $R_3$, while the theta function forces $\tau$ to lie in the neighborhood of zero $[0,\,\tau_{\mbox{\tiny MAX}}]$. Furthermore, we will use the same change of variables adopted in Section~\ref{subsubsec:frag_gluon_unsub}, defined in Eq.~\eqref{eq:change_var}.

\bigskip

The integration over the sub-region $R_1$ is totally analogous to Eq.~\eqref{eq:gluon_gW_R1_unapprox}. 
The $2$-jet limit of this sub-region corresponds to the left, vertical edge of the triangle in Fig.~\ref{fig:phase_space_real}, which coincides with having the gluon collinear to the antiquark except in the vertex, where the gluon turns soft.
Therefore, it is reasonable to expect that the result of the integration over $R_1$ will be related to the backward thrust function $J_B$ and to the soft function $S$, defined in Eqs.~\eqref{eq:B_thrust} and~\eqref{eq:S_thrust}, respectively.
It is given by:
\begin{align}
&\frac{\alpha_S}{4 \pi}\left(-g_{\mu \nu} 
\widehat{W}_{f,\,R_1}^{\mu\,\nu}{}^{\;[1]}\right) =
H_{0,\,f} \, 
\frac{\alpha_S}{4\pi} \, 
4 \, C_F \,S_\epsilon \, 
\left( \frac{\mu}{Q} \right)^{2\epsilon} \, 
\delta\left( \tau - (1-z) \right) \,
\theta \left( \tau_{\mbox{\tiny MAX}} - \tau \right) 
\, \times 
\notag \\
&\quad \times \, 
\theta\left(z - \frac{2}{3}\right) \, 
(1-z)^{-\epsilon} \,
\int_{\frac{1-z}{z}}^{2-\frac{1}{z}} \, 
d \alpha \, \alpha^{-\epsilon} \, (1-\alpha)^{-\epsilon}  \,
\times \notag \\
&\quad \times \,
\left[ 
\left( 
\frac{1}{1-z} \, \frac{1-\alpha}{\alpha} -\epsilon
\right) +
\frac{1-\epsilon}{2} \,
\left( 
\frac{1-z}{z} \, \frac{1}{\alpha} +
\frac{z}{1-z} \, \alpha
\right) 
\right]=
\notag \\
&\quad=
H_{0,\,f} \, 
\frac{\alpha_S}{4\pi} \, 
4 \, C_F \,S_\epsilon \, 
\left( \frac{\mu}{Q} \right)^{2\epsilon} \, 
\delta\left( \tau - (1-z) \right) \,
\theta\left(\frac{1}{3}-\tau\right) \,
\theta \left( \tau_{\mbox{\tiny MAX}} - \tau \right) \, 
\times \notag \\
&\quad \times \,
\bigg[
\tau^{-1-\epsilon}\,I_{1,\,-1}(\epsilon, \,\tau)
-\epsilon \,\tau^{\epsilon} \,
I_{0,\,0}(\epsilon, \,\tau)+
\notag \\
&\quad+
\frac{1-\epsilon}{2} \,
\left( 
\frac{\tau^{1-\epsilon}}{1-\tau} \,
I_{1,\,0}(\epsilon, \,\tau) +
(1-\tau) \, \tau^{-1-\epsilon} \,
I_{-1,\,0}(\epsilon, \,\tau)
\right)
\bigg],
\quad \mbox{Re~}{\epsilon}<0,
\label{eq:ferm_gW_R1_unapprox}
\end{align}
where the integrals $I_{a,\,b}$ have been defined in Eq.~\eqref{eq:R1_int_def}.
The $2$-jet limit of the previous expression is found by following the same argument presented in Section~\ref{subsubsec:frag_gluon_unsub} and in particular by using Eqs.~\eqref{eq:R1_int_def_smallt} and~\eqref{eq:delta_R1}. This gives:
\begin{align}
&\frac{\alpha_S}{4 \pi}\left(-g_{\mu \nu} 
\widehat{W}_{f,\,R_1}^{\mu\,\nu}{}^{\;[1]}\right)
\stackrel{\tau_{\mbox{\tiny MAX}} \to 0}{\sim}
H_{0,\,f} \, \delta( 1-z ) \,
\frac{\alpha_S}{4\pi} \, 
4 \, C_F \,S_\epsilon \, 
\left( \frac{\mu}{Q} \right)^{2\epsilon} \,
\theta 
\left( \tau_{\mbox{\tiny MAX}} - \tau \right)
\times
\notag \\
&\quad \times \, 
\bigg[
\frac{1}{\epsilon} \,\tau^{-1-2\epsilon} +
\tau^{-1-\epsilon} \, 
\left(
B(2-\epsilon,\,-\epsilon) + 
\frac{1-\epsilon}{2} \, B(1-\epsilon,\,2-\epsilon)
\right) +
\mathcal{O}\left( \tau^{-\epsilon}\right)
\bigg] = 
\notag \\
&\quad=
H_{0,\,f} \,
\theta 
\left( \tau_{\mbox{\tiny MAX}} - \tau \right)\,
\delta( 1-z ) \,
\frac{\alpha_S}{4 \pi}\,
\left[
\frac{1}{2} \, S^{[1]}(\epsilon,\,\tau) +
J_B^{[1]}(\epsilon,\,\tau)
+
\mathcal{O}\left( \tau^{-\epsilon}\right)
\right],
\quad \mbox{Re~}{\epsilon}<0.
\label{eq:ferm_gW_R1_approx}
\end{align}
As expected, the $2$-jet limit of sub-region $R_1$ is dominated by the backward thrust function $J_B$ and by the soft thrust function $S$. 
Notice that only half of the soft function appears in the final result, since ``half" of the vertex is shared with sub-region $R_2$.
The other projection (with respect to $k_{1,\mu}k_{1,\nu}$) is suppressed in the $2$-jet limit, since it is proportional to the integral $\tau^{-\epsilon}\,I_{0,\,1} \sim \mathcal{O}(\tau^{-\epsilon})$  that vanishes if $\mbox{Re~}{\epsilon}<0$.

Let's now consider the contribution of sub-region $R_2$. Its contribution to the $2$-jet limit of the process corresponds to the lower edge of the triangle in Fig.~\ref{fig:phase_space_real}, that coincides with the configuration in which the gluon is collinear to the fragmenting quark, except in the vertex, where the gluon turns soft.
Hence, the final result should be related to the thrust jet function of the fragmenting quark $J_{q/q}$ and to the soft thrust function $S$, defined in Eqs.~\eqref{eq:jetferm_thrust} and~\eqref{eq:S_thrust}, respectively.
It is given by:
\begin{align}
&\frac{\alpha_S}{4 \pi}\left(-g_{\mu \nu} 
\widehat{W}_{f,\,R_2}^{\mu\,\nu}{}^{\;[1]}\right) =
H_{0,\,f} \, 
\frac{\alpha_S}{4\pi} \, 
4 \, C_F \,S_\epsilon \, 
\left( \frac{\mu}{Q} \right)^{2\epsilon} \,
\theta 
\left( \tau_{\mbox{\tiny MAX}} - \tau \right) \,
\times
\notag \\
&\quad \times \,
\left[
\theta\left( \frac{2}{3} - z \right) \, \int_0^{\frac{1}{2}} + \,
\theta\left( z - \frac{2}{3} \right) \, \int_0^{\frac{1-z}{z}}
\right] \,
d \alpha \, \alpha^{-\epsilon} \, (1-\alpha)^{-\epsilon} \,
\delta\left(\alpha - \frac{\tau}{z} \right)\,
\times
\notag \\
&\quad \times \,
\left[
\left( 
\frac{1}{1-z} \, \frac{1-\alpha}{\alpha} -\epsilon
\right) +
\frac{1-\epsilon}{2} \,
\left( 
\frac{1-z}{z} \, \frac{1}{\alpha} +
\frac{z}{1-z} \, \alpha
\right)
\right] =
\notag \\
&\quad=
H_{0,\,f} \, 
\frac{\alpha_S}{4\pi} \, 
4 \, C_F \,S_\epsilon \, 
\left( \frac{\mu}{Q} \right)^{2\epsilon} \,
\theta 
\left( \tau_{\mbox{\tiny MAX}} - \tau \right) \,
\times
\notag \\
&\quad \times
\left[
\theta\left( \frac{2}{3} - z \right) \, 
\theta\left( z - 2\tau \right) + \,
\theta\left( z - \frac{2}{3} \right) \, 
\theta\left( z - \tau \right) - \,
\theta\left( z - \frac{2}{3} \right) \, 
\theta\left( z - (1-\tau) \right)
\right] \,\times
\notag \\
&\quad \times \,
 \bigg[
z^{\epsilon} \, (1-z)^{-1-\epsilon} \, \tau^{-1-\epsilon} \,
\left( 1 - \frac{\tau}{z} \right)^{1-\epsilon}-
\epsilon \, 
z^{-1+\epsilon} \, (1-z)^{-\epsilon} \,
\tau^{-\epsilon}\,
\left( 1 - \frac{\tau}{z} \right)^{-\epsilon}+
\notag \\
&\quad +
\frac{1-\epsilon}{2} 
\bigg( 
z^{-1+\epsilon} (1-z)^{1-\epsilon} 
\tau^{-1-\epsilon}
\left( 1 - \frac{\tau}{z} \right)^{-\epsilon}+
z^{-1+\epsilon}  (1-z)^{-1-\epsilon} 
\tau^{1-\epsilon} 
\left( 1 - \frac{\tau}{z} \right)^{-\epsilon}
\bigg)
\bigg].
\label{eq:ferm_gW_R2_unapprox}
\end{align}
The $2$-jet limit is obtained by expanding in powers of $\tau$ all the non-divergent terms in $\tau = 0$ and by using Eqs.~\eqref{eq:R2_theta_appr_1} and~\eqref{eq:R2_theta_appr_2} to properly approximate the theta functions. This gives:
\begin{align}
&\frac{\alpha_S}{4 \pi}\left(-g_{\mu \nu} 
\widehat{W}_{f,\,R_2}^{\mu\,\nu}{}^{\;[1]}\right)
\stackrel{\tau_{\mbox{\tiny MAX}} \to 0}{\sim}
H_{0,\,f} \, 
\frac{\alpha_S}{4\pi} \, 
4 \, C_F \,S_\epsilon \, 
\left( \frac{\mu}{Q} \right)^{2\epsilon} \,
\theta 
\left( \tau_{\mbox{\tiny MAX}} - \tau \right) \,
\theta(1-z) \,
\times
\notag \\
&\quad \times
\Bigg\{
z^{\epsilon}
\left(
\left[ (1-z)^{-1-\epsilon} + 
\frac{1-\epsilon}{2} \, 
\frac{(1-z)^{1-\epsilon}}{z} 
\right] \,\tau^{-1-\epsilon}+
\mathcal{O}(\tau^{-\epsilon})
\right)+
\notag \\
&\quad+
\theta\left(z-(1-\tau)\right) \,
\bigg((1-z)^{-1-\epsilon} \,
\tau^{-1-\epsilon}+
\mathcal{O}(\tau^{-\epsilon})+
\mathcal{O}((1-z)^{-\epsilon})
\bigg)
\Bigg\}.
\label{eq:ferm_gW_R2_approx}
\end{align}
Eq.~\eqref{eq:R2_theta_appr_2} allows us to select out the terms that are divergent as $z$ approaches $1$. Furthermore, in this approximation the limit $\tau_{\mbox{\tiny MAX}} \to 0$ coincides with the limit $z \to 1$, therefore the last line of Eq.~\eqref{eq:ferm_gW_R2_approx} has to be proportional to $\delta(1-z)$. 
In fact, integration with a test function $T(z)$ gives:
\begin{align}
&\theta 
\left( \tau_{\mbox{\tiny MAX}} - \tau \right) \,
\int_0^1 \, dz \, T(z)
\theta(1-z) \, \theta\left(z-(1-\tau)\right) \,
(1-z)^{-1-\epsilon} =
\notag \\
&\quad=
\theta 
\left( \tau_{\mbox{\tiny MAX}} - \tau \right) \,
\left[
-T(1) \, \frac{\tau^{-\epsilon}}{\epsilon} +
\mathcal{O}\left( \tau^{1-\epsilon} \right)
\right],
\quad \mbox{Re~}{\epsilon}<0.
\label{eq:R2_theta_appr_test}
\end{align}
Therefore, the vanishing cut-off limit can be written as:
\begin{align}
&\theta 
\left( \tau_{\mbox{\tiny MAX}} - \tau \right) \,
\theta(1-z) \, \theta\left(z-(1-\tau)\right) \,
(1-z)^{-1-\epsilon} 
\stackrel{\tau_{\mbox{\tiny MAX}} \to 0}{\sim}
\notag \\
&\quad
\stackrel{\tau_{\mbox{\tiny MAX}} \to 0}{\sim}
-\frac{\tau^{-\epsilon}}{\epsilon}\,\delta(1-z)\,
\theta 
\left( \tau_{\mbox{\tiny MAX}} - \tau \right).
\label{eq:R2_theta_appr_3}
\end{align}
The final result is:
\begin{align}
&-g_{\mu \nu} 
\widehat{W}_{f,\,R_2}^{\mu\,\nu}{}^{\;[1]}
\stackrel{\tau_{\mbox{\tiny MAX}} \to 0}{\sim}
H_{0,\,f} \,
\theta 
\left( \tau_{\mbox{\tiny MAX}} - \tau \right) \, 
\times
\notag \\
&\quad \times
\left[
\frac{1}{2}\,\delta(1-z)\,
S^{[1]}(\epsilon;\,\tau)+
J_{q/q}^{[1]}(\epsilon;\,\tau,\,z) +
\mathcal{O}\left( \tau^{-\epsilon} \right)
\right],
\quad \mbox{Re~}{\epsilon}<0,
\label{eq:ferm_gW_R2_approx_2}
\end{align}
where $J_{q/q}^{[1]}$ is the thrust jet function for the fragmenting fermion, defined in Eq.~\eqref{eq:jetferm_thrust}.
Notice that, as expected, the previous expression provides the missing half of the soft function when it is summed to the contribution of the sub-region $R_1$, presented in Eq.~\eqref{eq:ferm_gW_R1_approx}.
The other projection (with respect to $k_{1,\mu}k_{1,\nu}$) is suppressed in the $2$-jet limit, in fact its contribution is of order $\mathcal{O}\left( \tau^{-\epsilon }\right)$ and vanishes as $\mbox{Re~}{\epsilon} < 0$.

Finally, sub-region $R_3$ does not contribute to the $2$-jet limit. Its contribution can be computed similarly to that of sub-region $R_2$, but in this case thrust must satisfy the condition $\alpha = 1 - {\tau}/{z}$ (as in Eq.~\eqref{eq:gluon_gW_R3_unapprox}). As a consequence, the final result will be suppressed by powers of order $\mathcal{O}\left( \tau^{-\epsilon }\right)$, vanishing as $\mbox{Re~}{\epsilon} < 0$.

Therefore, the full contribution of the real emission to the unsubtracted final state tensor is given by:
\begin{align}
&-g_{\mu \nu} 
\widehat{W}_{f,\,R}^{\mu\,\nu}{}^{\;[1]}
(\epsilon;\,z,\,\tau;\,\tau_{\mbox{\tiny MAX}}\to 0)
\stackrel{\tau_{\mbox{\tiny MAX}} \to 0}{\sim}
H_{0,\,f} \,
\theta 
\left( \tau_{\mbox{\tiny MAX}} - \tau \right) \,
\times \notag \\
&\quad \times \,
\left\{
\delta(1-z)\,
\left[
S^{[1]}(\epsilon;\,\tau)+
J_B^{[1]}(\epsilon;\,\tau)
\right]+
J_{q/q}^{[1]}(\epsilon;\,\tau,\,z) +
\mathcal{O}\left( \tau^{-\epsilon} \right)
\right\},
\quad \mbox{Re~}{\epsilon}<0.
\label{eq:W_ferm_real_1}
\end{align}
Following the same argument presented in Section~\ref{subsubsec:frag_gluon_unsub}, the cut-off $\tau_{\mbox{\tiny MAX}}$ can be related to a cut-off on the transverse momentum of the fragmenting parton, as in Eq.~\eqref{eq:lambda_tmax}.
With this choice, we can drop the theta function anytime it appears multiplied by $\delta(1-z)$. Therefore, we will rewrite the result of Eq.~\eqref{eq:W_ferm_real_1} as:
\begin{align}
&-g_{\mu \nu} 
\widehat{W}_{f,\,R}^{\mu\,\nu}{}^{\;[1]}
(\epsilon;\,z,\,\tau;\,{\lambda^2}/{Q^2} \to 0)
=
H_{0,\,f} \,
\times \notag \\
&\quad \times \,
\left\{
\delta(1-z)\,
\left[
S^{[1]}(\epsilon;\,\tau)+
J_B^{[1]}(\epsilon;\,\tau)
\right]+
J_{q/q}^{[1],\,(\lambda)}(\epsilon;\,\tau,\,z)
\right\},
\quad \mbox{Re~}{\epsilon}<0,
\label{eq:W_ferm_real_2}
\end{align}
where:
\begin{align}
J_{q/q}^{[1],\,(\lambda)}(\epsilon;\,\tau,\,z) = 
J_{q/q}^{[1]}(\epsilon;\,\tau,\,z)\,
\theta\left(
\frac{\lambda^2}{Q^2}-\frac{1-z}{z} \tau
\right).
\label{eq:jetferm_thrust_lambda}
\end{align}
Notice that this result is perfectly consistent with the intuitive prediction about the functions that participate to the $2$-jet limit of the real emission unsubtracted final state tensor.

\bigskip

The whole unsubtracted final state tensor is the sum of the virtual emission  (Eq.~\eqref{eq:W_ferm_virtual}) and the real emission term (Eq.~\eqref{eq:W_ferm_real_2})).
This gives:
\begin{align}
\widehat{W}_f^{\mu\,\nu}{}^{\;[1]}
(\epsilon;\,z,\,\tau;\,{\lambda^2}/{Q^2} \to 0) =
H_T^{\mu \nu}(z) \, \widehat{F}_{1,\,f}^{[1]}
(\epsilon;\,z,\,\tau;\,{\lambda^2}/{Q^2} \to 0),
\label{eq:W_ferm}
\end{align}
where:
\begin{align}
&\widehat{F}_{1,\,f}^{[1]}
(\epsilon;\,z,\,\tau;\,{\lambda^2}/{Q^2} \to 0) =
\notag \\
&\quad=
\frac{H_{0,\,f}}{2} \,
\left[
\delta(1-z) \left(
\delta(\tau) \, V^{[1]}(\epsilon) +
S^{[1]}(\epsilon;\,\tau)+
J_B^{[1]}(\epsilon;\,\tau)
\right) + 
J_{q/q}^{[1],\,(\lambda)}(\epsilon;\,\tau,\,z)
\right]
\label{eq:F1_ferm}
\end{align}
The $\epsilon$-expansion of $V^{[1]}$ is given in Eq.~\eqref{eq:vertex_1loop_2}, while $S^{[1]}$ and $J_B^{[1]}$ are easily expanded by inserting Eq.~\eqref{eq:thrust_distr_exp} in 
Eqs.~\eqref{eq:S_thrust} and~\eqref{eq:B_thrust}, respectively.
On the other hand, the expansion of $J_{q/q}^{[1],\,(\lambda)}$ is not straightforward, due to the non-trivial interplay between the theta function, that involves the $\lambda$ cut-off , and the distributions in $\tau$ and $z$ in $J_{q/q}^{[1]}$.
We have already considered the product of the theta function with the plus distribution $\left({1}/{\tau}\right)_+$ in Eq.~\eqref{eq:plus_theta_int}.
In this case, we also have to face the problem of multiplying the theta function by the product of two plus distributions, $\left({1}/{(1-z)}\right)_+$  and 
$\left({1}/{\tau}\right)_+$.
This particular combination originates from the $\epsilon$-expansion of:
\begin{align}
&(1-z)^{-1-\epsilon} \,
\tau^{-1-\epsilon} \,
\theta \left( \frac{\lambda^2}{Q^2} - \frac{1-z}{z} \tau\right) = 
\frac{1}{\epsilon^2} \, 
\delta(\tau) \,\delta(1-z) -
\notag \\
&\quad-
\frac{1}{\epsilon} \, 
\left[
\delta(\tau) \, 
\left( \frac{1}{1-z} \right)_+
+ \delta(1-z) \, 
\left( \frac{1}{\tau} \right)_+
\right] +
\delta(\tau)\,
\left( \frac{\log{(1-z)}}{1-z}+ \right)_+
+
\notag \\
&\quad+
\delta(1-z)\,
\left( \frac{\log{\tau}}{\tau}\right)_+ +
\left( \frac{1}{1-z} \right)_+ \,
\left( \frac{1}{\tau} \right)_+ \,
\theta \left( \frac{\lambda^2}{Q^2} - \frac{1-z}{z} \tau\right)
+ \mathcal{O}\left(\epsilon\right),
\label{eq:zthrust_distr_exp_lambda_1}
\end{align}
which follows from a direct expansion of the l.h.s. by using Eqs.~\eqref{eq:thrust_distr_exp} and~\eqref{eq:z_distr_exp}. 
The last term can be rewritten by using the result of Eq.~\eqref{eq:plus_theta_int}:
\begin{align}
&\left( \frac{1}{1-z} \right)_+ \,
\left( \frac{1}{\tau} \right)_+ \,
\theta \left( r^2 - \frac{1-z}{z} \tau\right) =
\notag \\
&\quad=
\frac{1}{1-z} \, 
\log{\left(\frac{z}{1-z}r^2\right)} \,
\theta\left(
\frac{1}{1+r^2} - z
\right) \,
\delta(\tau) + 
\mathcal{O}\left(r^2\right),
\label{eq:zplus_theta_1}
\end{align}
where the ``+" label can be dropped since $z$ never reaches $1$ as long as $r = {\lambda}/{Q}$ is different from zero. 
The integration with a test function $T(z)$ allows us to write the previous result in terms of distributions in $z$:
\begin{align}
&\int_0^{\frac{1}{1+r^2}} \, dz \, T(z) \,
\frac{1}{1-z} \, 
\log{\left(\frac{z}{1-z}r^2\right)} =
\notag \\
&\quad=
T(1) \,
\int_0^{\frac{1}{1+r^2}} \, dz \,
\frac{1}{1-z} \, 
\log{\left(\frac{z}{1-z}r^2\right)} + 
\notag \\
&\quad+
\sum_{n=1}^{\infty} \frac{(-1)^n}{n!} \,
T^{(n)}(1) \,
\int_0^{\frac{1}{1+r^2}} \, dz \,
(1-z)^{n-1} \,
\log{\left(\frac{z}{1-z}r^2\right)} = 
\notag \\
&\quad=
T(1) \,
\left[
-\frac{\pi^2}{6} - 
\frac{1}{2} \, \left( \log{r^2} \right)^2 + 
\mathcal{O}\left( r^2 \right)
\right] +
\notag \\
&\quad+
\sum_{n=1}^{\infty} \frac{(-1)^n}{n!} \,
T^{(n)}(1) \,
\left[
\frac{1}{n} \, \log{r^2} + 
\mathcal{O}\left( r^2 \right)
\right] -
\sum_{n=1}^{\infty} \frac{(-1)^n}{n!} \,
T^{(n)}(1) \,c_n,
\label{eq:zplus_theta_int}
\end{align}
where the coefficients $c_n$ are rational numbers.
The second step of the previous equation has been found by exploiting the analyticity properties of the test function $T(z)$.
In the first term in the last line of Eq.~\eqref{eq:zplus_theta_int} we recognize the action of the plus distribution, $\left({1}/{(1-z)}\right)_+$, while the last term  in  Eq.~\eqref{eq:zplus_theta_int} can be neglected in the small-$r$ limit, as we only consider the dominant (divergent) parts.
Finally we have:
\begin{align}
&\left( \frac{1}{1-z} \right)_+ \,
\left( \frac{1}{\tau} \right)_+ \,
\theta \left( r^2 - \frac{1-z}{z} \tau\right) =
\notag \\
&\quad=
\left[
-\frac{\pi^2}{6} - 
\frac{1}{2} \, \left( \log{r^2} \right)^2 
\right]\,
\delta(1-z) \, \delta(\tau) + 
\log{r^2} \, 
\left( \frac{1}{1-z} \right)_+ \,
\delta(\tau) +
\mathcal{O}\left(r^2\right).
\label{eq:zplus_theta_2}
\end{align}
Notice that since the theta function is not symmetric under the interchange of $\tau$ and $z$, the final result of the above equation is not symmetric under $\tau \leftrightarrow z$ either.
Finally, by using Eqs.~\eqref{eq:thrust_distr_exp_lambda_2},~\eqref{eq:zthrust_distr_exp_lambda_1} and~\eqref{eq:zplus_theta_2}, we obtain the following $\epsilon$-expansion for $J_{q/q}^{[1],\,(\lambda)}$:
\begin{align}
&\frac{\alpha_S}{4 \pi}
J_{q/q}^{[1],\,(\lambda)}(\epsilon;\,\tau,\,z) =
\frac{\alpha_S}{4\pi}\,2 C_F\,S_\epsilon\,
\Bigg\{
\frac{1}{\epsilon^2} \delta(\tau)\,\delta(1-z) +
\notag \\
&\quad+
\frac{1}{\epsilon} 
\left[
\delta(\tau)
\left(
2\,\delta(1-z)\,\log{\frac{\mu^2}{Q^2}}+
1-\frac{1}{z} - \frac{2}{(1-z)_+}
\right) -
2 \delta(1-z) \,\left( \frac{1}{\tau} \right)_+
\right]+
\notag \\
&\quad+
\delta(\tau) \,
\Bigg[
\left( 
-\frac{\pi^2}{3} 
- \left(\log{\frac{\lambda^2}{Q^2}}\right)^2
+ \left(\log{\frac{\mu^2}{Q^2}}\right)^2
\right)\,\delta(1-z) +
\frac{1-z}{z}\,
\left( 1-\log{\frac{\mu^2}{\lambda^2}}\right)-
\notag \\
&\quad-
2\frac{\log{z}}{1-z} -
\frac{2}{(1-z)_+}
\left( 
- \log{\frac{\lambda^2}{Q^2}}
+ \log{\frac{\mu^2}{Q^2}}
\right) +
2 \left( \frac{\log{(1-z)}}{1-z} \right)_+
\Bigg]-
\notag \\
&\quad-
2 \, \delta(1-z) 
\left[
\left( \frac{1}{\tau} \right)_+
\log{\frac{\mu^2}{Q^2}} -
\left( \frac{\log{\tau}}{\tau} \right)_+
\right]
\Bigg\},
\quad \mbox{Re~}{\epsilon}<0.
\label{eq:jetferm_thrust_lambda_exp}
\end{align}

\bigskip

%
\subsubsection{Subtraction Mechanism \label{subsubsec:frag_quark_sub}}

\bigskip

The unsubtracted final state tensor of Eq.~\eqref{eq:W_ferm} describes the fragmentation of a quark at partonic level. 
In the final cross section, the TMD FF of the quark encodes part of this same information; hence, the overlapping content has to be properly removed from $\widehat{W}_f^{\mu\nu}$, similarly to what we did in Section~\ref{subsubsec:frag_gluon_sub}.
In this case, however, the procedure is complicated by the fact that the quark-from-quark TMD FF is itself a subtracted quantity, to avoid overlapping with the soft momentum region (see Ref.~\cite{Collins:2011zzd} and \cite{Boglione:2020cwn}).
Furthermore, it needs to be renormalized by a proper UV counterterm and such renormalization affects the subtracted final state tensor, which will need an UV counterterm as well.
The subtraction will be performed in $b_T$-space, where the TMDs are explicitly defined in terms of operators (see Ref.~\cite{Boglione:2020cwn} and Ref.~\cite{Collins:2011zzd}).

\bigskip

In momentum space, the partonic version of the 1-loop quark from quark TMD FF is given by (see Ref.~\cite{Collins:2011zzd}):
\begin{align}
&\frac{\alpha_S}{4 \pi}
D_{q/q}^{[1]}(\epsilon;\,z,\,k_T,\,\zeta) = 
\frac{\alpha_S}{4\pi}\,2 C_F \,S_\epsilon\,
\frac{\Gamma(1-\epsilon)}{\pi^{1-\epsilon}}\,
\mu^{2\epsilon}\,\frac{1}{k_T^2}\,\times
\notag \\
&\quad \times \,
\left[
\left(\frac{2}{z\,(1-z)_+} + 
\frac{(1-\epsilon)\,(1-z)}{z^2}
\right)\,\theta(1-z) + 
\delta(1-z)\,\log{\frac{\zeta}{k_T^2}}
\right].
\label{eq:quark_TMD_kT}
\end{align}
Here, the term proportional to $\delta(1-z)$ is the subtraction term of the partonic TMD FF that removes the overlapping with the soft momentum region.
It involves a rapidity cut-off $\zeta$ that acts as a lower bound for the rapidity of the gluon emitted by the fragmenting quark.
Its presence ensures that the gluon is actually collinear, with low transverse momentum and large rapidity.

Following the argument of Section~\ref{subsubsec:frag_gluon_unsub}, the Fourier transform of Eq.~\eqref{eq:quark_TMD_kT} will be performed only up to $k_T = \lambda$, in order to match the same collinear momentum region described by the unsubtracted final state tensor and, in particular, by $J_{q/q}^{[1],\,(\lambda)}$, defined in Eq.~\eqref{eq:jetferm_thrust_lambda}.
The incomplete Fourier transform of ${1}/{k_T^2}$ was computed in Eq.~\eqref{eq:FT_kT2_cutoff}.
The incomplete Fourier transform of ${1}/{k_T^2}\,\log{{\zeta}/{k_T^2}}$ is given by:
\begin{align}
&\int d^{2-2\epsilon} \vec{k}_T \,
e^{i \vec{k}_T \cdot \vec{b}_T} \,
\mu^{2\epsilon} \, 
\frac{1}{k_T^2} \,
\log{\frac{\zeta}{k_T^2}}
\theta \left( \lambda^2 - k_T^2 \right) = 
\notag \\
&\quad =
\frac{\pi^{1-\epsilon}}{\Gamma(1-\epsilon)}\,
\left( \frac{\mu^2}{\lambda^2} \right)^{\epsilon}
\Bigg[
\left( 
\frac{\Gamma(-\epsilon)}{\Gamma(1-\epsilon)}
\right)^2\,
{}_2 F_3\left(
-\epsilon,\,-\epsilon;\,
1-\epsilon,\,1-\epsilon,\,1-\epsilon;\,
-\frac{\lambda^2\,b_T^2}{4}
\right) +
\notag \\
&\quad+
\log{\frac{\zeta}{\lambda^2}}\,
\frac{\Gamma(-\epsilon)}{\Gamma(1-\epsilon)}\,
{}_1 F_2\left(
-\epsilon;\,
1-\epsilon,\,1-\epsilon;\,
-\frac{\lambda^2\,b_T^2}{4}
\right)
\Bigg] =
\notag \\
&\quad=
\frac{\pi^{1-\epsilon}}{\Gamma(1-\epsilon)}\,
\bigg[
\frac{1}{\epsilon^2} - 
\frac{1}{\epsilon}\,\log{\frac{\zeta}{\lambda^2}}-\log{\frac{\zeta}{\lambda^2}}\,
\log{\frac{\mu^2}{\lambda^2}} + 
\frac{1}{2}\,
\left( \log{\frac{\mu^2}{\lambda^2}} \right)^2+
\notag \\
&\quad
+\mathcal{O}\left( 
\frac{\lambda^2}{Q^2},\,
\frac{\lambda^2}{Q^2}\,
\log{\frac{\lambda^2}{Q^2}}
\right) +
\mathcal{O}\left( \epsilon \right)
\bigg]
, \quad \mbox{Re~}{\epsilon}<0.
\label{eq:FT_logkT2_cutoff}
\end{align}
The incomplete Fourier transform of \eqref{eq:quark_TMD_kT} follows straightforwardly. 
The result gives the definition of the quark-from-quark TMD FF, equipped with the cut-off $\lambda$, analogously to Eq.~\eqref{eq:gluon_TMD_cutoff}.
We have:
\begin{align}
&\frac{\alpha_S}{4 \pi}
\widetilde{D}_{q/q,\,(0)}^{[1],\,(\lambda)}
(\epsilon;\,z,\,\zeta) = 
-\frac{\alpha_S}{4 \pi} \, 
2  C_F \, S_\epsilon \,
\delta(1-z)\,
\left[
-\frac{1}{\epsilon^2}+
\frac{1}{\epsilon}
\left(
-\frac{3}{2} + \log{\frac{\zeta}{\mu^2}}
\right)
\right] -
\notag \\
&\quad
-\frac{\alpha_S}{4 \pi} \, 
2  C_F \, \frac{S_\epsilon}{\epsilon}
\left[
\frac{1}{z^2} + \frac{1}{z} +
\frac{2}{(1-z)_+} + \frac{3}{2}\delta(1-z)
\right] +
\notag \\
&\quad+
\frac{\alpha_S}{4 \pi} \, 
2  C_F\, S_\epsilon \,
\Bigg\{
\frac{1}{2}\delta(1-z)\,
\left[
\left(\log{\frac{\mu^2}{\lambda^2}}\right)^2 -
2 \log{\frac{\mu^2}{\lambda^2}}\,
\log{\frac{\zeta}{\lambda^2}}
\right] +
\notag \\
&\quad +
\frac{1}{z^2}-\frac{1}{z} -
\left(
\frac{1}{z^2}+\frac{1}{z}+\frac{2}{(1-z)_+}
\right)\,\log{\frac{\mu^2}{\lambda^2}}
\Bigg\}
, \quad \mbox{Re~}{\epsilon}<0.
\label{eq:quark_TMD_cutoff_bare}
\end{align}
The label ``0" reminds that the quantity in the previous expression needs to be renormalized with an UV counterterm. 
This is the same UV counterterm that renormalizes the quark-from-quark TMD FF obtained by performing a complete Fourier transform, running over the full spectrum of transverse momenta.
At 1-loop it is given by (see Ref.~\cite{Collins:2011zzd}):
\begin{align}
\frac{\alpha_S}{4 \pi}
Z_q^{[1]}(\epsilon;\,z,\,\zeta) = 
\frac{\alpha_S}{4 \pi} \, 
2 \, C_F \, S_\epsilon \,
\left[
-\frac{1}{\epsilon^2} + 
\frac{1}{\epsilon}
\left(
-\frac{3}{2} + \log{\frac{\zeta}{\mu^2}}
\right)
\right]\,
\delta(1-z).
\label{eq:UV_TMD}
\end{align}
Notice that this UV counterterm cancels the first line of Eq.~\eqref{eq:quark_TMD_cutoff_bare}.
The remaining pole in the second line is the collinear divergence associated to the TMD FF.
Finally, the renormalized version of Eq.~\eqref{eq:quark_TMD_cutoff_bare} is simply:
\begin{align}
&\frac{\alpha_S}{4 \pi}
\widetilde{D}_{q/q}^{[1],\,(\lambda)}
(\epsilon;\,z,\,\zeta) = 
\frac{\alpha_S}{4 \pi}
\left(\widetilde{D}_{q/q,\,(0)}^{[1],\,(\lambda)}
(\epsilon;\,z,\,\zeta) + 
Z_q^{[1]}(\epsilon;\,z,\,\zeta)
\right).
\label{eq:quark_TMD_cutoff}
\end{align}
which will have to be used in the subtraction mechanism.
The 1-loop version of Eq.~\eqref{eq:sub_mech_W} for the fragmenting fermion case gives the corresponding subtracted, renormalized, final state tensor:
\begin{align}
&
\left.
\widehat{W}_f^{\mu \nu,\,[1]}
(z,\,\tau,\,\lambda,\,\zeta)
\right \rvert_{\mbox{\small sub, R}}= 
\widehat{W}_{f}^{\mu \nu,\,[1]}
(\epsilon;\,z,\,\tau;\,{\lambda^2}/{Q^2} \to 0) - \notag \\
&-
\sum_k \, 
\int_z^1 \frac{d\widehat{z}}{\widehat{z}}\;
\widehat{W}_k^{\mu\nu,\,[0]}
({z}/{\widehat{z}},\tau)
\,\left[ 
\widehat{z} \, 
\widetilde{D}_{f/k}^{[1],\,(\lambda)} 
(\epsilon; \, \widehat{z},\,\zeta)
\right],
\label{eq:sub_mech_W_1loop_ferm}
\end{align}
where the unsubtracted $\widehat{W}_{f}^{\mu \nu,\,[1]}(\epsilon;\,z,\,\tau;\,{\lambda^2}/{Q^2} \to 0)$ is given in Eq.~\eqref{eq:W_ferm}, while the lowest order $\widehat{W}_k^{\mu\nu,\,[0]}$ can be found in Appendix~\ref{app:lo_partonic_xs}.
Notice that the partonic TMD FF $\widetilde{D}_{f/k}^{[1],\,(\lambda)}$ is the identity matrix in flavor space.
The contraction with the metric tensor $g_{\mu\nu}$ simplifies the computation. 
With the help of Eq.~\eqref{eq:F1_ferm}, we have:
\begin{align}
&\frac{\alpha_S}{4 \pi}\,
\left.
\widehat{F}_{1,\,f}^{[1]}
(z,\,\tau,\,\lambda,\,\zeta)
\right \rvert_{\mbox{\small sub, R}}= 
\notag \\
&\quad=
\frac{\alpha_S}{4 \pi}
\left\{
\widehat{F}_{1,\,f}^{[1]}
(\epsilon;\,z,\,\tau;\,{\lambda^2}/{Q^2} \to 0) - \sum_k \, 
\int_z^1 \frac{d\widehat{z}}{\widehat{z}}\;
\widehat{F}_{1,\,k}^{[0]}
({z}/{\widehat{z}},\tau )
\,\left[ 
\widehat{z} \, 
\widetilde{D}_{f/k}^{[1],\,(\lambda)} 
(\epsilon; \, \widehat{z},\,\zeta)
\right] 
\right\}=
\notag \\
&\quad=
\frac{H_{0,\,f}}{2} \,
\frac{\alpha_S}{4 \pi}\,
\bigg[
\delta(1-z) \left(
\delta(\tau) \, V^{[1]}(\epsilon) +
S^{[1]}(\epsilon;\,\tau)+
J_B^{[1]}(\epsilon;\,\tau)
\right) + 
\notag \\
&\quad+
J_{q/q}^{[1],\,(\lambda)}(\epsilon;\,\tau,\,z)-
z \, \widetilde{D}_{q/q}^{[1],\,(\lambda)}
(\epsilon;\,z,\,\zeta) \,\delta(\tau)
\bigg] = 
\notag \\
&\quad=
\frac{H_{0,\,f}}{2} \,
\frac{\alpha_S}{4 \pi} \, 
2 \, C_F \,
\Bigg\{
\delta(1-z)\,
\Bigg[
\delta(\tau)\,
\Bigg(
-\frac{9}{2} + \frac{\pi^2}{3} -
\frac{3}{2}\,
\log{\frac{\mu^2}{Q^2}}-
\left(\log{\frac{\mu^2}{Q^2}}\right)^2
\notag \\
&\quad+
\log{\frac{\zeta}{\mu^2}}\,
\log{\frac{\mu^2}{\lambda^2}} +
2\,\log{\frac{\mu^2}{Q^2}}\,
\log{\frac{\mu^2}{\lambda^2}}-
\frac{1}{2}
\left(\log{\frac{\mu^2}{\lambda^2}}\right)^2
\Bigg) -
\frac{3}{2} \left(\frac{1}{\tau}\right)_+ - 
4 \left(\frac{\log{\tau}}{\tau}\right)_+
\Bigg]+ 
\notag \\
&\quad+
2 \delta(\tau)\,
\left[
-\frac{\log{z}}{1-z} + 
\left(\frac{\log{(1-z)}}{1-z}\right)_+
\right]
\Bigg\},
\label{eq:F1_quark_sub}
\end{align}
where we used Eqs.~\eqref{eq:vertex_1loop_2},~\eqref{eq:S_thrust},~\eqref{eq:B_thrust},~\eqref{eq:jetferm_thrust_lambda_exp} and~\eqref{eq:quark_TMD_cutoff} to expand in powers of $\epsilon$ all terms\footnote{In the final result we have removed $S_\epsilon$.
This fixes the renormalization scheme as the $\overline{\mbox{MS}}$ scheme.} that contribute to the subtracted, renormalized structure function $\widehat{F}_{1,\,f}^{[1]}$.
Notice that the final result is a \emph{finite} quantity, since all the divergences have been canceled.
Finally, the subtracted, renormalized partonic cross section for the fragmenting fermion follows trivially from Eq.~\eqref{eq:F1_quark_sub}:
\begin{align}
&\left.\frac{d \widehat{\sigma}_f^{[1]}}{dz \, dT} \right \rvert_{\mbox{\small sub, R}} =  
\sigma_B \, z \, 
\left.
\widehat{F}_{1,\,f}^{[1]}
(z,\,1-T,\,\lambda,\,\zeta )
\right \rvert_{\mbox{\small sub, R}}.
\label{eq:sub_xs_ferm}
\end{align}

\bigskip

The result of Eq.~\eqref{eq:sub_xs_ferm} can easily be generalized to all orders.
%
\begin{figure}[t]
\centering
\includegraphics[width=14cm]{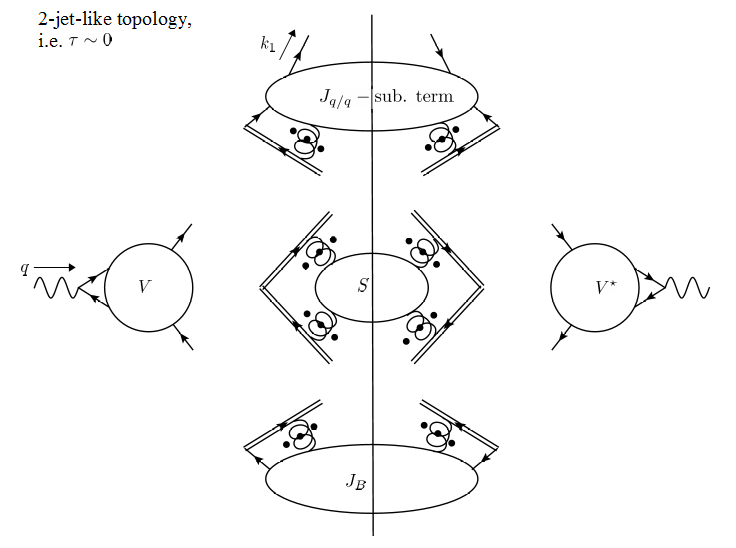}
\caption{Factorization of the partonic final state tensor $\widehat{W}^{\mu\nu}_j$ in a $2$-jet-like topology. Every blob can be computed in perturbation theory.  All terms appearing in Eq.~\eqref{eq:F1_quark_sub} can easily be recognized: the virtual vertex $V(\epsilon) = V \, V^\star$, the soft thrust function $S(\epsilon;\,\tau) = S$, the backward thrust function $J_B(\epsilon;\,\tau) = J_B$ and the jet thrust function equipped with the subtraction mechanism described in Section~\ref{subsubsec:frag_quark_sub}.}
\label{fig:W_fact_2jet}
\end{figure}
%
A $2$-jet-like topology of the final state implies that the partonic final state tensor can be factorized as depicted in Fig.~\ref{fig:W_fact_2jet}. This means:
\begin{align}
&
\left.
\widehat{W}_f^{\mu \nu}
(z,\,\tau,\,\lambda,\,\zeta)
\right \rvert_{\mbox{\small sub, R}}=
H_T^{\mu\nu}(z) \,
\frac{H_{0,\,f}}{2} \,
V(\epsilon)\,
\times \notag \\
&\quad \times
\int d\tau_1 \, d\tau_2 \, d\tau_S \,
\left.J_{q/q}^{(\lambda)}
(\epsilon;\,\tau_1,\,z,\,\zeta)
\right \rvert_{\mbox{\small sub, R}}
J_B(\epsilon;\,\tau_2)\,
S(\epsilon;\,\tau_S)\,
\delta(\tau-\tau_1-\tau_2-\tau_S),
\label{eq:all_orders_Wferm}
\end{align}
where the subtracted, renormalized thrust jet function for the fragmenting fermion is related to its unsubtracted analogue by:
\begin{align}
J_{q/q}^{(\lambda)}(\epsilon;\,\tau,\,z) = 
\int_z^1 \, 
\frac{d\widehat{z}}{\widehat{z}}\,
\left.J_{q/q}^{(\lambda)}
(\epsilon;\,\tau,\,{z}/{\widehat{z}},\,\zeta)
\right \rvert_{\mbox{\small sub, R}} \,
\left[ 
\widehat{z} \, \widetilde{D}_{q/q}^{(\lambda)}
(\epsilon;\,\widehat{z},\,\zeta)
\right].
\label{eq:Jqq_unsub_gen}
\end{align}
This relation can be inverted by using an inverse Mellin transform:
\begin{align}
\left.J_{q/q}^{(\lambda)}
(\epsilon;\,\tau,\,z,\,\zeta)
\right \rvert_{\mbox{\small sub, R}} = 
\frac{1}{2\pi i} \,
\int_{c-i\infty}^{c+i\infty}\,ds \, z^{-s}\,
\dfrac{
J_{q/q}^{(\lambda)}(\epsilon;\,\tau,\,s)}
{\widetilde{D}_{q/q}^{(\lambda)}
(\epsilon;\,s+1,\,\zeta)},
\label{eq:Jqq_sub_gen}
\end{align}
for a suitable real $c$.
Notice that in Eq.~\eqref{eq:all_orders_Wferm}, all the non-trivial $z$ dependence is  encoded in the subtracted, renormalized thrust jet function $J_{q/q}^{(\lambda)}$, i.e. in the contribution 
that describes the radiation collinear to the fragmenting fermion.
Finally, the partonic cross section at all order can be written as:
\begin{align}
&\left.\frac{d \widehat{\sigma}_f(\mu,\,\lambda,\,\zeta)}
{dz \, dT} \right \rvert_{\mbox{\small sub, R}} = \sigma_B \, z \, N_C \, e_f^2 \,
V(\epsilon)\,
\times \notag \\
&\quad \times
\int d\tau_1 \, d\tau_2 \, d\tau_S \,
\left.J_{q/q}^{(\lambda)}
(\epsilon;\,\tau_1,\,z,\,\zeta)
\right \rvert_{\mbox{\small sub, R}}
J_B(\epsilon;\,\tau_2)\,
S(\epsilon;\,\tau_S)\,
\delta(1-T-\tau_1-\tau_2-\tau_S),
\label{eq:all_orders_partonicxs}
\end{align}
This result closely resembles the structure of the thrust distribution associated to $\epm$ events, see e.g. Refs.~\cite{Catani:1991kz,Sterman:1995fz,GehrmannDeRidder:2007hr,Becher:2008cf}. Clearly, this structure will be transferred to the final cross section (see Eq.~\eqref{eq:xs_final_allorders}).
In particular, the (finite part\footnote{Since the subtracted, renormalized partonic cross section is a finite quantity, we can drop all the dependence on $\epsilon$ in the l.h.s of Eq.~\eqref{eq:all_orders_partonicxs}.} of the) functions $V$, $J_B$ and $S$ are well-known objects that have been widely studied in the past, and they have been resummed to obtain the correct small-$\tau$ behaviour of the $\epm$ thrust distribution, see e.g. Refs.~\cite{Catani:1992ua,Monni:2011gb,Schwartz:2013pla}. 
An analogous resummation of the \emph{new} function $J_{q/q}^{(\lambda)}$, subtracted and renormalized, is presently not known, and its computation goes beyond the purposes of this paper.
In order to try to bypass this difficulty, in the next Section we will propose a CSS-like resummation procedure for the whole partonic cross section, by deriving the proper evolution equations for each of the scales appearing in our final result.
We will omit the labels ``sub" and ``R" since, from now on, we will always refer to subtracted and renormalized quantities.

\bigskip

\subsubsection{Evolution and Resummation \label{subsubsec:evo_resumm}}

\bigskip

Since the UV counterterm that renormalizes the subtracted partonic cross section is exactly the inverse of the UV counterterm of the TMD FF for a fragmenting fermion, the subtracted, renormalized partonic cross section obeys the following RG-evolution equation:
\begin{align}
\frac{\partial}{\partial \log{\mu}}\,
\log{
\frac{d \widehat{\sigma}_f
(\mu,\,\lambda,\,\zeta)}
{dz \, dT}
}= 
-\gamma_D\left(\alpha_S(\mu),\,
{\zeta}/{\mu^2}\right),
\label{eq:RG_evo_partonicxs}
\end{align}
where $\gamma_D$ is the anomalous dimension of the TMD FF associated to a fragmenting fermion, see Eq.~\eqref{eq:RG_evo}. 
At 1-loop, it is given by:
\begin{align}
\gamma_D\left(\alpha_S(\mu),\,
{\zeta}/{\mu^2}\right) 
\overset{\text{1-loop}}{=}
\frac{\alpha_S(\mu)}{4 \pi} \, 
2 \, C_F \,
\left(
3 - 2\log{\frac{\zeta}{\mu^2}}
\right).
\label{eq:gamma_D_1loop}
\end{align}
Then, the Eq.~\eqref{eq:RG_evo_partonicxs} can easily be verified at 1-loop by using the result of Eq.~\eqref{eq:sub_xs_ferm}.
The derivative with respect to the rapidity cut-off $\zeta$ plays the role of the CS-evolution equation for the TMDs, see Eq.~\eqref{eq:CS_evo}.
It is given by:
\begin{align}
\frac{\partial}{\partial \log{\sqrt{\zeta}}}\,
\log{
\frac{d \widehat{\sigma}_f
(\mu,\,\lambda,\,\zeta)}
{dz \, dT}} = 
\frac{1}{2}\,
\widehat{K}\left( \alpha_S(\mu),\,{\mu^2}/{\lambda^2}\right).
\label{eq:CS_evo_partonicxs}
\end{align}
Notice that, analogously to the CS-evolution for the TMDs, the kernel $\widehat{K}$ does not depend on the rapidity cut-off.
At 1-loop, we can compute $\widehat{K}$ from Eq.~\eqref{eq:sub_xs_ferm}:
\begin{align}
\widehat{K}\left( \alpha_S(\mu),\,{\mu^2}/{\lambda^2}\right) 
\overset{\text{1-loop}}{=}
\frac{\alpha_S(\mu)}{4 \pi} \, 
8  C_F \,
\log{\frac{\mu^2}{\lambda^2}}.
\label{eq:partonicK_1loop}
\end{align}
The analogy with the soft kernel $\widetilde{K}$ is evident if we consider the RG-evolution of $\widehat{K}$. In fact, by combining Eqs.~\eqref{eq:RG_evo_partonicxs} and~\eqref{eq:CS_evo_partonicxs} with Eq.~\eqref{eq:gammaD_evo}, we have:
\begin{align}
\frac{\partial}{\partial \log{\mu}} \,
\widehat{K}\left( \alpha_S(\mu),\,{\mu^2}/{\lambda^2}\right) = 
\gamma_K\left(\alpha_S(\mu)\right),
\label{eq:RG_evo_partonicK}
\end{align}
which has to be compared with Eq.~\eqref{eq:Ktilde_evo}.
The function $\gamma_K$ is the anomalous dimension of the soft kernel (see Eq.~\eqref{eq:Ktilde_evo}).
At 1-loop:
\begin{align}
\gamma_K^{[1]} = 16 C_F,
\label{eq:gammaK_1loop}
\end{align}
in agreement with Eq.~\eqref{eq:partonicK_1loop}.
Notice that, while the TMD depends only on two energy scales ($\mu$ and $\zeta$), the subtracted, renormalized partonic cross section depends also on $\lambda$, the cut-off limiting the transverse momentum range of the fragmenting parton.
Therefore, we define the following $\lambda$-evolution equation:
\begin{align}
\frac{\partial}{\partial \log{\lambda}}\,
\log{
\frac{d \widehat{\sigma}_f(\mu,\,\lambda)}
{dz \, dT}
}= 
G\left(\alpha_S(\mu),\,
{\mu^2}/{Q^2},\,{\zeta}/{\mu^2},\,
{\mu^2}/{\lambda^2}\right),
\label{eq:lambda_evo_partonicxs}
\end{align}
where $G$ is the evolution kernel.
At 1-loop, its value can be computed directly from Eq.~\eqref{eq:sub_xs_ferm}:
\begin{align}
G\left(\alpha_S(\mu),\,
{\mu^2}/{Q^2},\,{\zeta}/{\mu^2},\,
{\mu^2}/{\lambda^2}\right) 
\overset{\text{1-loop}}{=}
-\frac{\alpha_S(\mu)}{4 \pi} \, 
4 \, C_F \,
\left(
2 \log{\frac{\mu^2}{Q^2}} + 
\log{\frac{\zeta}{\mu^2}}-
\log{\frac{\mu^2}{\lambda^2}}
\right).
\label{eq:G_1loop}
\end{align}
This kernel is RG-invariant. In fact, as the r.h.s of Eq.~\eqref{eq:RG_evo_partonicxs} does not depend on $\lambda$, we can easily write:
\begin{align}
\frac{\partial}{\partial \log{\mu}}
G\left(\alpha_S(\mu),\,
{\mu^2}/{Q^2},\,{\zeta}/{\mu^2},\,
{\mu^2}/{\lambda^2}\right) = 0.
\label{eq:RG_evo_G}
\end{align}
Finally, the CS-evolution for $G$ can be found by combining Eqs.~\eqref{eq:CS_evo_partonicxs} and~\eqref{eq:lambda_evo_partonicxs}:
\begin{align}
\frac{\partial}{\partial \log{\sqrt{\zeta}}}
G\left(\alpha_S(\mu),\,
{\mu^2}/{Q^2},\,{\zeta}/{\mu^2},\,
{\mu^2}/{\lambda^2}\right) = 
\frac{1}{2} \,
\frac{\partial}{\partial \log{\lambda}}
\widehat{K}\left( \alpha_S(\mu),\,{\mu^2}/{\lambda^2}\right).
\label{eq:CS_evo_G}
\end{align}

\bigskip

The evolution equations presented in Eqs.~\eqref{eq:RG_evo_partonicxs},~\eqref{eq:CS_evo_partonicxs} and~\eqref{eq:lambda_evo_partonicxs} can be solved in order to obtain an expression for the partonic cross section that resums the dependence on $\mu$, $\zeta$ and $\lambda$.
This is particularly important for the transverse momentum cut-off. In fact, all the results obtained so far are valid in the limit of ${\lambda^2}/{Q^2} \to 0$, that corresponds to the  $2$-jet limit of the final state topology.
On the other hand, a perturbative expansion can only be trusted when the parameter of the expansion is small. Here, even if $\alpha_S$ is a small number, the same does not necessarily hold for the product $\alpha_S \log{{\lambda^2}/{Q^2}}$, especially in the limit of vanishing cut-off.
As a consequence, a resummed formula is necessary not only to obtain an all-order expression for the partonic cross section, but also to provide an adequate description of the $2$-jet-like topology.
We have:
\begin{align}
&\frac{d \widehat{\sigma}_f
(\mu,\,\lambda,\,\zeta)}
{dz \, dT} = 
\left.\frac{d \widehat{\sigma}_f}
{dz \, dT}\right\rvert_{
\substack{\text{ref.}}}\,
\mbox{exp}
\Bigg\{
\int_\mu^Q \,\frac{d\mu'}{\mu'}\,
\gamma_D\left( \alpha_S(\mu'),\,{\zeta}/{(\mu')^2}\right)
\Bigg\} \times
\notag \\
&\quad\times
\mbox{exp}
\Bigg\{
\frac{1}{4}\,
\widehat{K}\left( \alpha_S(Q),\,1\right)\,
\log{\frac{\zeta}{Q^2}} -
\int_\lambda^Q \,\frac{d\lambda'}{\lambda'}\,
G\left(\alpha_S(Q),\,
1,\,{\zeta}/{Q^2},\,
{Q^2}/{(\lambda')^2}\right)
\Bigg\},
\label{eq:xs_ferm_resumm}
\end{align}
where we have used the RG-invariance of the kernel $G$. The label ``ref." indicates that the energy scales are fixed to their the reference values: $\mu = Q$, $\zeta = Q^2$ and $\lambda = Q$. With this choice, all logarithms involving energy scale ratios vanish in the structure function $\widehat{F}_{1,\,f}$ of Eq.~\eqref{eq:F1_quark_sub}. 
Therefore, the cross section computed at the reference scales can be expanded in powers of $\alpha_S(Q)$, which now can be considered a small parameter, and the perturbative expansion is reliable to any order.
The expression to NLO is given by the LO result of Eq.~\eqref{eq:lo_partonic_xs} added to the 1-loop result presented in Eq.~\eqref{eq:sub_xs_ferm} in which all logs are set to zero. This gives:
\begin{align}
&\left.\frac{d \widehat{\sigma}_f^{NLO}}
{dz \, dT}\right\rvert_{\substack{\text{ref.}}} =
\sigma_B \, z \, 
\left( 
\widehat{F}_{1,\,f}^{[0]}
(z,\,\tau)+
\frac{\alpha_S(Q)}{4\pi}
\left.\widehat{F}_{1,\,f}^{[1]}
(z,\,\tau)\right\rvert_{\substack{\text{ref.}}}
\right) = 
\notag \\
&\quad=
\sigma_B \,
e_f^2 \,N_C \,
\Bigg( 
\delta(1-z) \, \delta(\tau) 
+
\notag \\
&\quad+
\frac{\alpha_S(Q)}{4 \pi} \, 
2 \, C_F \,
\Bigg\{
\delta(1-z)\,
\left[
\delta(\tau)\,
\left(
-\frac{9}{2} + \frac{\pi^2}{3}
\right) -
\frac{3}{2} \left(\frac{1}{\tau}\right)_+ - 
4 \left(\frac{\log{\tau}}{\tau}\right)_+
\right]+ 
\notag \\
&\quad+
2 \left[
-\frac{z}{1-z}\,\log{z} -
\log{(1-z)}+
\left(\frac{\log{(1-z)}}{1-z}\right)_+
\right] \,\delta(\tau)
\Bigg\}
\Bigg).
\label{eq:partonic_xs_NLO_ref}
\end{align}
It is important to notice that all terms containing the non-trivial dependence on thrust $\tau$ are multiplied by $\delta(1-z)$, while all terms containing the non-trivial dependence on $z$ are multiplied by $\delta(\tau)$.
In Eq.~\eqref{eq:xs_ferm_resumm}, all the dependence on $\lambda$ has been confined in the $G$-term of the exponent in the second line.
Now the limit ${\lambda}/{Q} \to 0$ is not an hazard anymore: in fact, as the cut-off decreases, the partonic cross section becomes more and more suppressed.
This exponent can easily be computed in perturbation theory, since the strong coupling $\alpha_S$ is evaluated at the scale $Q$ and, consequently, it is not involved in the integration. 
We can use the 1-loop expressions for $\widehat{K}^{[1]}$ and $G^{[1]}$ in Eqs.~\eqref{eq:partonicK_1loop} and~\eqref{eq:G_1loop} to obtain:
\begin{align}
&\mbox{exp}
\Bigg\{
\frac{1}{4}\,
\widehat{K}\left( \alpha_S(Q),\,1\right)\,
\log{\frac{\zeta}{Q^2}} -
\int_\lambda^Q \,\frac{d\lambda'}{\lambda'}\,
G\left(\alpha_S(Q),\,
1,\,{\zeta}/{Q^2},\,
{Q^2}/{(\lambda')^2}\right)
\Bigg\} 
\overset{\text{1-loop}}{=}
\notag \\
&\quad\overset{\text{1-loop}}{=}
\mbox{exp}
\left\{
-\frac{\alpha_S(Q)}{4 \pi}\, C_F \,
\left[
2\log{\frac{\zeta}{Q^2}}\,
\log{\frac{\lambda^2}{Q^2}}+
\left(\log{\frac{\lambda^2}{Q^2}}\right)^2
\right]
\right\}
\label{eq:lambda_suppr_xspartonic}
\end{align}
Therefore, for any $\zeta \neq 0$, the partonic cross section goes to zero when the $\lambda$ cut-off vanishes.
As observed in the final part of Section~\ref{subsubsec:frag_gluon_sub}, this cut-off is related to the measured value of thrust.
In particular, since 
$\tau_{\mbox{\tiny meas.}} = {\lambda^2}/{Q^2}$
the suppression is associated to the topology of the final state. 
This correctly implies that for  perfectly pencil-like event ($\tau_{\mbox{\tiny meas.}} = 0$), the partonic cross section vanishes.
Clearly, this is a very unrealistic case, in which the radius of the jet is zero and the fragmentation process develops along the direction of the fragmenting quark. The probability of finding a $\tau_{\mbox{\tiny meas.}} = 0$ event is then exactly zero.

Eq.~\eqref{eq:xs_ferm_resumm} represents the exact expression of the partonic cross section, to all orders. In particular, the exponent in the second line of 
Eq.~\eqref{eq:xs_ferm_resumm} correctly expresses its all-order resummation with respect to $\lambda$ (and $\zeta$). Instead, 
in Eq.~\eqref{eq:lambda_suppr_xspartonic} the exponent is truncated to 1-loop accuracy in the perturbative expansion. This has some effects on the log counting.
In fact, a rigorous NLL computation would also include terms of order $\alpha_S^2$ that we explicitly neglected in the NLO computation of the partonic cross section. 
In particular, the correlated emission of two gluons, each going in one of the two  hemispheres defined by the thrust axis, can produce non-global logarithms (NGLs) of ${\lambda^2}/{Q^2}$.  
A proper treatment of such contributions goes beyond the purpose of this paper, since our partonic cross section is computed up to NLO, i.e. considering only 
single gluon emissions.
The non-global effects of $\epm \to H\,X$ have been investigated in Refs.~\cite{Kang:2020yqw} and~\cite{Makris:2020ltr}. General, non-global observables have been studied e.g. in Refs.~\cite{Dasgupta:2001sh,Banfi:2002hw,Larkoski:2015zka,Becher:2017nof}.
The inclusion of the contributions beyond the NLO in the resummed partonic cross section generates difficulties associated to the determination of the resummed expression for the subtracted, renormalized function $J_{q/q}^{(\lambda)}$ (see the discussion in the end of the previous Section).
The resummation proposed above allows to bypass such difficulties by providing a result that can be trusted at fixed order in perturbation theory.

\bigskip

\section{TMD FF for the fragmenting quark \label{sec:nll_tmd}}

\bigskip

The long-distance behavior of the cross section for $\epm \to H\,X$ presented in Eq.~\eqref{eq:xs_final_general} is encoded in the unpolarized TMD FF, which provides a 3D description of the fragmentation of a parton of type $j$ 
into a spinless hadron $H$. 
TMD FFs depend on the transverse momentum of the parton with respect to the direction of the produced hadron and reflect the dynamics of hadronizing partons.
In the computation of the partonic cross section carried out in Section~\ref{sec:partonic_xs}, we have found that only the TMDs associated to fragmenting fermions survive in the sum over parton-types in Eq.~\eqref{eq:xs_final_general}, as the contribution of the fragmenting gluon is suppressed in a $2$-jet-like topology, see Eq.~\eqref{eq:sub_xs_gluon}.
Therefore, in the following we will define the TMD FF of a fragmenting fermion of flavor $f$, providing a review of all the explicit expressions relevant to its computation to NLL accuracy. 

The TMD FFs are defined in the Fourier conjugate space of the transverse momentum $\vec{k}_T$ of the fragmenting parton. The conjugate variable is commonly denoted as $\vec{b}_T$.
Furthermore, they are equipped with a rapidity cut-off $y_1$ which constrains the rapidity of the parton described by the TMDs.
We will use the following definition for $D_1$ (see Ref.~\cite{Collins:2011zzd} and Ref.~\cite{Boglione:2020cwn}):
\begin{align}
&\widetilde{D}_{1,\,H/j}
(z,\, b_T; \, \mu, \, y_P - y_1) = 
\notag \\
&\quad=
Z_j(\mu,\, y_P - y_1) 
Z_2 \left(\alpha_S(\mu)\right) 
\lim_{y_{u_2} \to -\infty}
\dfrac{\widetilde{D}_{1,\,H/j}^{(0),\,
\mbox{\small unsub}} 
(z, \,b_T; \, \mu, \, y_P - y_{u_2})}
{\ftsoft{2}^{(0)} (b_T; \, \mu, \,y_1 - y_{u_2})} \,,
\label{eq:TMDFF_def_1}
\end{align}
where $y_P$ is the rapidity of the detected hadron $H$.
This definition holds for any fragmenting parton of type $j$.
The function $Z_j$ is the UV counterterm that renormalizes the TMD FF. Its dependence on $j$ originates from the color representation of the fragmenting parton (gluon or fermion).
Eq.~\eqref{eq:TMDFF_def_1} shows how the overlapping between the TMD and the soft momentum region is removed by a subtraction mechanism similar to that used in Section~\ref{sec:partonic_xs} to cancel the double counting between the hard and the collinear momentum region.
In the case of a fragmenting fermion of flavor $f$, at numerator we have the unsubtracted TMD FF defined as:
\begin{align}
&\widetilde{D}_{1,\,H/f}^{(0),\,
\mbox{\small unsub}} 
(z, \,b_T; \, \mu, \, y_P - y_{u_2})=
\notag \\
&\quad=
\dfrac{1}{z} \,
\sum\limits_{X} \, 
\langle P\,(H),\,X;\,\mbox{out} | \overline{\psi}_f(-{x}/{2}) \, W_q(-{x}/{2}, \infty;\, 
n_1(y_1)\,)^\dagger | 0 \rangle 
\times \notag \\
&\quad\times \, 
\langle 0 | W_q({x}/{2}, \infty;\, w_{-}\,) \, \psi_f({x}/{2})|  
P\,(H),\,X;\,\mbox{out} \rangle \, \vert_{\mbox{\small NO S.I.}},
\label{eq:TMDFF_unsub}
\end{align}
where $x = ( 0,\,x^-,\,\vec{b}_T )$ and $P$ is the momentum of the outgoing hadron $H$.
The (renormalized) quark field is $\psi_f$, while the operators $W_q$ are Wilson lines associated to a fermion field along the light-like minus direction $w_{-}$.
The label ``NO S.I." indicates that the Wilson line self energies must not be considered, while the label ``(0)" refers to the fact that the definition of Eq.~\eqref{eq:TMDFF_unsub} holds for the bare unsubtracted TMD, defined with bare fields. 
For this reason, we multiplied the whole expression by 
$Z_2$, which is the wave-function renormalization factor for the quark field. 
This fixes the UV counterterm $Z_q$ as the renormalization factor of the bare TMD, defined with bare fields.
Notice that since the Wilson lines in $\widetilde{D}_{1,\,H/f}^{(0)}$ are defined along a light-like direction, the unsubtracted TMD is affected by unregulated rapidity divergences.
The denominator in Eq.~\ref{eq:TMDFF_def_1} is a $2$-h soft factor\footnote{See the classification in Ref.~\cite{Boglione:2020cwn}.} and it represents the overlapping contribution with the soft momentum region. 
Its subtraction removes the double counting and regulates the rapidity divergences.
In fact, $\ftsoft{2}$ represents all the soft gluons emitted by the fragmenting fermion along the jet direction, i.e. all the gluons with a small transverse momentum and a rapidity no  larger than the rapidity cut-off $y_1$.
The $2$-h soft factor is a color singlet. Then $\ftsoft{2}$ is defined as the coefficient of the identity matrix in color space:
\begin{align}
&\ftsoft{2}^{(0)}
(b_T;\, \mu, \,y_1 - y_{u_2}) =
\frac{\mbox{Tr}_C}{N_C} \,
\langle 0 | 
W(-{\vec{b}_T}/{2}, \,\infty; \, n_1(y_1)\,)^\dagger \, 
W({\vec{b}_T}/{2}, \,\infty; \, n_1(y_1)\,) \,
\times \notag \\
&\quad \times \,  
W({\vec{b}_T}/{2}, \,\infty; \, w_{-}\,)^\dagger \, 
W(-{\vec{b}_T}/{2}, \,\infty; \, w_{-}\,) 
| 0 \rangle \, \vert_{\mbox{\small NO S.I.}},
\label{eq:S2h_def}
\end{align}
where the trace is over all color indices.
Also in this case, the label ``(0)" reminds that Eq.~\eqref{eq:S2h_def} defines the bare soft factor, in terms of bare fields.
Finally, let us point out that the TMD FF $D_1$ as defined above represents all the collinear radiation of the jet initiated by a fermion of flavor $f$ that fragments into the hadron $H$. By collinear, we mean that all partons described by the TMD FF have a small transverse momentum $k_T$ and a large rapidity $y$ that lies in the range $y_1 \leq y \leq y_P$.
The TMD FF $D_1$ defined in Eq.~\eqref{eq:TMDFF_def_1} obeys the following evolution equations:
\begin{align}
&\dfrac{\partial \log{\widetilde{D}_{1,\,H/j}
(z,\, b_T; \, \mu, \,\zeta)}}
{\partial \log{\mu}} = 
\gamma_D 
\left(\alpha_S(\mu),\,{\zeta}/{\mu^2} \right) ,
\quad &\mbox{RG-evolution;}\label{eq:RG_evo}
\\
&\dfrac{\partial \log{\widetilde{D}_{1,\,H/j}
(z,\, b_T; \, \mu, \,\zeta)}}
{\partial \log{\sqrt{\zeta}}} = 
\frac{1}{2} \widetilde{K}(b_T;\,\mu), 
\quad &\mbox{CS-evolution,}\label{eq:CS_evo}
\end{align}
where, as usual, we defined:
\begin{align}
\zeta = 
2 (k^+)^2 \, e^{-2y_1} = 
\frac{M_H^2}{z_h^2} \, e^{2(y_P-y_1)},
\label{eq:zeta_def}
\end{align}
and $k^+$ is the plus component of the momentum of the fragmenting parton.
The function $\gamma_D$ is the anomalous dimension of the TMD FF, while $\widetilde{K}$ is the rapidity-independent kernel of the CS-evolution. They, in turn, solve the following equations:
\begin{align}
&\frac{\partial 
\gamma_D 
\left(\alpha_S(\mu),\,{\zeta}/{\mu^2} \right)}
{\partial \log{\sqrt{\zeta}}} = -\frac{1}{2}\gamma_K\left(\alpha_s(\mu)\right) ; 
\label{eq:gammaD_evo}\\
&\frac{d  \widetilde{K}(b_T;\,\mu)}{d \log{\mu}} = - 
\gamma_K\left(\alpha_s(\mu)\right),
\label{eq:Ktilde_evo}
\end{align}
where $\gamma_K$ is the anomalous dimension of the soft kernel $\widetilde{K}$.
The solution to Eqs.~\eqref{eq:RG_evo} and~\eqref{eq:CS_evo} is given by~\cite{Aybat:2011zv,Aybat:2011ge}:
\begin{align}
&\widetilde{D}_{1,\,H/f}
(z,\, b_T; \, \mu, \,\zeta) = 
\underbrace{
\frac{1}{z^2}\, 
\sum_k \, \int_z^1 \, \frac{d\rho}{\rho}\,
d_{H/k}({z}/{\rho},\,\mu_b)\,
\left[
\rho^2 \,
\mathcal{C}_{k/f}
\left(\rho,\,\alpha_S(\mu_b)\right)
\right]
}_{\mbox{TMD at reference scale}} \,
\times
\notag \\
&\quad\times
\underbrace{
\mbox{exp}
\left\{
\frac{1}{4} \, \widetilde{K}(b_T^\star;\,\mu_b) \, \log{\frac{\zeta}{\mu_b^2}} 
+
\int_{\mu_b}^{\mu} \frac{d \mu'}{\mu'} \, \left[ \gamma_D(\alpha_S(\mu'),\,1) - \frac{1}{4} \, 
\gamma_K(\alpha_S(\mu')) \, \log{\frac{\zeta}{\mu'^2}}\right]
\right\}
}_{\mbox{Perturbative Sudakov Factor}} \,
\times
\notag \\
&\quad\times
\underbrace{
\left(M_D\right)_{j,\,H}(z,\,b_T) 
\mbox{ exp} \left\{
-\frac{1}{4} \, g_K(b_T) \, \log{\frac{z_h^2 \, \zeta}{M_H^2}}
\right \}
}_{\mbox{Non-Perturbative content}} .
\label{eq:TMDFF_def_2}
\end{align}
In the previous expression, the reference scales are $\mu = \mu_b$ and $\zeta=\mu_b^2$, where $\mu_b$ is defined as:
\begin{align}
&\mu_b = \frac{2 e^{-\gamma_E}}{b_T^\star(b_T)},
\label{eq:mub_def}
\end{align}
where the $b_T^\star$ prescription allows to separate the perturbative small-$b_T$ behavior of the TMD from its non-perturbative large $b_T$ content. In fact, $b_T^\star(\vec{b}_T)$ is the same of $\vec{b}_T$ at small $b_T$, while at large $b_T$ it is no larger than a certain $b_{\mbox{\scriptsize max}}$:
\begin{align}
&\vec{b}_T^\star \left(b_T\right) = \dfrac{\vec{b}_T}{\sqrt{1 + {b_T^2}/{b_{\rm{max}}^2}}}.
\label{bstar_def}
\end{align}
We also introduce a minimum value $b_{\mbox{\scriptsize min}}$, that allows to recover the collinear FFs by integrating over the transverse momentum of the fragmenting parton. Therefore we will adopt the modified $b_T^\star$ prescription defined as:
\begin{align}
&\vec{b}_T^\star\left(b_c (b_T)\right) = 
\vec{b}_T^\star\left(\sqrt{b_T^2+b^2_{\mbox{\tiny min}}}\right).
\label{bstarmod_def}
\end{align}
In this paper, we use standard choices for the minimum and maximum value of $b_T$ involved in the previous equations. In particular, we set $b_{\mbox{\scriptsize max}} = 1$ GeV$^{-1}$ and $b_{\mbox{\scriptsize min}} = {C_1}/{Q}$, where $C_1 = 2 e^{-\gamma_E}$.

\bigskip

It is important to underline that the TMD definition adopted here, Eq.~\eqref{eq:TMDFF_def_1}, differs from the usual definition commonly used in the factorized cross sections of $2$-h class processes like SIDIS and $\epm \to H_1\,H_2\,X$, with the two hadrons almost back to back (see Refs.~\cite{Boglione:2020cwn} and, e.g., Ref.~\cite{Collins:2011zzd}). 
In such processes, the final cross section presents a soft factor connecting the target and the detected hadrons, in SIDIS, and the two back-to-back hadrons, in $\epm \to H_1\,H_2\,X$.
This soft factor cannot be fully computed in perturbative QCD, since it encodes non-perturbative information about the soft radiation that flows through the two collinear parts.
But it cannot be directly extracted from experimental data either, since it always appears in connection to the two collinear parts.
Notice that, despite the similarities, this soft factor \emph{is not} the same object that we encountered in the derivation of the partonic cross section (see e.g. Eq.~\eqref{eq:all_orders_partonicxs} or Fig.~\ref{fig:W_fact_2jet}). 
In fact, as shown to 1-loop accuracy, $S(\epsilon;\,\tau)$ is totally predicted by perturbative QCD, see Eq.~\eqref{eq:S_thrust}.
To overcome the difficulties induced by the presence of a non-trivial soft term in the final cross section, the prescription of Ref.~\cite{Collins:2011zzd} would include part of this soft factor in the definition of the TMDs. This has the advantage of freeing the cross section of the explicit  presence of the soft factor, but creates a hadron-class dependence in the TMDs, lowering their degree of universality.  
In Ref.~\cite{Boglione:2020cwn} we investigated the relationship between this currently accepted definition, referred to as ``square root definition", and the ``factorization definition", adopted in this paper and free of any external soft contribution.
The two definitions are totally equivalent at small-$b_T$, in the perturbative region, but they show a rather different large-$b_T$, non-perturbative, behavior. 
It is not by chance that the ``square root definition" is obtained by multiplying the TMD defined in Eq.~\eqref{eq:TMDFF_def_1} by the square root of $M_S(b_T)$, the non-perturbative function that models the large-$b_T$ behavior of the soft factor:
\begin{align} 
&\widetilde{D}_{1,\,H/f}^{\mbox{\small sqrt}}
(z,\, b_T; \, \mu, \,\zeta)
=  \sqrt{M_S(b_T)} \times 
\widetilde{D}_{1,\,H/f}
(z,\, b_T; \, \mu, \,\zeta),
\label{eq:def_comparison}
\end{align}
The ``square root definition" is optimal for cross sections corresponding to the $2$-h class (Drell-Yan, SIDIS, $\epm \to H_1\,H_2\,X$) in which two collinear parts appear, each associated to one of the two reference hadrons. 
However, this definition lowers the degree of universality of TMDs, since it includes some extra soft physics information typical of the $2$-h class.
In contrast, the ``factorization definition" of Eq.~\eqref{eq:TMDFF_def_1} allows to define a totally universal TMD, which can be applied equally well to processes like $\epm \to H\,X$ that do not belong to the $2$-h class.
Notice that Eq.~\eqref{eq:def_comparison} is of crucial importance from a phenomenological point of view, as it relates the TMDs obtained from data analyses based on the square 
root definition (widely used in the last decade) to the TMDs 
extracted using the factorization definition. This implies that all previous work on the extraction of polarized and unpolarized TMDs could still be easily exploited in global analyses. 
The universality breaking effects generated in processes belonging to different hadron classes,
as addressed in Ref.~\cite{Boglione:2020cwn}, have also been investigated in Ref.~\cite{delCastillo:2020omr}. Here TMD factorization for 3-h class processes, like dijet and heavy-meson pair production in SIDIS, is presented in a SCET framework, where the introduction of a new soft function is required.

\bigskip

In the following, we will focus on the main three ingredients of Eq.~\eqref{eq:TMDFF_def_2} separately. In particular, in Section~\ref{subsec:tmd_refscale} we will review the NLO expression for the TMD FF at the reference scale, in Section~\ref{subsec:tmd_sud} we will compute the perturbative Sudakov factor at NLL precision and finally, in Section~\ref{subsec:tmd_nonpert} we will consider the non perturbative content of the TMD FF.

\bigskip

\subsection{Operator Product Expansion at NLO} \label{subsec:tmd_refscale}

\bigskip

The first term in Eq.~\eqref{eq:TMDFF_def_2} is the TMD FF at the reference scales $\mu = \mu_b$ and $\zeta = \mu_b^2$, where $\mu_b$ has been defined in Eq.~\eqref{eq:mub_def}. 
It is related to the short-distance, small-$b_T$-behavior of $D_1$ and therefore computable in perturbation theory. 
The application of the factorization procedure to the TMD itself shows that $D_1$ can be expressed in this region as an Operator Product Expansion (OPE), in which the basis of the operators is given by the collinear FFs $d_{h/k}$.
The Wilson coefficients $\mathcal{C}_{k/f}$ of the OPE are fully computable in perturbative QCD and can be expanded in powers of $\alpha_S(\mu_b)$.
Having set the scales to the reference values, no dangerous logarithms will affect the perturbative expansion, that can be considered reliable at any order.
In particular, at LO the Wilson coefficients are just delta functions:
\begin{align}
\mathcal{C}_{k / f}^{[0]}(z) = 
\delta_{k,\,f} \, \delta(1-z).
\label{eq:wils_LO}
\end{align}
To 1-loop we have the following expressions, see e.g. Refs.~\cite{Collins:2011zzd,Echevarria:2016scs,Collins:2017oxh}:
\begin{align}
&z^2 \, \mathcal{C}_{k / f}^{[1]}(z) =
z^2 \, \delta_{k,\,f} \, 
\mathcal{C}_{q / q}^{[1]}(z) =
\delta_{k,\,f} \, 2 C_F \, \left\{
1-z 
+ \left[
4 \,\left(\frac{1}{1-z} \right)_+ - 2 (1+z)
\right] \, \log{z}
\right\};\label{eq:wils_qq_1loop}
\\
&z^2 \, \mathcal{C}_{g / f}^{[1]}(z) =
z^2 \, \mathcal{C}_{g / q}^{[1]}(z) =
2 C_F \, \left[
z 
+ \left(
4 \, \frac{1-z}{z} + \frac{2}{z}
\right) \, \log{z}
\right]
.
\label{eq:wils_gq_1loop}
\end{align}
Notice that the plus prescription in Eq.~\eqref{eq:wils_qq_1loop} can be dropped.
Then, to NLO, the OPE is given by:
\begin{align}
&\sum_k \, \int_z^1 \, \frac{d\rho}{\rho}\,
d_{H/k}({z}/{\rho},\,\mu_b)\,
\left[
\rho^2 \,
\mathcal{C}_{k/f}^{NLO}
\left(\rho,\,\alpha_S(\mu_b)\right)
\right] = 
\notag \\
&\quad=
\sum_k \, \int_z^1 \, \frac{d\rho}{\rho}\,
d_{H/k}({z}/{\rho},\,\mu_b)\,
\left[
\rho^2 \,
\left(
\mathcal{C}_{k / f}^{[0]}(\rho) + 
\frac{\alpha_S(\mu_b)}{4\pi}\,
\mathcal{C}_{k / f}^{[1]}(\rho)
\right)
\right]=
\notag \\
&\quad=
d_{H/f}(z,\,\mu_b) + 
\frac{\alpha_S(\mu_b)}{4\pi}\,
\bigg\{
\int_z^1 \, \frac{d\rho}{\rho}\,
d_{H/f}({z}/{\rho},\,\mu_b)\,
\left[
\rho^2 \, 
\mathcal{C}_{q / q}^{[1]}(\rho)
\right] + 
\notag \\
&\quad+
\int_z^1 \, \frac{d\rho}{\rho}\,
d_{H/g}({z}/{\rho},\,\mu_b)\,
\left[
\rho^2 \, 
\mathcal{C}_{g / q}^{[1]}(\rho)
\right]
\bigg\},
\label{eq:OPE_NLO}
\end{align}
where we used Eqs.~\eqref{eq:wils_LO},~\eqref{eq:wils_qq_1loop} and~\eqref{eq:wils_gq_1loop}.

\bigskip

%
\subsection{Perturbative Sudakov Factor at NLL} \label{subsec:tmd_sud}

\bigskip

The second term in Eq.~\eqref{eq:TMDFF_def_2} is the perturbative part (small-$b_T$) of the Sudakov factor, which originates from the resummation of the TMD with respect to the scales $\mu$ and $\zeta$. The exponent cannot be expressed as a fixed order expansion, since the logarithms of the energy scales multiply $\alpha_S$ at any order. Therefore, the proper way to expand this quantity consists in counting the power of such logs and performing a NLL expansion (this guarantees  consistency with the NLO precision of the OPE, see Section~\ref{subsec:tmd_refscale}).
This operation is easily done by separating out the part that depends on the rapidity cut-off $\zeta$ from the rest of the exponent appearing in the perturbative Sudakov factor. 
This gives:
\begin{align}
&\mbox{exp} 
\left\{
\frac{1}{4} \, \widetilde{K}(b_T^\star;\,\mu_b) \, \log{\frac{\zeta}{\mu_b^2}} 
+
\int_{\mu_b}^{\mu} \frac{d \mu'}{\mu'} \, \left[ \gamma_D(\alpha_S(\mu'),\,1) - \frac{1}{4} \, 
\gamma_K(\alpha_S(\mu')) \, \log{\frac{\zeta}{\mu'^2}}\right]
\right\} = 
\notag \\
&\quad=
\mbox{exp} 
\left\{
\frac{1}{4} \, \widetilde{K}(b_T^\star;\,\mu_b) \, \log{\frac{\mu^2}{\mu_b^2}} 
+
\int_{\mu_b}^{\mu} \frac{d \mu'}{\mu'} \, 
\left[ \gamma_D(\alpha_S(\mu'),\,1) - \frac{1}{4} \, 
\gamma_K(\alpha_S(\mu')) \, \log{\frac{\mu^2}{\mu'^2}}\right]
\right\}\times
\notag \\
&\quad \times
\mbox{exp} 
\left\{
\frac{1}{4} \, 
\log{\frac{\zeta}{\mu^2}} \,
\left[
\widetilde{K}(b_T^\star;\,\mu_b) - 
\int_{\mu_b}^{\mu} \frac{d \mu'}{\mu'} \,
\gamma_K(\alpha_S(\mu'))
\right]
\right\}.
\label{eq:sud_sep}
\end{align}
The recipe to obtain the previous quantity at NLL accuracy is the following:
\begin{itemize}
\item The anomalous dimension $\gamma_K$ of the soft kernel is expanded up to 2-loops.
The 1-loop coefficient can be found in Eq.~\eqref{eq:gammaK_1loop}, while at 2-loops we have, see Refs.~\cite{Echevarria:2016scs,Collins:2017oxh}:
\begin{align}
&\gamma_K^{[2]} =
2 \, C_A \, C_F \,
\left(
\frac{536}{9} - \frac{8 \pi^2}{3}
\right) -
\frac{160}{9} \, C_F \, n_f,
\label{eq:gammaK_2loop}
\end{align}
where $n_f$ is the total number of fermion fields that we consider.
\item All the other quantities in the exponent are expanded up to 1-loop. Their expressions are given by, see Refs.~\cite{Echevarria:2016scs,Collins:2017oxh}:
\begin{align}
&\widetilde{K}^{[1]} = 0;
\label{eq:K_1loop} \\
&\gamma_D^{[1]} = 6 C_F,
\label{eq:gammaC_1loop}
\end{align}
where Eq.~\eqref{eq:K_1loop} refers to the 1-loop coefficient of the soft kernel computed at the reference scale.
\end{itemize}
Therefore, the $\zeta$-independent part of the Sudakov at NLL is:
\begin{align}
&\mbox{exp} 
\left\{
\frac{1}{4} \, \widetilde{K}(b_T^\star;\,\mu_b) \, \log{\frac{\mu^2}{\mu_b^2}} 
+
\int_{\mu_b}^{\mu} \frac{d \mu'}{\mu'} \, 
\left[ \gamma_D(\alpha_S(\mu'),\,1) - \frac{1}{4} \, 
\gamma_K(\alpha_S(\mu')) \, \log{\frac{\mu^2}{\mu'^2}}\right]
\right\} 
\stackrel{NLL}{=}
\notag \\
&\quad \stackrel{NLL}{=}
\mbox{exp} 
\left\{
\log{\frac{\mu}{\mu_b}}\,g_1 (x) + g_2(x)
\right\},
\label{eq:sud_norap_NLL}
\end{align}
where:
\begin{align}
x = \frac{\alpha_S(\mu)}{4\pi} \, \log{\frac{\mu}{\mu_b}},
\label{eq:x_def}
\end{align}
and the functions $g_1$ and $g_2$ are given by:
\begin{align}
&g_1(x) = 
\frac{\gamma_K^{[1]}}{4 \beta_0} \,
\left[
1 + \frac{\log{(1-2\beta_0\,x)}}{2\beta_0\,x}
\right];
\label{eq:g1_NLL}\\
&g_2(x) =
\frac{\gamma_K^{[1]}}{4 \beta_0}\,
\frac{\beta_1}{\beta_0} \,
\left[
\frac{x}{1-2\beta_0\,x} + 
\frac{1}{2\beta_0} \left(
\log{(1-2\beta_0\,x)}+
\frac{1}{2}\log{(1-2\beta_0\,x)}^2
\right)
\right] -
\notag \\
&\quad-
\frac{\gamma_K^{[2]}}{8 \beta_0^2}\,
\left[
\frac{2 \beta_0 \,x}{1-2\beta_0\,x} + 
\log{(1-2\beta_0\,x)}
\right] -
\frac{\gamma_D^{[1]}}{2 \beta_0}\,\log{(1-2\beta_0\,x)},
\label{eq:g2_NLL}
\end{align}
where $\beta_0$ and $\beta_1$ are the coefficients of the beta functions up to $2$ loop:
\begin{align}
&\beta_0 = \frac{11}{3}C_A - \frac{2}{3}n_f,
\label{eq:beta0}\\
&\beta_1 =  \frac{34}{3}C_A^2 -
\frac{11}{3}C_A \, n_f - 2 C_F n_f. 
\label{eq:beta1}
\end{align}
This result is in agreement with the Sudakov factor computed e.g. in Ref.~\cite{Koike:2006fn}.
On the other hand, the term depending on the rapidity cut-off $\zeta$ is given by:
\begin{align}
&\mbox{exp} 
\left\{
\frac{1}{4} \, 
\log{\frac{\zeta}{\mu^2}} \,
\left[
\widetilde{K}(b_T^\star;\,\mu_b) - 
\int_{\mu_b}^{\mu} \frac{d \mu'}{\mu'} \,
\gamma_K(\alpha_S(\mu'))
\right]
\right\}\stackrel{NLL}{=}
\notag \\
&\quad \stackrel{NLL}{=}
\mbox{exp} 
\left\{
\frac{1}{4} \, 
\log{\frac{\zeta}{\mu^2}} \,
\left[
g_2^{K}(x) + 
\frac{1}{\log{\frac{\mu}{\mu_b}}}\,g_3^{K}(x)
\right]
\right\},
\label{eq:sud_rap_NLL}
\end{align}
where $x$ has been defined in Eq.~\eqref{eq:x_def}, while the functions $g_2^{K}$ and $g_3^{K}$ are:
\begin{align}
&g_2^{K}(x) = 
\frac{\gamma_K^{[1]}}{2 \beta_0} \,
\log{(1-2\beta_0\,x)};
\label{eq:g2K_NLL}\\
&g_3^{K}(x) =
\frac{x^2}{1-2\beta_0\,x} \,
\left[
\gamma_K^{[1]} \,
\frac{\beta_1}{\beta_0} - 
\gamma_K^{[2]} + 
\frac{1}{x}\,\widetilde{K}^{[1]}
\right].
\label{eq:g3K_NLL}
\end{align}

\bigskip

\subsection{Non-Perturbative content} \label{subsec:tmd_nonpert}

\bigskip

The last term in Eq.~\eqref{eq:TMDFF_def_2} encodes the whole non-perturbative content of the TMD fragmentation function,  $D_1$. Clearly, this cannot be predicted by perturbative QCD and hence it has to be extracted from experimental data, through a phenomenological analysis.
It involves two functions.
The first is $g_K$, which describes the long-distance behavior of the soft kernel $\widetilde{K}$, defined as:
\begin{align}
&g_K(b_T) = 
\widetilde{K}(b_T^\star;\,\mu) -
\widetilde{K}(b_T;\,\mu).
\label{eq:gK_def}
\end{align}
In most phenomenological applications it is assumed to behave quadratically:
\begin{align}
&g_K(b_T) = a \, b_T^2.
\label{eq:gK_fit}
\end{align}
with $a \sim 0.01 \div 0.1$ GeV$^2$.

The other non-perturbative function is the model $M_D$, which represents the very heart of the TMD, since it identifies $D_1$ uniquely. 
Besides $b_T$, it may depend on the flavor of the fragmenting parton, on its light-cone momentum fraction $z$ and also on the detected hadron $H$ (in the following, however, we will assume that it is just a function of $b_T$, dropping any other dependence).
Its functional form is arbitrary, as long as it goes to $1$ as $b_T \to 0$ and goes to zero in the large-$b_T$-limit, $b_T \to \infty$. It is commonly modelled as a Gaussian or as a (second kind) Bessel function, which in the Fourier conjugate momentum space becomes a power-law,  
reminding of a squared propagator for an on-shell fermion.
This will be discussed in more detail in Section~\ref{sec:pheno}.

\bigskip

\section{Final Cross Section \label{sec:final_xs}}

\bigskip

We are now ready to assemble the expression for the final cross section of $\epm \to H\,X$ in the $2$-jet limit, as sketched in Eq.~\eqref{eq:xs_final_general} in its general factorized form.
It is differential in the fractional energy of the detected hadron $z_h$, defined in Eq.~\eqref{eq:zh_def}, in the transverse momentum $P_T$ of the detected hadron with respect to the thrust axis, and also in thrust, $T$, as defined in Eq.~\eqref{eq:thrust_def},
The factorization procedure allows to separate out the short-distance contributions, encoded in the partonic cross section (Section~\ref{sec:partonic_xs}), from the long-distance behavior, described by the TMD FFs (Section~\ref{sec:nll_tmd}).
Referring to the very general expression presented in Eq.~\eqref{eq:xs_final_general}, we should keep in mind that:
\begin{itemize}
\item The contribution of the fragmenting gluon is suppressed by orders of $\mathcal{O}\left({\lambda^2}/{Q^2}\right)$ in the $2$-jet limit (see Section~\ref{subsubsec:frag_gluon_sub}).
Therefore, the sum over parton types, $j$, runs only over the flavors $f$ of the fragmenting fermions.
\item The final cross section is RG-invariant.
In fact, the anomalous dimension of the partonic cross section, Eq.~\eqref{eq:RG_evo_partonicxs}, is exactly equal and opposite to the anomalous dimension of the TMD FF associated to a fragmenting fermion, Eq.~\eqref{eq:RG_evo}.
Therefore, we are allowed to set $\mu = Q$.
\item The final cross section depends on two cut-offs, $\lambda$ and $\zeta$, so far considered as being independent.
However, they have been introduced for intimately related reasons: they both ensure that the partonic cross section describes all particles involved in the process, \emph{except} those that are \emph{collinear} i.e. with low transverse momentum ($k_T \leq \lambda$) and a very large rapidity ($y \geq y_1$), which are included in the TMD FF.
In fact, a neat and consistent separation of the partonic cross section from the TMD fragmentation function guarantees the successful factorization of the cross section.
Furthermore, both cut-offs have to be intended in the limit in which they vanish. 
On one hand, $\zeta \to 0$ (i.e. $y_1 \to +\infty$, according to Eq.~\eqref{eq:zeta_def}) is the limit in which the factorization procedure of Ref.~\cite{Collins:2011zzd} can be applied.
On the other hand, $\lambda \to 0$ corresponds to the $2$-jet limit for the topology of the final state of the process, according to the relation ${\lambda^2}/{Q^2} = \tau_{\mbox{\tiny meas.}}$, as we discussed at the end of Section~\ref{subsubsec:frag_gluon_sub}. 
The above considerations suggest that $\lambda$ and $\zeta$ are closely related.
The subtraction mechanism presented in Section~\ref{subsubsec:frag_quark_sub} was built in such a way that the range of transverse momentum of the collinear emission covered by the unsubtracted hard part of Eq.~\eqref{eq:W_ferm} would exactly match the $k_T$-range covered by the subtraction term of Eq.~\eqref{eq:quark_TMD_cutoff}. 
In fact, in both cases the collinear gluon can have a transverse momentum $k_T$ no larger than $\lambda$.
As far as rapidity is concerned, however, the subtraction mechanism needs to be applied with special care.
Let's consider the 1-loop computation of the partonic cross section associated to a fragmenting fermion.
On one hand, the subtraction term describes a collinear gluon that, when $\tau = 0$, has a rapidity larger than $y_1$, linked to $\zeta$ by the definition of Eq.~\eqref{eq:zeta_def}.
On the other hand, the rapidity of the collinear gluon described by the unsubtracted hard part depends on $k_T$ and $\tau$,  and when $\tau = 0$ it cannot be smaller than $y_\lambda$, defined exactly as $y_1$ but with $\zeta = \lambda^2$.
Therefore, in order to cover the same range of rapidities in both terms, we have to set $y_1 = y_\lambda$, which means $\zeta = \lambda^2$.
Any different value of $\zeta$ would lead to an incorrect separation of the collinear and the hard contributions in the final cross section.
In fact, if $\zeta > \lambda^2$, then $y_1 < y_\lambda$ and the hard part would also include the contribution of the collinear particles with a rapidity in the range $y_1 \leq y \leq y_\lambda$, which is already covered by the TMD FFs. Conversely, if $\zeta < \lambda^2$, then $y_1 > y_\lambda$ and the contribution of the collinear particles with rapidity in the range $y_\lambda \leq y \leq y_1$ would not be present at all in the final result. Consequently, the only possible  choice is $\zeta=\lambda ^2$.
\end{itemize}
Such considerations lead us to the following formula:
\begin{align}
&\frac{d \sigma}
{dz_h \, dT_{\mbox{\tiny meas.}} \, dP_T^2}
=
\pi 
\int \, \frac{d^2\vec{b}_T}{(2\pi)^2} 
\times \notag \\
&\quad \times
\sum_f \,
\int_{z_h}^1 \,\frac{d z}{z} \, 
\frac{d \widehat{\sigma}_f
(Q,\,\lambda,\,\lambda^2)}
{d {z_h}/{z} \, dT_{\mbox{\tiny meas.}}} \,
\widetilde{D}_{1,\,H/f}
(z,\,b_T,\,Q,\,\lambda^2)\,
e^{i \frac{\vec{P}_T}{z} \cdot \vec{b}_T} \,
\left[
1+\mathcal{O}\left( \frac{M_H^2}{Q^2}\right)
\right] ,
\label{eq:xs_final_1}
\end{align}
where $\lambda$ is related to the measured value of thrust, $T_{\mbox{\tiny meas.}}$.
Alternatively, the final cross section can be written in terms of the soft and collinear contributions encoded in the partonic cross section, as in Eq.~\eqref{eq:all_orders_partonicxs}. 
Therefore, by using the final, all-order, results of Section~\ref{subsubsec:frag_quark_sub} we have:
\begin{align}
&\frac{d \sigma}
{dz_h \, dT_{\mbox{\tiny meas.}} \, dP_T^2}
=
\pi \, \sigma_B \, N_C \, V 
\int d\tau_1 \, d\tau_2 \, d\tau_S \,
J_B(\tau_2)\,S(\tau_S)\,
\times \notag \\
&\quad \times
\int_{z_h}^1 \,d z
 \left.J_{q/q}^{(\lambda)}
(\tau_1,\,{z_h}/{z},\,\lambda^2)
\right \rvert_{\mbox{\small sub, R}}
\,\sum_f \, e_f^2 \,
\int \, \frac{d^2\vec{b}_T}{(2\pi)^2}
\widetilde{D}_{1,\,H/f}
(z,\,b_T,\,Q,\,\lambda^2)\,
e^{i \frac{\vec{P}_T}{z} \cdot \vec{b}_T}
\times \notag \\
&\quad \times \,
\delta(1-T_{\mbox{\tiny meas.}}-\tau_1-\tau_2-\tau_S) \,
\left[
1+\mathcal{O}\left( \frac{M_H^2}{Q^2}\right)
\right] ,
\label{eq:xs_final_allorders}
\end{align}
As we pointed out at the end of Section~\ref{subsubsec:frag_quark_sub}, the structure of this result is similar to the well-known thrust distribution associated to $\epm$ events, with the exception of the contributions associated to the radiation of partons collinear to the thrust axis, which give the non-trivial dependence on $z_h$ and $P_T$, encoded in the second line of Eq.~\eqref{eq:xs_final_allorders}.
Notice that here we explicitly indicated that the thrust refers to its measured value. This does not exactly coincide with the variable $T = 1-\tau$ that we used in Section~\ref{sec:partonic_xs} in computing the partonic cross section, since the thrust distribution calculated at partonic level does not encode the information about the non-perturbative hadronization process.
According to Ref.~\cite{Catani:1991kz}, the observed distribution of ``hadronic" thrust, $T_{\mbox{\tiny meas.}}$, is  related to its partonic counterpart, $T$, by a correlation function $C(T_{\mbox{\tiny meas.}},\,T)$. 
Therefore, the partonic cross section written in terms of $T_{\mbox{\tiny meas.}}$ will be obtained as:
\begin{align}
&\frac{d \widehat{\sigma}_f
(Q,\,\lambda,\,\lambda^2)}
{d z \, dT_{\mbox{\tiny meas.}}} = 
\int_0^1\, dT \, 
C(T_{\mbox{\tiny meas.}},\,T)\,
\frac{d \widehat{\sigma}_f
(Q,\,\lambda,\,\lambda^2)}
{d z \, dT} =
\notag \\
&=
\mbox{exp}
\Bigg\{
\frac{1}{4}\,
\widehat{K}\left( \alpha_S(Q),\,1\right)\,
\log{\frac{\lambda^2}{Q^2}} -
\frac{1}{2}\,
\int_{\lambda^2}^{Q^2} \,\frac{d(\lambda')^2}{(\lambda')^2}\,
G\left(\alpha_S(Q),\,
1,\,{\lambda^2}/{Q^2},\,
{Q^2}/{(\lambda')^2}\right)
\Bigg\} 
\times \notag \\
&\quad\times\,
\int_0^1\, dT \, 
C(T_{\mbox{\tiny meas.}},\,T)\,
\left.\frac{d \widehat{\sigma}_f}
{dz \, dT}\right\rvert_{
\substack{\text{ref.}}},
\label{eq:partonic_xs_Tmeas}
\end{align}
having used the expression for the partonic cross section as obtained in Eq.~\eqref{eq:xs_ferm_resumm}.
Monte Carlo simulations show that the correlation function is sharply peaked around $T_{\mbox{\tiny meas.}} \sim T$ and that it is not strongly sensitive to the parton distribution~\cite{Catani:1991kz}.
Therefore, a formula suitable for phenomenology (that clearly requires an expression written in terms of functions of $T_{\mbox{\tiny meas.}}$) can be obtained by approximating the correlation function with a (smooth) delta function and then neglecting all the contributions proportional to $C(T_{\mbox{\tiny meas.}},\,1)$, associated to strictly pencil-like events.
In other words, the partonic cross section written in terms of $\tau_{\mbox{\tiny meas.}}$ can be obtained from the result of the fixed order computation, expressed in terms of distributions of $\tau$, by simply neglecting all contributions proportional to $\delta(\tau)$, by removing all the ``plus" that label the plus-distributions and finally by replacing $\tau$ with its hadronic counterpart.
At NLO this operation gives:
\begin{align}
&\frac{d \widehat{\sigma}^{NLO}_f
(Q,\,\lambda,\,\lambda^2)}
{d z \, dT_{\mbox{\tiny meas.}}} = 
\int_0^1\, dT \, 
C(T_{\mbox{\tiny meas.}},\,T)\,
\frac{d \widehat{\sigma}^{NLO}_f
(Q,\,\lambda,\,\lambda^2)}
{d z \, dT} =
\notag \\
&=
-\sigma_B \,
e_f^2 \,N_C \,
\frac{\alpha_S(Q)}{4 \pi} \, 
C_F \,
\delta(1-z)\,
\frac{3+
8\log{\tau_{\mbox{\tiny meas.}}}}
{\tau_{\mbox{\tiny meas.}}}\,
\mbox{exp}
\left\{
-\frac{\alpha_S(Q)}{4 \pi}\,3 C_F \,
\left(\log{\frac{\lambda^2}{Q^2}}\right)^2
\right\}
\label{eq:partonic_xs_Tmeas_NLO}
\end{align}
Of course, this is a rather rough approximation which reduces the potential of a formula like Eq.~\eqref{eq:partonic_xs_NLO_ref}. 
In particular, neglecting the $\tau = 0$ terms not only eliminates the contributions of pencil-like events, but it also  drastically modifies the behaviour of the cross section in the region where $T$ is very close to 1.
Furthermore, by applying this approximation we loose part of the information on the $z_h$-dependence of the cross section, since all $z$-distributions that are coupled to $\delta(\tau)$ are removed. 
As a consequence, we expect that the dependence on $z_h$ will be compromised, especially at large values of thrust.
The $z_h$-dependent neglected part, associated with the last line of Eq.~\eqref{eq:partonic_xs_NLO_ref}, involves two terms: an integral from $z_h$ to $1$ and a contribution dominated by $\log{(1-z_h)}^2$. 
Therefore, the discrepancy due to the brute approximation of Eq.~\eqref{eq:partonic_xs_Tmeas_NLO} should be more important at small and large values of $z_h$, 
where these two terms have a non negligible impact on the final cross section. We will come back to this point in the next Section.

Finally, we require 
${\lambda^2}/{Q^2}=\tau_{\mbox{\tiny meas.}}$, and 
we remove the label ``meas", which has now become redundant.
At NLO, NLL accuracy, the cross section for $\epm \to H,\,X$ differential in $z_h$, $T$ and $P_T$ is then given by:
\begin{align}
&\frac{d \sigma^{\text{NLO, NLL}}}
{dz_h \, dT \, dP_T^2}
=
\notag \\
&\quad=
-\sigma_B \,\pi \, N_C \,
\frac{\alpha_S(Q)}{4 \pi} \, 
C_F \,
\frac{3+8\log{(1-T)}}{1-T}\,
\mbox{exp}
\left\{
-\frac{\alpha_S(Q)}{4 \pi}\,3 C_F \,
\left(\log{(1-T)}\right)^2
\right\} \,
\times \notag \\
&\quad \times \,
\sum_f e_f^2 \,
\int \, \frac{d^2\vec{b}_T}{(2\pi)^2} \,
e^{i \frac{\vec{P}_T}{z_h} \cdot \vec{b}_T}\,
\widetilde{D}^{\text{NLL}}_{1,\,H/f}
(z_h,\,b_T,\,Q,\,(1-T)\,Q^2)\,
\left[
1+\mathcal{O}\left( \frac{M_H^2}{Q^2}\right)
\right]
\label{eq:xs_final_NLO_NLL}
\end{align}
where, according to Section~\ref{sec:nll_tmd}:
\begin{align}
&\widetilde{D}^{\text{NLL}}_{1,\,H/f}
(z_h,\,b_T,\,Q,\,(1-T)\,Q^2) = 
\frac{1}{z_h^2}\,
\bigg(
d_{H/f}(z,\,\mu_b) + 
\notag \\
&\quad+
\frac{\alpha_S(\mu_b)}{4\pi}
\int_{z_h}^1 \, \frac{d\rho}{\rho}\,
\left[
d_{H/f}({z_h}/{\rho},\,\mu_b)
\,\rho^2 \, 
\mathcal{C}_{q / q}^{[1]}(\rho) +
d_{H/g}({z_h}/{\rho},\,\mu_b)
\,\rho^2 \, 
\mathcal{C}_{g / q}^{[1]}(\rho) 
\right]
\bigg) 
\times \notag \\
&\quad \times
\mbox{exp} 
\left\{
\log{\frac{Q}{\mu_b}}\,g_1 (x) + g_2(x)
\right\}
\mbox{exp} 
\left\{
\frac{1}{4} \, 
\log{(1-T)} \,
\left[
g_2^{K}(x) + 
\frac{1}{\log{\frac{Q}{\mu_b}}}\,g_3^{K}(x)
\right]
\right\} 
\times \notag \\
&\quad \times
M_D(b_T) 
\mbox{ exp} \left\{
-\frac{1}{4} \, g_K(b_T) \, 
\log{\left(\frac{z_h^2 \, Q^2}{M_H^2}\,(1-T)\right)}
\right \},
\label{eq:tmd_NLL}
\end{align}
with $x$ defined as in Eq.~\eqref{eq:x_def}. For a convenient and straightforward application of Eq.~\eqref{eq:tmd_NLL} we recall that the Wilson coefficients $\mathcal{C}_{q / q}^{[1]}$ and $\mathcal{C}_{g / q}^{[1]}$ are presented in Eqs.~\eqref{eq:wils_qq_1loop} and~\eqref{eq:wils_gq_1loop}, the functions $g_1$, $g_2$ and $g_2^{K}$, $g_3^{K}$ contributing to the Sudakov are computed in Eqs.~\eqref{eq:g1_NLL},~\eqref{eq:g2_NLL},~\eqref{eq:g2K_NLL} and~\eqref{eq:g3K_NLL}, and the non-perturbative functions $g_K$ and $M_D$ are defined in Eqs.~\eqref{eq:gK_fit} and~\eqref{eq:model_fit}.
Notice that the cross section in  Eq.~\eqref{eq:xs_final_NLO_NLL} is \emph{not} resummed in thrust. 
A proper resummation in $T$ 
is beyond the purpose of this paper.
Such resummation must also include a correct treatment of the dependence on $z_h$, by considering the terms that have been neglected in the approximation used in Eq.~\eqref{eq:partonic_xs_Tmeas_NLO}.
Clearly, this is strictly connected to the difficulties in finding a fully resummed expression for the subtracted, renormalized function $J_{q/q}^{(\lambda)}$ (see the discussion at the end of Section~\ref{subsubsec:frag_quark_sub}) and, ultimately, for the whole second line of Eq.~\eqref{eq:xs_final_allorders}.

\bigskip

Very recently, the factorization of the $e^+e^- \to HX$ cross section, as measured by the BELLE Collaboration (Ref.~\cite{Seidl:2019jei}), has been investigated in two papers, both based on the SCET formalism.
In Ref.~\cite{Kang:2020yqw}, the authors propose to integrate out the thrust $T$ and to reproduce the experimental cross section by combining all the measured thrust bins, within the range $[0.5-1.0]$.
A cross section differential in 
$z_h$, $P_T$ and $T$, is presented in Ref.~\cite{Makris:2020ltr}; it results from matching three different kinematical regions, each associated with a different factorized expression for the final cross section. The phenomenological application of this formula is not shown.

\section{A basic phenomenological application \label{sec:pheno}}

In this conclusive Section we provide a basic but realistic example of the strong phenomenological potential of Eq.~\eqref{eq:xs_final_NLO_NLL}. We stress that this is not a proper and complete data-analysis of the BELLE measurements, as in what follows there is no fit of the experimental data and no fine tuning of the free parameters.
However, the agreement of the predictions obtained from Eq.~\eqref{eq:xs_final_NLO_NLL} with the experimental data is astonishing, even without a proper fit.

As anticipated in Section~\ref{subsec:tmd_nonpert}, to model the non-perturbative part of the TMD FF we will exploit a  parameterization that, in the $b_T$ space, corresponds to a Bessel-K function, normalized in such a way that it is $1$ at $b_T=0$:
\be
M_D (b_T,\,m,\,p) = 
\frac{2^{2-p}}{\Gamma(p-1)} \, (b_T \, m)^{p-1} \, K_{p-1}(b_T \,m) 
\label{eq:model_fit}
\ee
where $K_{-1+p}$ is the modified Bessel function of the second kind.
This model was successfully applied in Ref.~\cite{Boglione:2017jlh} 
and, more recently, in Ref.~\cite{Boglione:2020cwn}.
It is known as the ``power law model", since its Fourier transform is given by:
\begin{align}
\mathcal{FT}\{M_D\} = 
\frac{\Gamma(p)}{\pi \, \Gamma(p-1)}
\frac{m^{2(p-1)}}
{\left(k_T^2 + m^2\right)^p},
\label{eq:model_fit_FT}
\end{align}
which mimics the power $p$  
of a function closely reminiscent of the propagator of an on-shell fermion (see dedicated discussion in Ref.~\cite{Collins:2016hqq}).

As mentioned above, we do not perform a proper fit but simply fix the two non perturbative  parameters of the $k_T$-distribution to reasonable and perfectly standard values, inspired by pure common sense: $m=1$ GeV, to represent a typical hadronic mass, and $p=2$ which translates, in $k_T$-space, in a $k_T$-distribution 
similar to the square of the Feynman propagator, see Eq.~\eqref{eq:model_fit_FT}.
One additional free parameter, $a$, is embedded in the non perturbative $g_K(b_T)$ function, as defined in Eq.~\eqref{eq:gK_def}. Its value depends on the choice of $b_{MAX}$; for $b_{MAX} = 1$ GeV$^{-1}$ it is expected to be in the range $0.01-0.1$ GeV, while considerably larger values are found at smaller $b_{MAX}$. Interesting discussions and reference values of $a$ can be found, for example, in Refs~\cite{Anselmino:2013lza,Aidala:2014hva,Boglione:2014oea,Boglione:2018dqd}. 
We set $a=0.05$ GeV$^{-1}$, as an intermediate value. There is clearly some degree of correlation among the three parameters $a$, $m$ and $p$, so that different values of $a$ can easily be compensated by slight variations of $m$ and $p$. Reassuringly, the final results are not particularly sensitive to the precise value of $a$, after $m$ and $p$ have been fixed. 
It is important to point out, here, that we do not need to introduce any overall normalization parameter. 

The collinear fragmentation functions used for the implementation of Eq.~\eqref{eq:xs_final_NLO_NLL} are the NNFF10\_NLO reference set~\cite{Bertone:2017tyb}, for $\pi^+$ production.

To study the $z_h$ and $P_T$ behaviour of the cross section at fixed values of thrust, $T$, we consider the data subset that the BELLE Collaboration has chosen to present in Fig. 9 of their article,~\cite{Seidl:2019jei}. The differential cross sections corresponding to pion production are shown as a function of $P_T$ for 15 $z_h$-bins, covering the range $[0.10 - 0.85]$ and one thrust bin, $0.85 < T < 0.9$ (we fix $T=0.875$). The error bars correspond to statistical and systematical errors added in quadrature, as reported by the BELLE Collaboration. For the BELLE experiment, $Q=10.58$ GeV, while the $P_T$ bins span a range from $0.06$ GeV to $2.5$ GeV. We expect our TMD description to hold at low $P_T$'s (more properly where the ratio ${P_T}/{z_h \, Q}$ is small). In our case $P_T$ should certainly be no larger than $1.2-1.5$ Gev/c, depending on the corresponding value of $z_h$.

\begin{figure}[t]
 \center
 \vspace*{-1cm}
 \includegraphics[width=15.00cm]{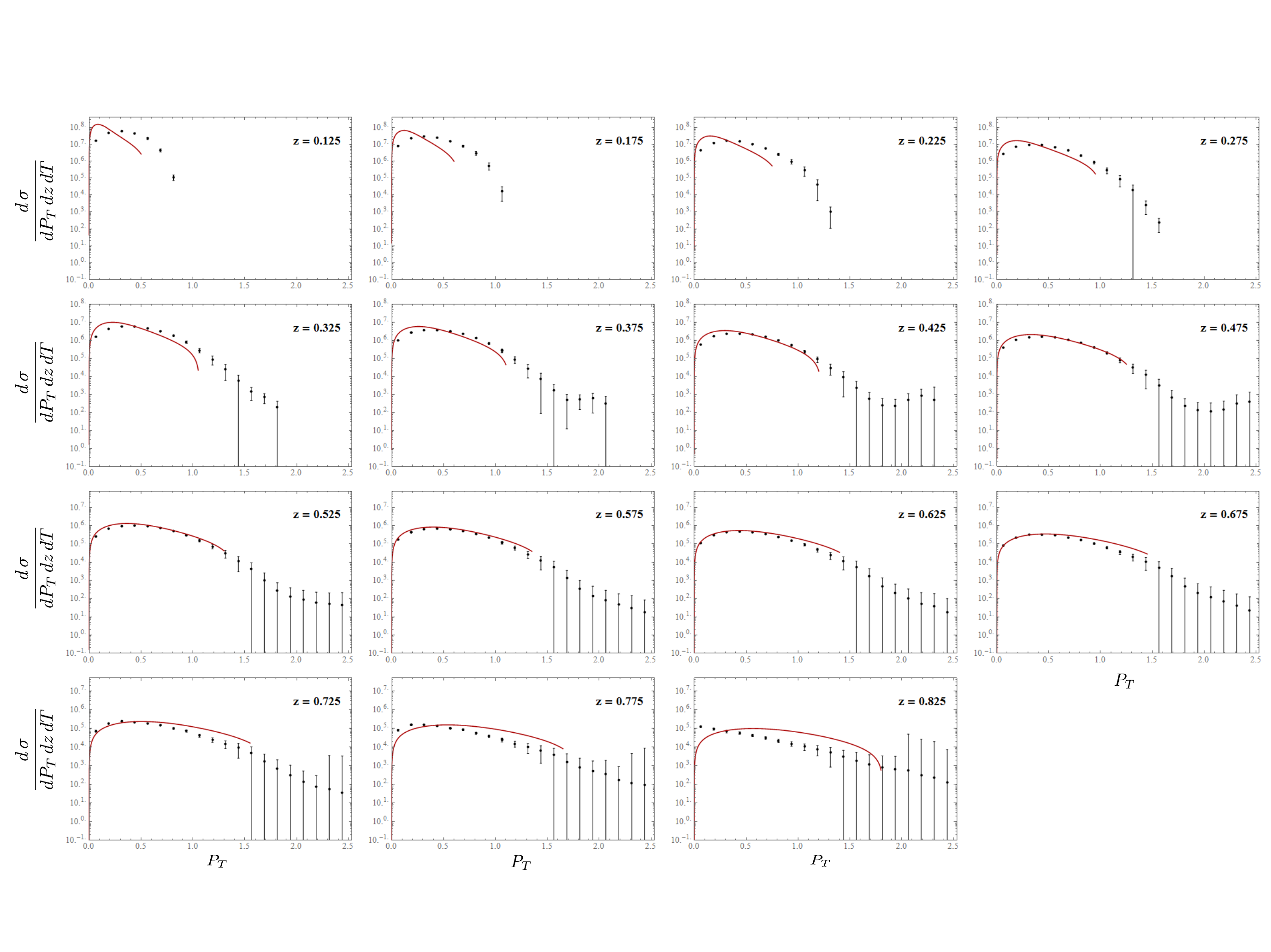}
 \vspace*{-1cm}
 \caption{The $e^+e^- \to hX$ cross section computed to NLO and NLL accuracy, corresponding to 15 $z_h$-bins as measured by the BELLE Collaboration. The central values of the $z_h$-bins is indicated in each panel. Thrust is fixed to one bin, corresponding to $0.85 < T < 0.9$. Uncertainty bands are not shown as no fit is involved in this computation. 
 The collinear fragmentation functions used for the implementation of Eq.~\eqref{eq:xs_final_NLO_NLL} are the NNFF10\_NLO reference set~\cite{Bertone:2017tyb}, for $\pi^+$ production. 
 Uncertainties due to the errors on the collinear fragmentation functions are not shown.}
 \label{fig:Belle-Canvas}
 \end{figure}

Fig.~\ref{fig:Belle-Canvas} shows the comparison between the BELLE experimental measurements and our predictions, represented by the solid (red) line, obtained by applying Eq.~\eqref{eq:xs_final_NLO_NLL}, giving the $e^+e^- \to HX$ cross section, to 1-loop and NLL accuracy, in the 2 jet limit. Our results are presented without uncertainty bands as there is no fitting involved in this calculation. The uncertainty induced by the error on the collinear 
FFs is not shown in this plot. 
The agreement of our results with experimental data is extraordinary, and gives strong support to the naturalness with which Eq.~\eqref{eq:xs_final_NLO_NLL} works.
In particular, our plots show how the full range of $z_h$ values explored by BELLE measurements can be described to an unprecedented level of quality. 
%
%

To analyse the $T$ behaviour of the $e^+e^- \to HX$ cross sections we will refer to a different subset of the BELLE experimental data, large enough to span the whole range of values of $T$, $z_h$ and $P_T$ where Eq.~\eqref{eq:xs_final_NLO_NLL} can safely be applied. In particular we will consider three bins in $T$, corresponding to  $0.70<T<0.80$, $0.80<T<0.85$ and $0.85<T<0.90$. In each of these T bins, we will consider three bins in $z_h$, corresponding to $0.35<z_h<0.40$, $0.55<z_h<0.6$ and $0.75<z_h<0.80$.  

\begin{figure}[t]
 \center
 \vspace*{-0.2cm}
 \includegraphics[width=15cm]{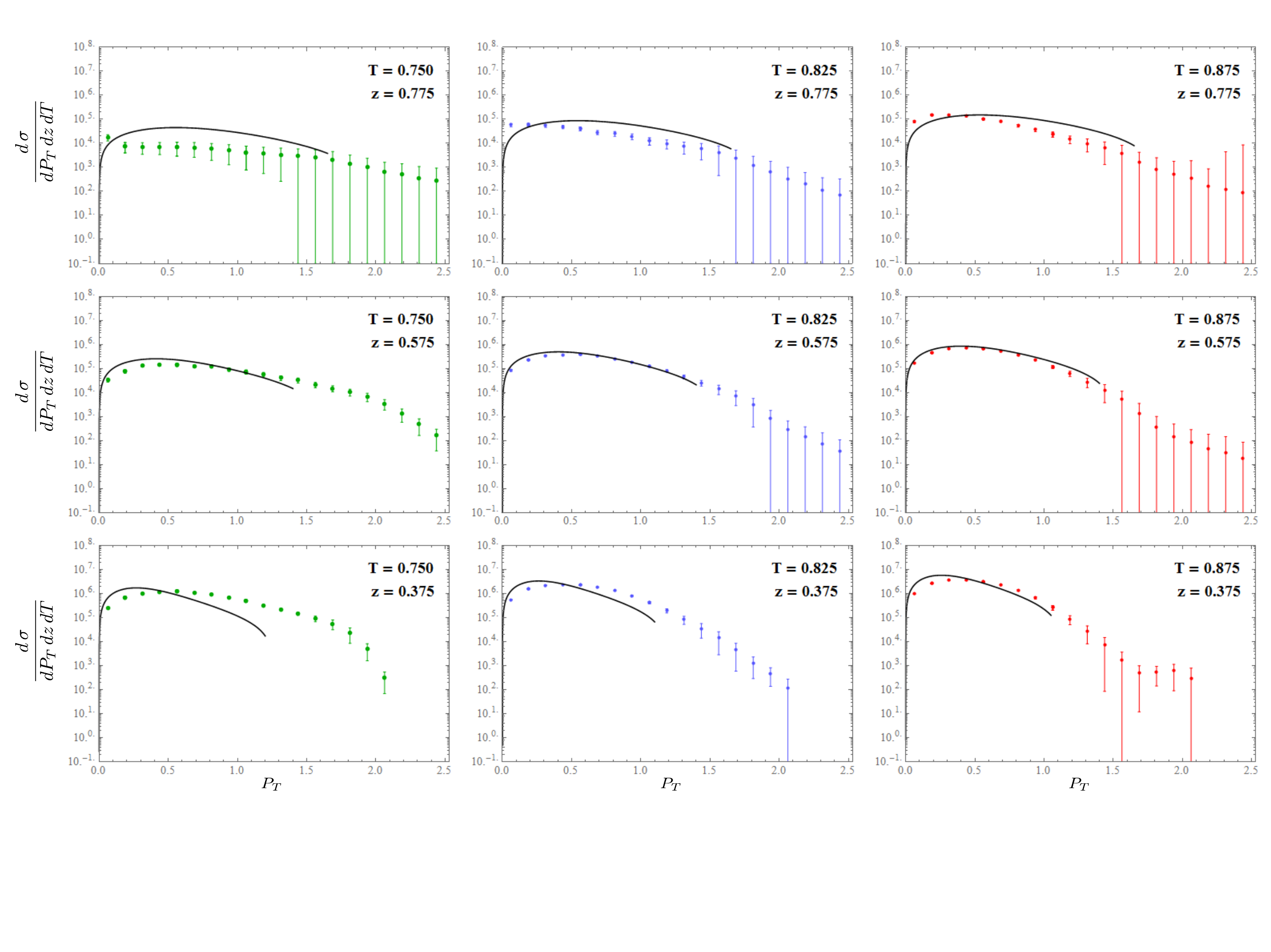}
 \vspace*{-1.8cm}
 \caption{$e^+e^- \to hX$ cross section computed to NLO and NLL accuracy, corresponding to 9 of the bins measured by the BELLE Collaboration. The value of $z_h$ increases by moving down in each column of panels: the upper row of panels corresponds to $z_h=0.375$, the middle row to $z_h=0.575$, the lowest row to $z_h=0.775$. Moving left to right, instead, corresponds to increasing values of thrust. In the leftmost column of panels $T=0.750$, in the central column $T=0.825$, in the rightmost column $T=0.875$.
Uncertainty bands are not shown as no fit is involved in this computation. 
 The collinear fragmentation functions used for the implementation of Eq.~\eqref{eq:xs_final_NLO_NLL} are the NNFF10\_NLO reference set~\cite{Bertone:2017tyb} for $\pi^+$ production.
 Uncertainties due to the errors on the collinear fragmentation functions are not shown.}
 \label{fig:pheno1}
 \end{figure}

The results obtained by using Eq.~\eqref{eq:xs_final_NLO_NLL} are shown in Fig.~\ref{fig:pheno1}, where they are compared to the BELLE experimental data corresponding to the nine $T$ and $z_h$ bins described above (individual plot legends clearly indicate their values in each panel). 
As in Fig.~\ref{fig:Belle-Canvas}, error bars corresponds to statistical and systematical errors added in quadrature. 
The solid lines correspond to our prediction for the $e^+e^- \to HX$ cross section to NLO and NLL accuracy, in the 2 jet limit: they are presented without uncertainty bands as there is no fitting involved in this calculation. The uncertainty induced by the error on the collinear unpolarized FF is not shown.   
As for the previous plots, the agreement with the experimental data is extremely good, especially if one considers that it has been obtained without a proper fit or a fine tuning of the parameters.
This confirms the remarkable success of Eq.~\eqref{eq:xs_final_NLO_NLL}, which can reproduce the experimentally measured $e^+e^- \to H\,X$ cross section in a perfectly natural way, and with a very mild dependence on the specific values of the free parameters.

\begin{figure}[t]
 \center
 \vspace*{-0.9cm}
 \includegraphics[width=15.0cm]{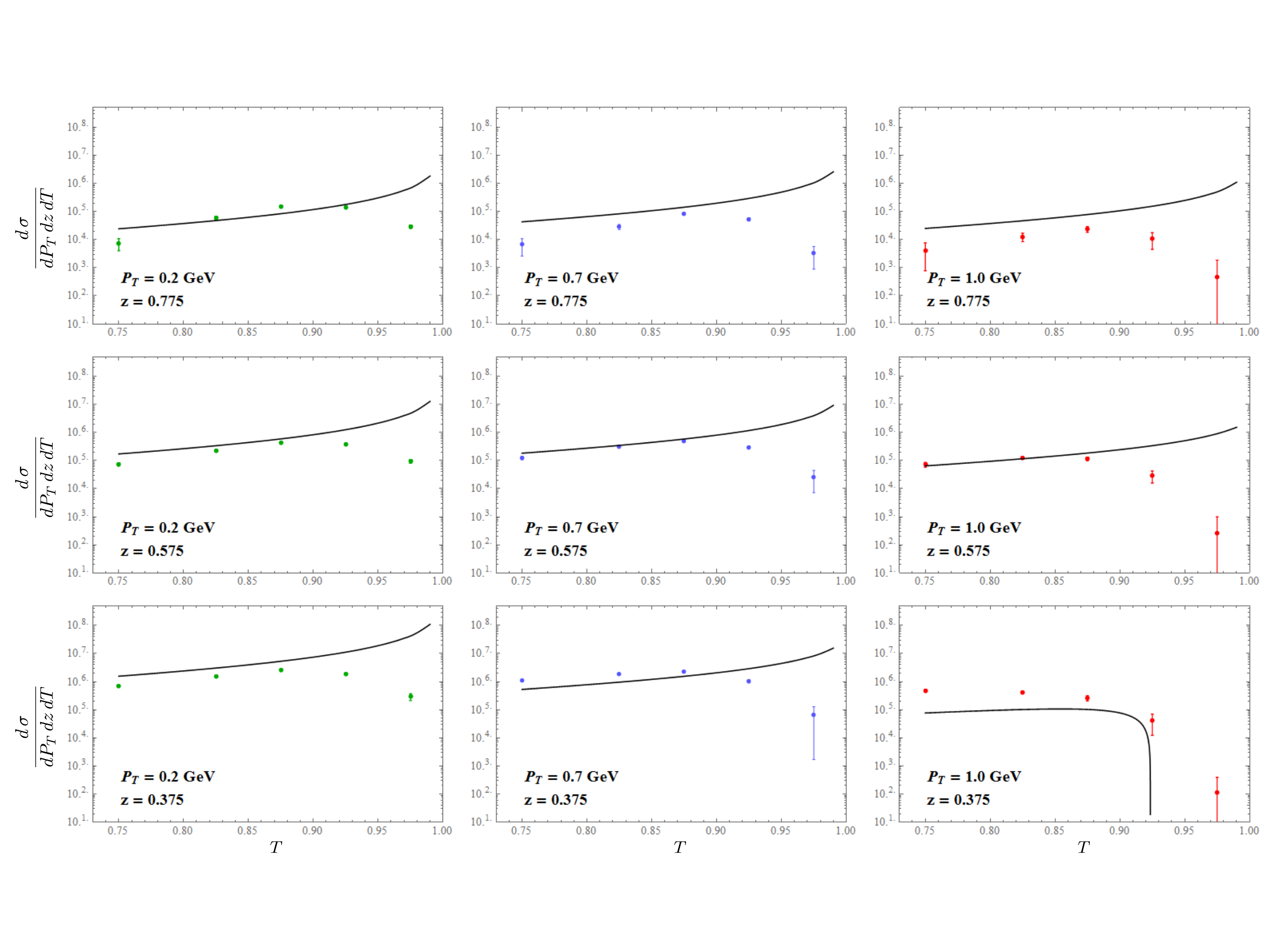}
 \vspace*{-1.1cm}
 \caption{$e^+e^- \to hX$ cross section computed to NLO and NLL accuracy plotted as a function of thrust, $T$, in three $z$-bins and three (representative) $P_T$-bins, as indicated in the plot legends. In the upper row of panels $z=0.775$, in the central row $z=0.575$ and in the lower row of panels $z=0.375$. 
 Uncertainty bands are not shown as no fit is involved in this computation. 
 The collinear fragmentation functions used for the implementation of Eq.~\eqref{eq:xs_final_NLO_NLL} are the NNFF10\_NLO reference set for $\pi^+$ production~\cite{Bertone:2017tyb}.Uncertainties due to the errors on the collinear fragmentation functions are not shown.}
 \label{fig:pheno2}
 \end{figure}

Outside of the considered range of $z_h$ and $T$ values, the quality of the description starts deteriorating slowly. 
This is more evident in Fig.~\ref{fig:pheno2} where our predictions of the $e^+e^- \to H\,X$ cross section are plotted as functions of thrust, $T$, in three $z_h$ bins and three representative $P_T$ bins (one very low, one central and one rather large value of $P_T$, within the TMD range). 
While the last $T$-bin is always greatly overestimated, the description of data deteriorates at extreme values of $z_h$. 
%
%
This is somehow expected. In fact, as we discussed in Section~\ref{sec:final_xs},
a combined resummation of Eq.~\eqref{eq:xs_final_NLO_NLL} in $T$ and $z_h$ would allow us to take into account terms which are presently neglected and restore a good agreement down to the lowest values of $z_h$.
Such a resummation is a project of its own, which we might consider undertaking in the future.

The high quality of the description of the BELLE experimental data as a function of thrust represents our most important achievement. For the first time after their publication, these valuable measurements can be fully explained, in their $z_h$, $P_T$ and $T$ behaviour, within a sound and well founded scheme of factorization, with minimal bias from non-perturbative free parameters.  

%

\section{Conclusions \label{sec:concl}}

In this paper we have constructed the factorized $e^+e^- \to HX$ cross section in the 2-jet limit, providing its explicit expression to NLO and NLL accuracy. Working in a CSS-like framework, factorization has been built, step by step, by means of power counting rules and applying an appropriate subtraction mechanism which allowed us to neatly separate the short distance, low-$b_T$ behaviour, from the long-distance behaviour. In this way, the cross section results in the convolution of the partonic cross section, which in our formalism includes all of the hard contributions, with a TMD fragmentation function, defined according to the scheme presented in Ref.~\cite{Boglione:2020cwn}. Here TMDs are 
purely collinear parts, 
universal and totally process independent.

The 2-jet limit strongly constrains the final state topology to an extreme configuration, in which the available phase space is strongly reduced, as shown in Figs.~\ref{fig:phase_space_real} and~\ref{fig:tmaxPlot}.
In our treatment, this is realized by introducing in the computation of the partonic cross section a cut-off $\tau_{\mbox{\tiny MAX}}$, ultimately related to the measured value of thrust. 
As a consequence, the TMD FFs appearing in the final cross section acquire a dependence on the thrust, through their rapidity cut-offs.

In Section~\ref{sec:pheno} we have performed a basic phenomenological analysis in which we compare the BELLE experimental data to the results obtained by using of our final formula, Eq.~\eqref{eq:xs_final_NLO_NLL}, which is differential in $z_h$ and $P_T$ as well as in thrust, $T$. This is the first phenomenological analysis of the thrust dependence of the $e^+e^- \to HX$ cross section. The agreement of our predictions with experimental data is excellent, even without performing a proper fit. In fact, in our calculation we fix the value of the three free parameters which regulate the non-perturbative behaviour of the TMD FF (one for $g_k$ and two for the model $M_D$) by completely general considerations, based on pure common sense.

The results obtained are very promising, and will represent the basis on which we will construct a more refined, global analysis, simultaneously considering the unpolarized cross sections from SIDIS and  $e^+e^- \to HX$ processes. This will be possible thanks to the formalism proposed in Ref.~\cite{Boglione:2020cwn}, that allows to define any TMD parton density in a completely universal way, separating out the soft and process dependent part, and finally restoring the possibility to perform global analyses including data from processes belonging to different hadron-classes.

\section*{Acknowledgements}

We are grateful to V. Bertone and E. Nocera for providing us with valuable information about the NLO Neural Network 
Fragmentation Function sets for pion production, and J.O. Gonzalez-Hernandez for useful comments and suggestions. \\    
This project has received funding from the European Union’s Horizon 2020 
research and innovation programme under grant agreement No 824093.

\newpage

\appendix


\section{Lowest Order Partonic Cross Section \label{app:lo_partonic_xs}}


\bigskip

\begin{figure}[t]
\centering
\includegraphics[width=3.5cm]{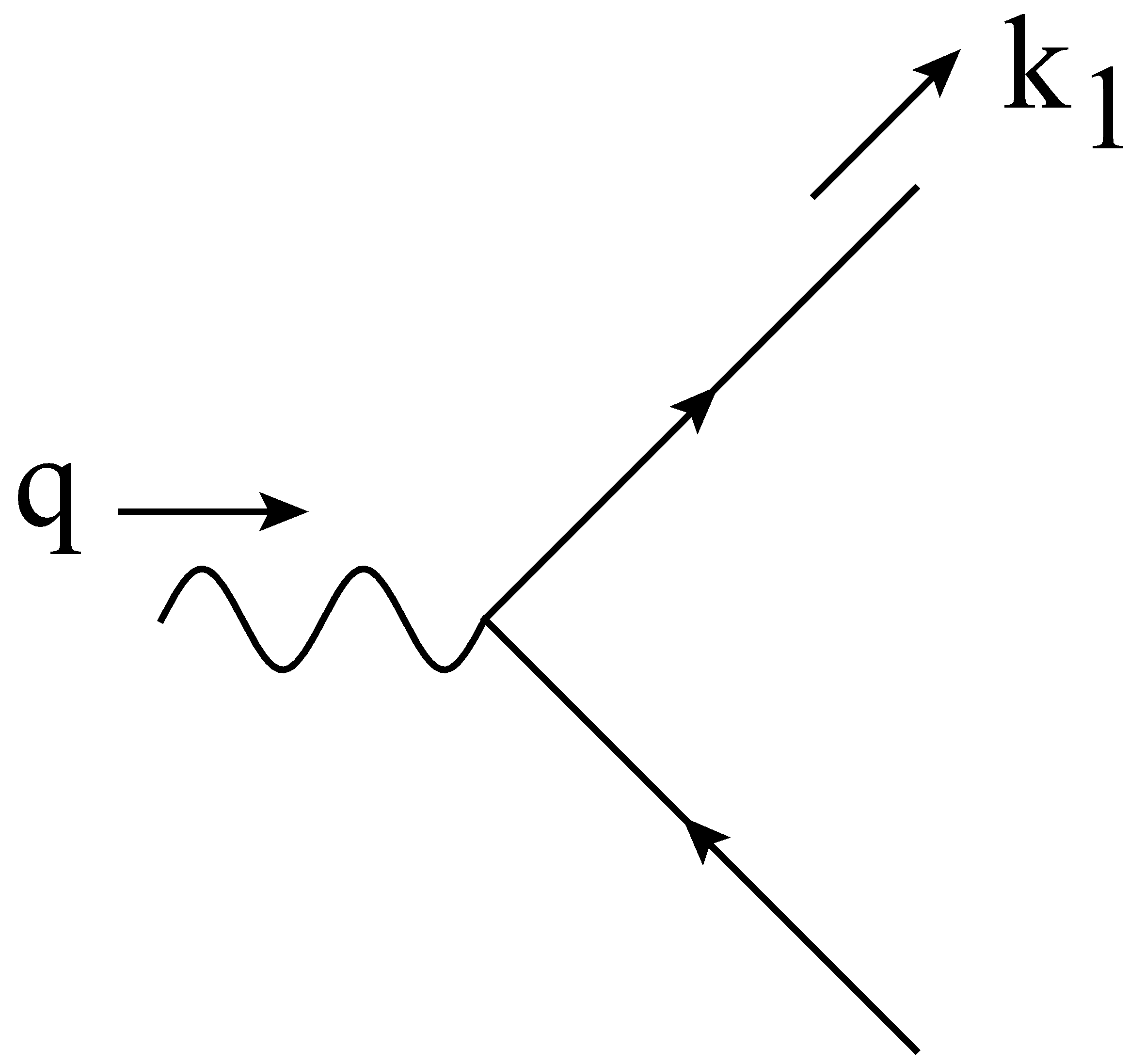}
\caption{The lowest order Feynman graph contributing to the partonic final state tensor.}
\label{fig:lo}
\end{figure}

In this section we will review the computation of the lowest order (LO) partonic cross section, in the $2$-jet limit.
Its final state tensor corresponds to the $\epm \to q \Bar{q}$ Feynman graph shown in Fig.~\ref{fig:lo}.
The LO subtraction mechanism is trivial. In fact, from Eq.~\eqref{eq:sub_mech_W} we have:
\begin{align}
&\left(
\widehat{W}_{f,\,R}^{\mu \, \nu}{}^{\mbox{\small , sub}}
(z,\,\tau )
\right)^{[0]} = 
\left(\widehat{W}_f^{\mu\,\nu}
{}^{\mbox{\small , unsub}} 
(z, \,\tau)
\right)^{[0]}.
\label{eq:sub_mech_W_lo}
\end{align}
Since here the final state topology can only be pencil-like, i.e. $\tau = 0$, there is no need of introducing a thrust cut-off.
%
%
For simplicity, we will assume that the fragmenting parton is a quark. Clearly a similar treatment would apply to the fragmentation of an antiquark.
As usual, we will perform the computation in the partonic analogue of the hadron frame, where the outgoing parton has momentum $k_1$ where only the plus component is non-vanishing (as in Eq.~\eqref{eq:k1_mom}).
As illustrated in Fig.~\ref{fig:lo}, the only matrix element we have to consider is given by:
\begin{align}
&M_f^{\mu\,\nu}{}^{\;[0]}(\epsilon; \,\mu, \,Q) = 
\begin{gathered}
\includegraphics[width=5.3cm]{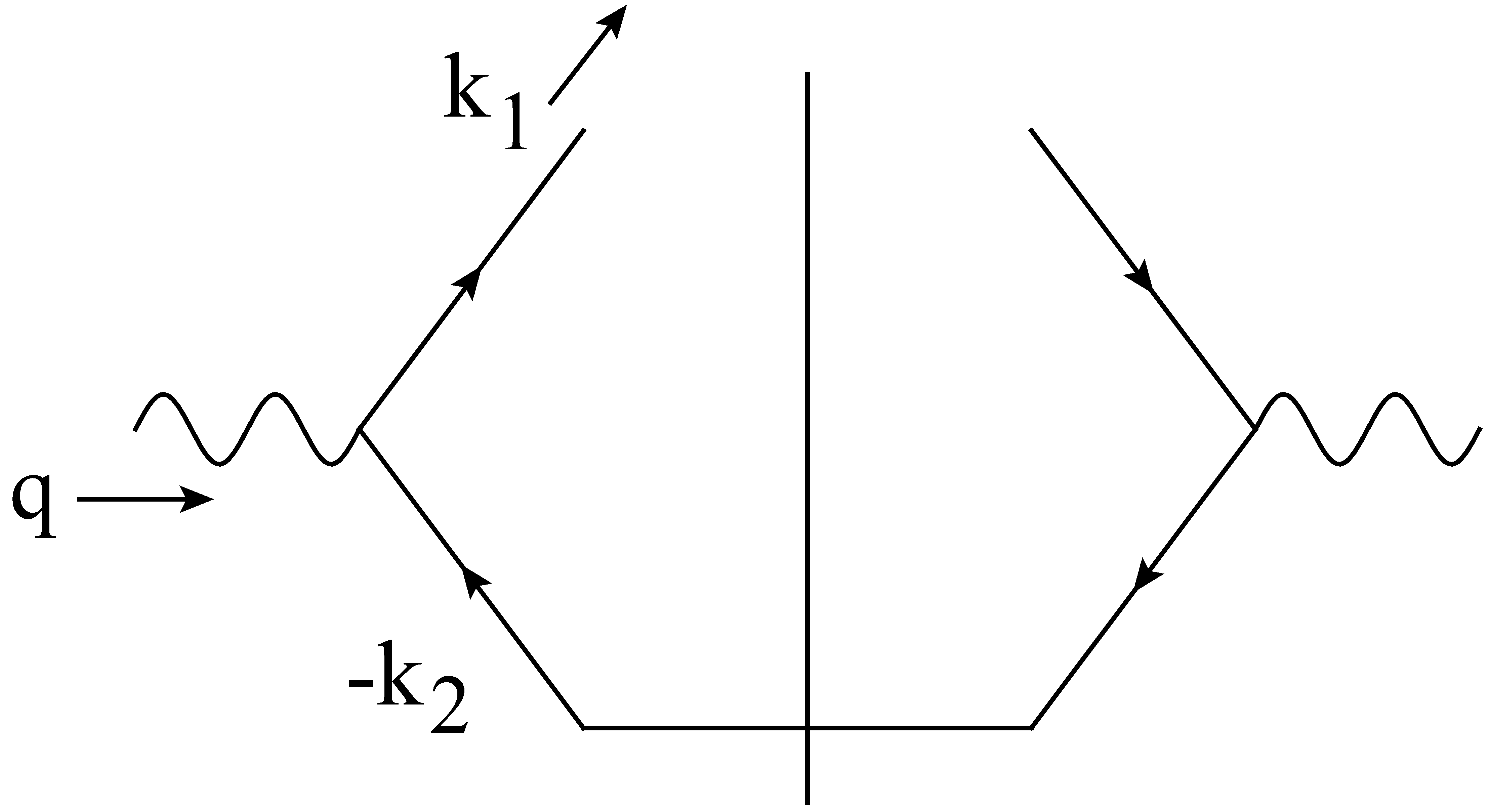}
\end{gathered}
=
\notag \\
&\quad=
e_f^2 \, \sum_{s_1,\,c} \, \Bar{u}_{c,\,f}(k_1,\,s_1) \,
\gamma^\mu \, \slashed{k}_2 \, \gamma^\nu \,
u_{c,\,f}(k_1,\,s_1) = 
e_f^2 \, N_C \, \mbox{Tr} \left\{ 
\slashed{k}_1 \, 
\gamma^\mu \, \slashed{k}_2 \, \gamma^\nu
\right\},
\label{eq:M_lo}
\end{align}
where we sum over all the spins and the colors of the fragmenting quark of flavour $f$ and  fractional electric charge $e_f$.
The projections onto the relevant Lorentz structures, as in Eqs.~\eqref{eq:W_proj1} and~\eqref{eq:W_proj2}, are:
\begin{align}
&-g_{\mu \nu}M_f^{\mu\,\nu}{}^{\;[0]} = 
8 \, e_f^2 \, N_C (1-\epsilon) \,
k_1 \cdot k_2 = 
4 \, e_f^2 \,  N_C (1-\epsilon) Q^2; 
\label{eq:g_M_lo}\\
&\frac{k_{1,\,\mu} \, k_{1,\,\nu}}{Q^2}
M_f^{\mu\,\nu}{}^{\;[0]} = 0,
\label{eq:k2_M_lo}
\end{align}
where we used momentum conservation $k_1 + k_2 = q$.
Since the projection of Eq.~\eqref{eq:k2_M_lo} is zero, the final state tensor will be written as in Eq.~\eqref{eq:k2prog_null_W} and the partonic cross section will only have the transverse projection.
The phase space integral only involves $k_2$ and it is straightforward:
\begin{align}
&-g_{\mu \nu}\widehat{W}_f^{\mu\,\nu}{}^{\;[0]}=
\frac{1}{4\pi} \,
\int \frac{d^D k_2}{(2\pi)^{D}} \, 
\left(
-g_{\mu \nu}M_f^{\mu\,\nu}{}^{\;[0]}
\right) \,
(2\pi)^{D} \, 
\delta^{(D)}\left( q - k_1 - k_2 \right) \,
\times
\notag \\
&\quad \times \,
(2\pi) \, \delta \left( k_2^2 \right) \, 
\theta(k_2^+) \,\theta(k_2^-) \,
\delta(\tau) = 
\notag \\
&\quad=
2 \, e_f^2 \,  N_C (1-\epsilon) \,
\delta \left( 1 - z \right) \,
\delta(\tau) = 
\notag \\
&\quad= 2 \widehat{F}_{1,\,f}^{[0]},
\label{eq:g_W_lo}
\end{align}
where the factor $1/(4\pi)$ is the normalization of the hadronic tensor fixed according to Ref.~\cite{Collins:2011zzd} and.~\cite{Boglione:2020cwn} and the last line is obtained using Eqs.~\eqref{eq:W_proj1} and~\eqref{eq:W_proj2}.
Finally:
\begin{align}
&\widehat{W}_f^{\mu\,\nu}{}^{\;[0]} (z,\,\tau) =
H_T^{\mu \nu}(z) \,
\widehat{F}_{1,\,f}^{[0]}(z,\,\tau),
\label{eq:W_lo_tensor}
\end{align}
where the transverse tensor $H_T^{\mu \nu}$ has been defined in Eq.~\eqref{eq:HT_tensor}.
The resulting partonic cross section follows from Eq.~\eqref{eq:xs_T} and is given by:
\begin{align}
\frac{d \widehat{\sigma}_f^{[0]}}{d z \, d T} = 
\sigma_B \, z \, \widehat{F}_{1,\,f}^{[0]}(z,\,\tau)
= 
\frac{4 \pi \alpha^2}{3 Q^2} \,
e_f^2 \,  N_C \,
\delta \left( 1 - z \right) \,
\delta(\tau)
\label{eq:lo_partonic_xs}
\end{align}
where the Born cross section $\sigma_B$ has been defined in Eq.~\eqref{eq:born_xs}.


\section{Thrust Divergent Functions \label{app:thrust_fun}}


\bigskip

In $\epm$ scattering, the topology of the final state is related to the value of thrust, $T$, defined in Eq.~\eqref{eq:thrust_def}.
In the following, we will focus on the quasi $2$-jet configuration, where $T$ can be considered sufficiently large.
In this case, the thrust axis $\vec{n}$ defines two opposite hemispheres: $S_n$, which points along the direction of the thrust axis, and $S_{\Bar{n}}$, which points backwards, along the direction $\vec{\Bar{n}}$ opposite to $\vec{n}$.
Then, Eq.~\eqref{eq:thrust_def} can be approximated by (see Ref.~\cite{Catani:1991kz} and~\cite{Catani:1992ua}):
\begin{align}
\tau=1-T = \frac{M_n^2}{Q^2} + \frac{M_{\Bar{n}}^2}{Q^2}.
\label{eq:thrust_2jet}
\end{align}
where $M_n$ and $M_{\Bar{n}}$ are the invariant masses of the two almost back-to-back jets. 
These invariant masses can be computed explicitly at partonic level, order by order in perturbation theory, by considering the full momentum flowing into each hemisphere. 
Clearly, the r.h.s. of Eq.~\eqref{eq:thrust_2jet} is trivially zero only when all partons emitted into each hemisphere are virtual. 
All other cases require at least one real emission.
Therefore, at 1-loop level the non-trivial configurations have three real particles in the final state.
A quasi $2$-jet configuration can be achieved either when one of them is soft or when two of them move collinearly.
In the following, we will review the 1-loop computation of the contribution of each of such final state configurations, obtained by applying the usual soft/collinear approximations to Feynman graphs (see Ref.~\cite{Collins:2011zzd}) and equipping them with a delta function that fixes the value of $\tau = 1-T$ according to Eq.~\eqref{eq:thrust_2jet}.
The thrust axis $\vec{n}$ will be identified with the $z$-axis of the frame used to compute the partonic cross section, i.e. the partonic analogue of the hadron frame, where the outgoing parton has zero transverse momentum.
All quantities will be computed in dimensional regularization, with $D = 4 - 2\epsilon$ being the spacetime dimension.

\bigskip


\subsection{Soft Thrust Function \label{app:soft_thrust_fun}}


\bigskip

The 1-loop soft thrust function $S(\epsilon,\,\tau)$ is obtained as a result of the application of the soft approximation ($T_S$ in Ref.~\cite{Collins:2011zzd}) to the Feynman graphs corresponding the 1-loop partonic cross section. The case of the fragmenting gluon is suppressed in this approximation, therefore it can be neglected. 
On the other hand, when the fragmenting parton is a fermion, the emitted gluon can be either virtual or real.
The virtual contribution vanishes in full dimensional regularization, hence the soft thrust function only originates from the real emission graphs.
In order to compute the r.h.s. of Eq.~\eqref{eq:thrust_2jet}, we will label the momenta of the three final state particles as in Section~\ref{subsubsec:frag_quark_unsub}, where $k_1$ refers to the fragmenting quark, $k_2$ to the antiquark and $k_3$ to the gluon.
In this case we have to use their soft approximations:
\begin{align}
&k_1^{(S)} = \left(q^+,\,0,\,\vec{0}_T \right), \\
&k_2^{(S)} = \left(0,\,q^-,\,\vec{0}_T \right), \\
&k_3^{(S)} = \left(l^+,\,l^-,\,\vec{l}_T \right) ,
\label{eq:soft_appr_kin}
\end{align}
where $q^+ = q^- = {Q}/{\sqrt{2}}$.
Notice that the soft approximation leaves $k_3$ unchanged in its functional form, but the size of all its components are now of order ${\lambda^2}/{Q}$, where $\lambda$ is a very low energy scale according to power counting.
Depending on the hemisphere in which the soft gluon has been emitted, we have two different, equally probable, configurations:
\begin{align}
&\tau = y_2^{(S)} = \frac{l^-}{q^-}, \mbox{ if } S_n \mbox{ emission, i.e. } l^+ > l^-; \\
&\tau = y_1^{(S)} = \frac{l^+}{q^+}, \mbox{ if } S_{\Bar{n}} \mbox{ emission, i.e. } l^- > l^+.
\label{eq:Sthrust_hemisph}
\end{align}
Therefore, by defining the usual light-like directions $w_1 = (1,\,0,\,\vec{0}_T)$ and $w_2 = (0,\,1,\,\vec{0}_T)$, the 1-loop soft thrust function is obtained as:
\begin{align}
&\frac{\alpha_S}{4 \pi}
S^{[1]}(\epsilon,\,\tau) = 
\int \frac{d^D l}{(2\pi)^D} \,
\frac{\mbox{Tr}_C}{N_C} \,
\Bigg[
\begin{gathered}
\includegraphics[width=3.6cm]{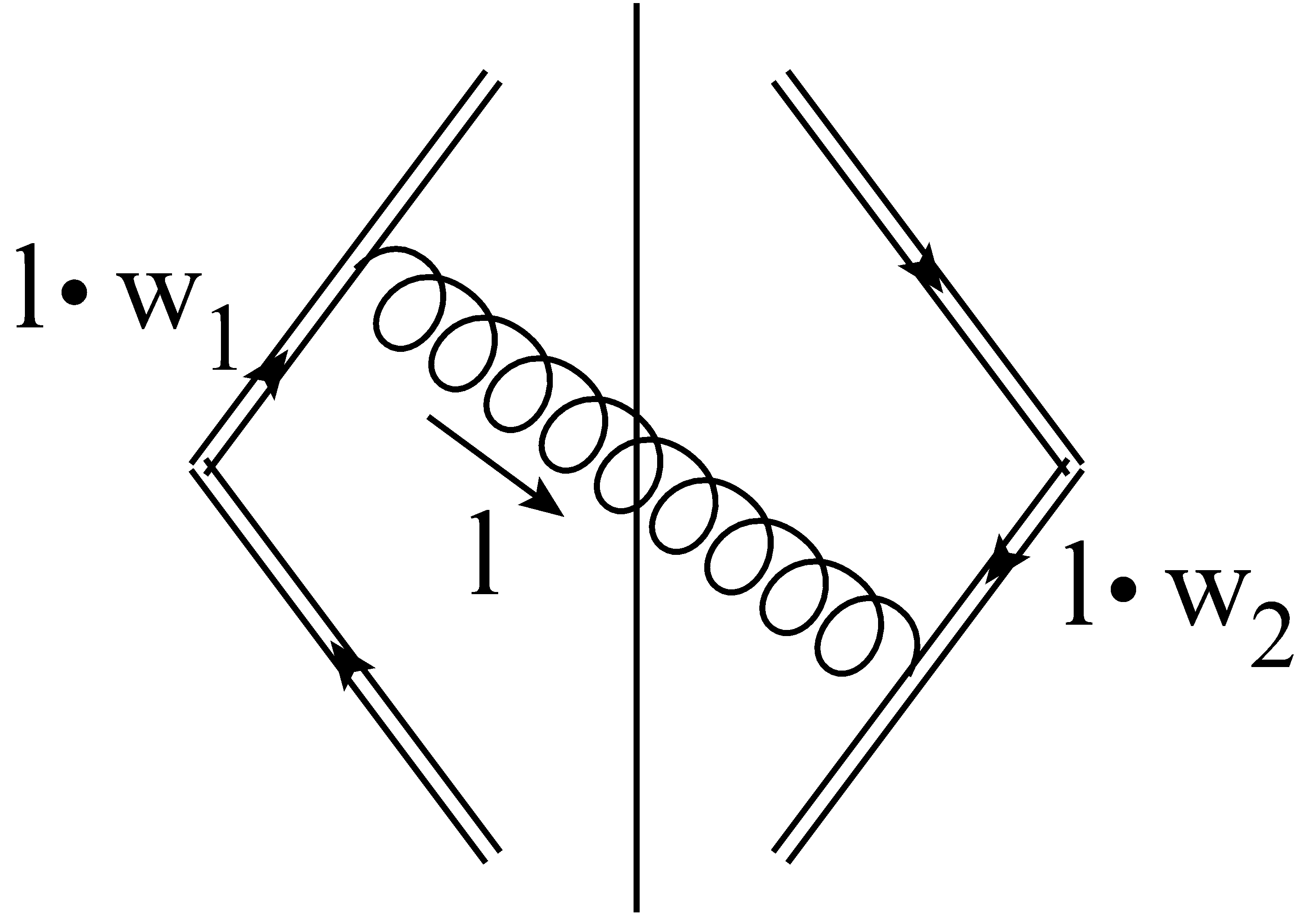}
\end{gathered}
+ h.c.
\Bigg] \,
(2 \pi) \delta_+ \left( l^2 \right) \,
\times 
\notag \\
&\quad \times \,
\left[
\delta \left(\tau - \frac{l^-}{q^-}\right) \,
\theta \left(l^+ - l^-\right)
+
\delta \left(\tau - \frac{l^+}{q^+}\right) \,
\theta \left(l^- - l^+\right)
\right] =
\notag \\
&\quad = 
\frac{\alpha_S}{4\pi} \, 8 \, C_F \, S_\epsilon \, 
\left( \frac{\mu}{Q} \right)^{2\epsilon} \, 
\tau^{-1-2\epsilon} \,\frac{1}{\epsilon}.
\label{eq:S_thrust}
\end{align}
Notice that all the divergences of $S(\epsilon,\,\tau)$ are regulated by dimensional regularization, and we did not encounter any unregulated rapidity divergence.
This is due to the explicit presence of $\tau$, that acts as a regulator. In fact, the expression in Eq.~\eqref{eq:S_thrust} is divergent if either $\epsilon$ or $\tau$ vanish.
The $\epsilon$-expansion of Eq.~\eqref{eq:S_thrust} follows from its integrability property.
In fact, $S^{[1]}(\epsilon,\,\tau)$ is integrable with respect to $\tau$, but its trivial expansion in powers of $\epsilon$ does not.
This problem is solved by considering an expansion in terms of $\tau$ {\it distributions}  instead of functions. Then, one can easily prove that:
\begin{align}
\tau^{-1-\epsilon} = 
-\frac{1}{\epsilon} \, \delta(\tau) +
\left(\frac{1}{\tau}\right)_+ - 
\epsilon \, 
\left(\frac{\log{\tau}}{\tau}\right)_+ +
\mathcal{O}\left(\epsilon^2\right), 
\quad \mbox{Re~}{\epsilon} < 0.
\label{eq:thrust_distr_exp}
\end{align}
%

\bigskip


\subsection{Backward Thrust Function \label{app:back_thrust_fun}}


\bigskip

The 1-loop backward thrust function $J_B(\epsilon,\,\tau)$ is obtained by applying the collinear approximation to the backward emission (analogue to $T_B$ in Ref.~\cite{Collins:2011zzd}) 
in the 1-loop Feynman graphs.
Here "backward" refers to the direction of emission, which goes along $\vec{\Bar{n}}$ (negative $z$-axis), opposite to the thrust axis. 
As for the previous case, the contribution of the fragmenting gluon is suppressed by power counting and virtual graphs are zero in full dimensional regularization.
In the following, we will assume that the fragmenting parton is a quark, hence the approximation concerns the configuration where the gluon is collinear to the antiquark. 
The case of a fragmenting antiquark is perfectly analogous.
The backward approximation applied to the momenta (labeled as usual) gives:
\begin{align}
&k_1^{(B)} = \left(q^+,\,0,\,\vec{0}_T \right), \\
&k_2^{(B)} = \left(\frac{l_T^2}{2(q^- - l^-)},\,q^- - l^-,\, -\vec{l}_T \right), \\
&k_3^{(B)} = \left(\frac{l_T^2}{2 l^-},\,l^-,\,\vec{l}_T\right),
\label{eq:B_appr_kin}
\end{align}
As in the previous case, the backward approximation leaves $k_3$ unchanged in its functional form, but resizes its components as $l^+ \sim {\lambda^2}/{Q}$, $l^- \sim Q$ and $l_T \sim \lambda$, where $\lambda$ is a very low energy scale according to power counting.
The emission of the collinear gluon along the direction of the fragmenting quark does not give any contribution, therefore we only have to consider the emission in the antiquark hemisphere:
\begin{align}
&\tau = y_1^{(B)} = 
\frac{l_T^2}{Q^2} \, \frac{q^-}{l^-} \, \frac{1}{1-\frac{l^-}{q^-}}, 
\mbox{ if } S_{\Bar{n}} \mbox{ emission, i.e. } l^- > l^+.
\label{eq:Bthrust_hemisph}
\end{align}
By defining $k = k_2^{(B)} + k_3^{(B)}$ as the total momentum entering into the antiquark hemisphere, the 1-loop backward thrust function is:
\begin{align}
&\frac{\alpha_S}{4 \pi}
J_B^{[1]}(\epsilon,\,\tau) = 
2 \, (2 \pi)^{D-1} \, 
\int \frac{d k^+ \, d^{D-2} \vec{k}_T}{(2\pi)^D} \,
\int \frac{d^D k_2}{(2\pi)^D} \,\times 
\notag \\
&\quad \times \,
\frac{\mbox{Tr}_C}{N_C} \, 
\frac{\mbox{Tr}}{4} \, 
\Bigg[ \gamma^- \, \Bigg(
\begin{gathered}
\includegraphics[width=4.6cm]{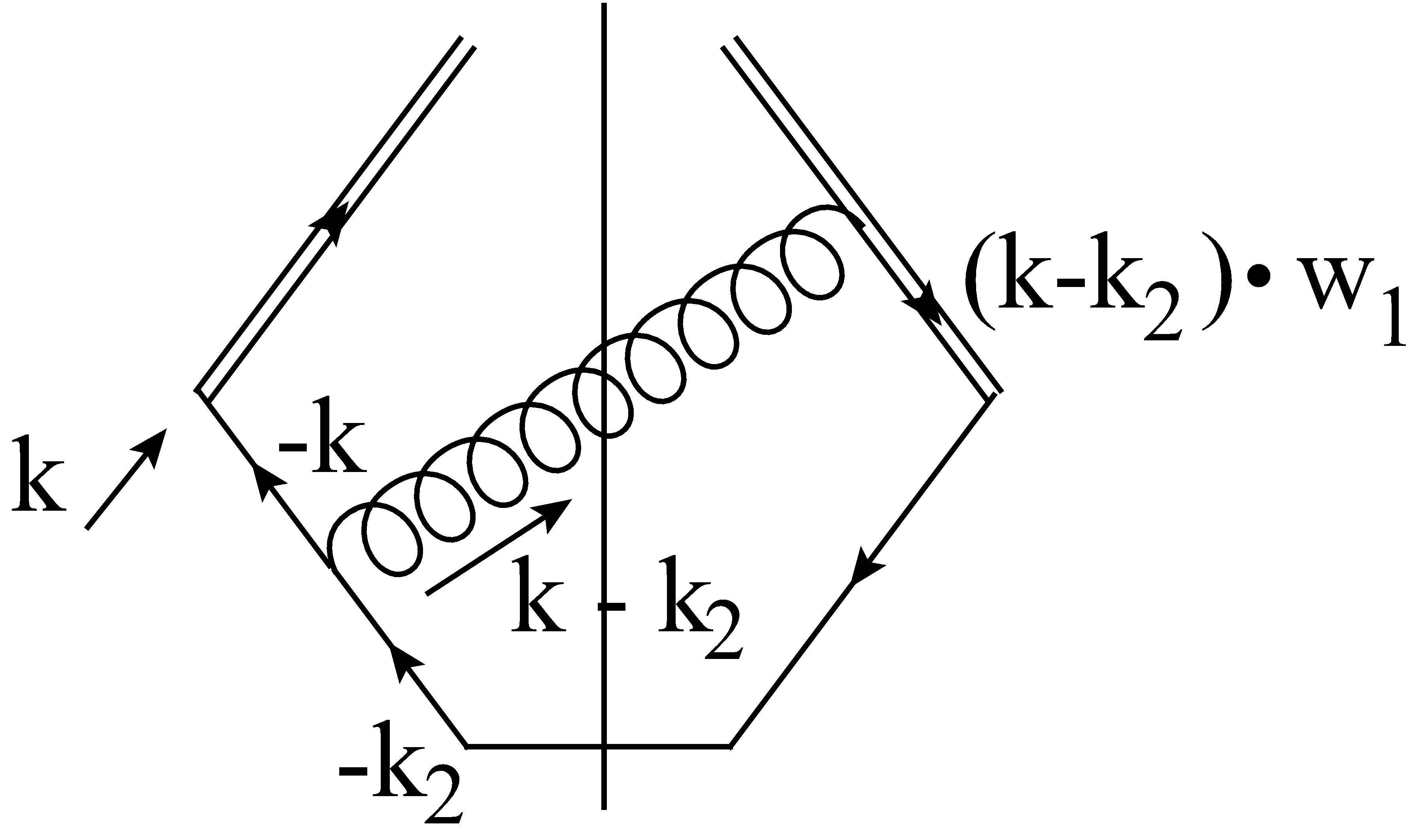}
\end{gathered}
+ h.c. + \,
\begin{gathered}
\includegraphics[width=3.3cm]{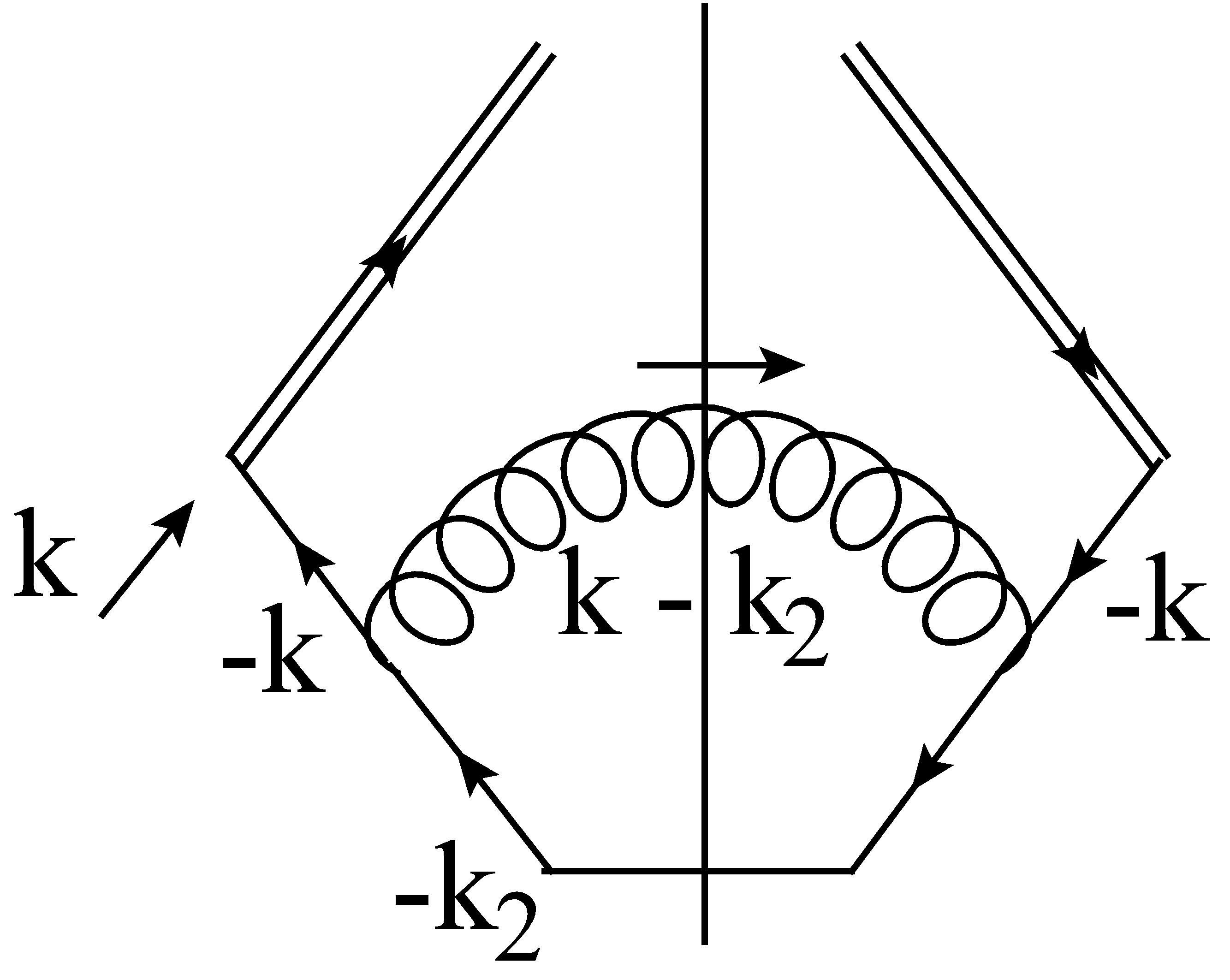}
\end{gathered} \Bigg)
\Bigg] \,\times 
\notag \\
&\quad \times \,
(2 \pi) \delta_+ \left( k_2^2 \right) \,
(2 \pi) \delta_+ \left( (k-k_2)^2 \right) \,
\delta^{(D-2)}( \vec{k}_T ) \,
\delta \left(\tau - \frac{k^+}{q^+}\right) \,
\theta \left(k^- - k_2^-\right)
= 
\notag \\
&\quad=
\frac{\alpha_S}{4 \pi} \, 4 \, C_F \, S_\epsilon \,
\left(\frac{\mu}{Q}\right)^{2\epsilon} \,
\tau^{-1-\epsilon} \,
\left[B(2-\epsilon,\,-\epsilon) + 
\frac{1-\epsilon}{2} \,B(1-\epsilon,\,2-\epsilon)  \right].
\label{eq:B_thrust}
\end{align}
Notice that in $J_B(\epsilon,\,\tau)$ the thrust acts as a regulator for the rapidity divergences, as in Eq.~\eqref{eq:S_thrust}. 
As a consequence, there is no overlapping with the soft momentum region and, contrary to the CSS scheme, no subtraction is required. 

\bigskip


\subsection{Jet Thrust Function \label{app:jet_thrust_fun}}


\bigskip

The 1-loop jet thrust function is obtained by applying the collinear approximation to the emission along the fragmenting parton direction (the analogue of $T_A$ in Ref.~\cite{Collins:2011zzd}) in all  Feynman graphs needed for the 1-loop partonic cross section.
The direction of the fragmenting parton coincides with the thrust axis $\vec{n}$ (positive $z$-axis).
The approximated momenta (labeled as usual) are:
\begin{align}
&k_1^{(J)} = \left(z q^+,\,0,\,\vec{0}_T \right), \\
&k_2^{(J)} = \left(0,\,q^-,\,\vec{0}_T \right), \\
&k_3^{(J)} = \left((1-z) q^+,\,\frac{l_T^2}{2 (1-z) q^+},\,\vec{l}_T \right),
\label{eq:J_appr_kin}
\end{align}
where $z$ represents the fractional energy associated to the partonic process:
\begin{align}
z = \frac{2 k_1 \cdot q}{q^2} = \frac{k_1^+}{q^+}.
\label{eq:z_def}
\end{align}
The emission in the hemisphere opposite to the direction of the fragmenting parton does not give any contribution. 
Therefore, the only relevant configuration is given by:
\begin{align}
&\tau = y_2^{(J)} = 
\frac{z}{1-z} \,\frac{l_T^2}{Q^2} ,
\mbox{ if } S_n \mbox{ emission, i.e. } l^+ > l^-.
\label{eq:Jthrust_hemisph}
\end{align}
In this case, the contribution of fragmenting gluons is not suppressed and needs to be computed explicitly.
By defining $k = k_1^{(J)} + k_3^{(J)}$ the total momentum entering the considered hemisphere, the 1-loop thrust jet function corresponding to a fragmenting gluon is given by:
\begin{align}
&\frac{\alpha_S}{4 \pi}
J_{g/q}^{[1]}(\epsilon;\,\tau, \, z) = 
\int \frac{d k^-}{(2\pi)^D} \, 
d^{2-2\epsilon} \vec{k}_T \,
\frac{\mbox{Tr}_C}{N_C} \,
\frac{\mbox{Tr}_D}{4} \,
\Bigg[
\gamma^+ \, 
\begin{gathered}
\includegraphics[width=2.8cm]{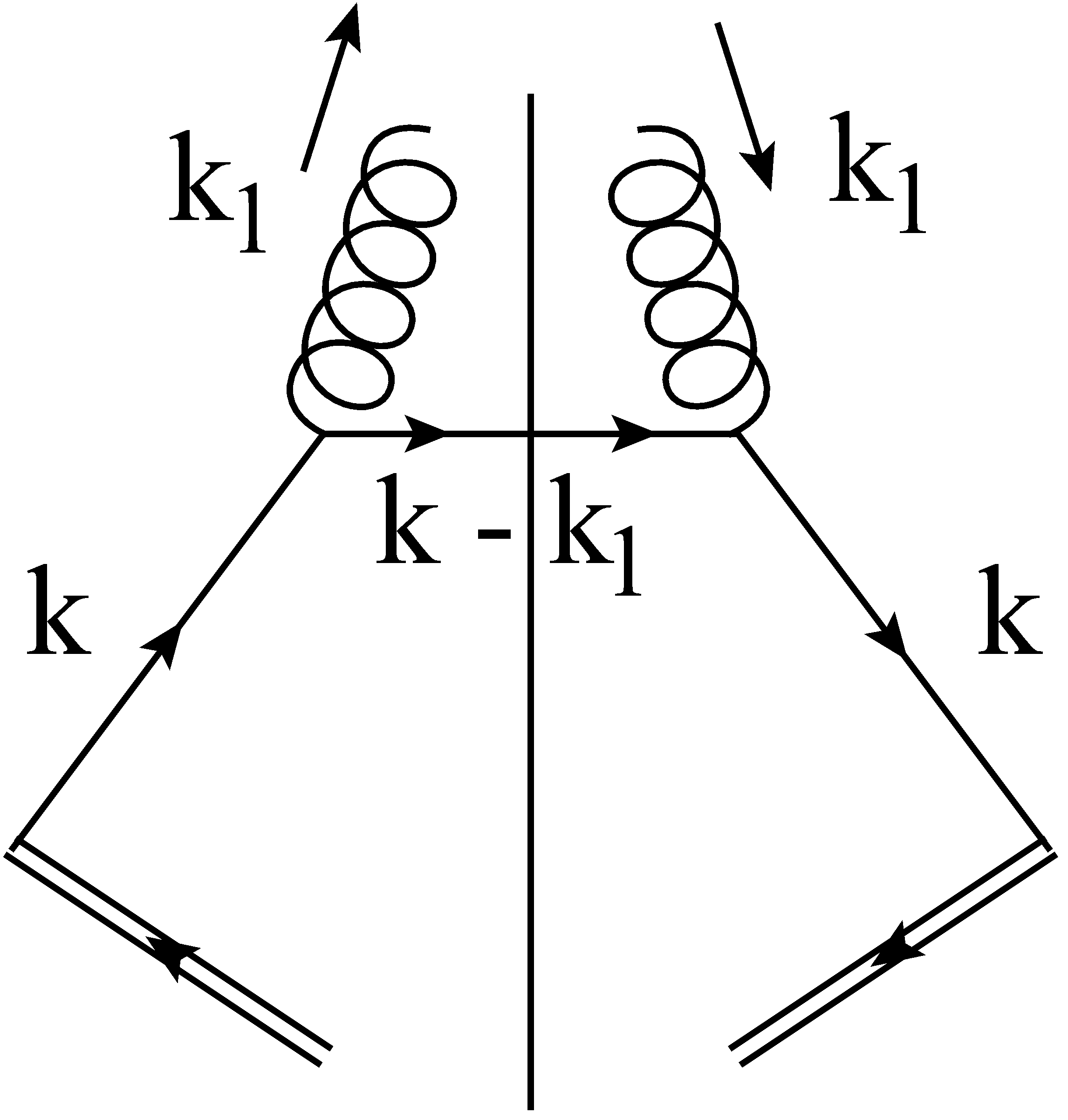}
\end{gathered}
\Bigg] \,
\times \notag \\
&\quad \times \,
(2\pi) \, \delta\left( (k-k_1)^2\right) \,
\delta\left(\tau - \frac{z}{1-z} \,\frac{k_T^2}{Q^2}\right) \,
\theta(1-z) = 
\notag \\
&\quad=
\frac{\alpha_S}{4 \pi} \, 
2 \, C_F \, S_\epsilon \,
\theta(1-z)\,
\frac{1+(1-z)^2-\epsilon z^2}{z^2} \, 
\times
\notag \\
&\quad \times \,
\frac{\Gamma(1-\epsilon)}{\pi^{1-\epsilon}} \,
\mu^{2 \epsilon} \, 
\int d^{2-2\epsilon} \vec{k}_T \,
\frac{1}{k_T^2} \,
\delta\left(\tau - \frac{z}{1-z} \,\frac{k_T^2}{Q^2}\right) = 
\notag \\
&\quad=
\frac{\alpha_S}{4 \pi} \, 
2 \, C_F \, S_\epsilon \,
\left( \frac{\mu}{Q} \right)^{2 \epsilon} \, 
\theta(1-z)\,
\left(\frac{z}{1-z}\right)^\epsilon \,
\frac{1+(1-z)^2-\epsilon z^2}{z^2} \,
\tau^{-1-\epsilon}.
\label{eq:jetgluon_thrust}
\end{align}
%
The case of a fragmenting fermion involves more Feynman graphs.
The contribution of the virtual emission is zero in full dimensional regularization.
The 1-loop jet thrust function associated to a fragmenting quark is given by:
\begin{align}
&\frac{\alpha_S}{4 \pi}
J_{q/q}^{[1]}(\epsilon;\,\tau, \, z) = 
\int \frac{d k^-}{(2\pi)^D} \,
d^{2-2\epsilon} \vec{k}_T \,
\times \notag \\
&\quad \times \,
\frac{\mbox{Tr}_C}{N_C} \,
\frac{\mbox{Tr}_D}{4} \,
\Bigg[ \gamma^+ \, \Bigg(
\begin{gathered}
\includegraphics[width=4.2cm]{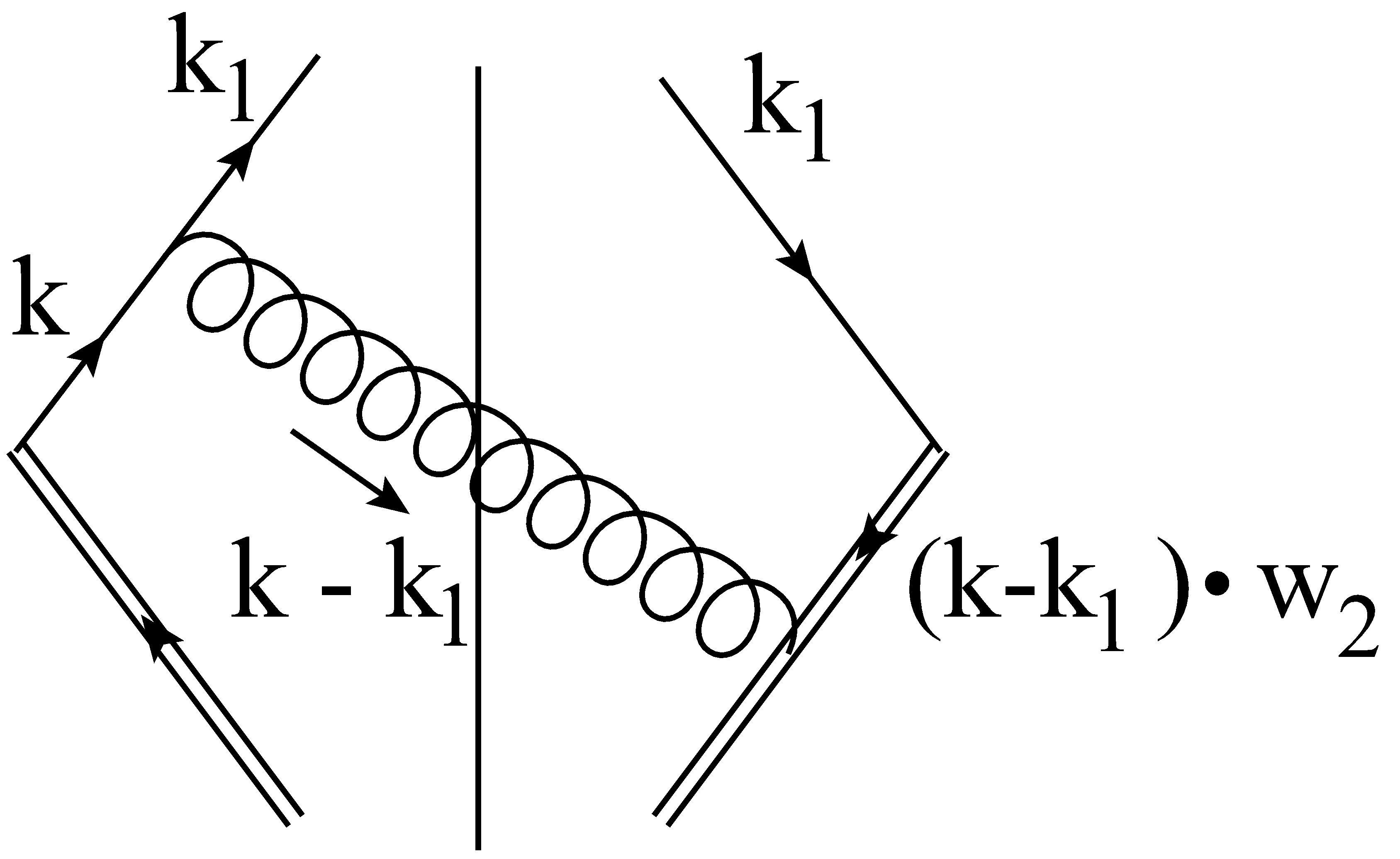}
\end{gathered}
+ h.c. + \,
\begin{gathered}
\includegraphics[width=2.9cm]{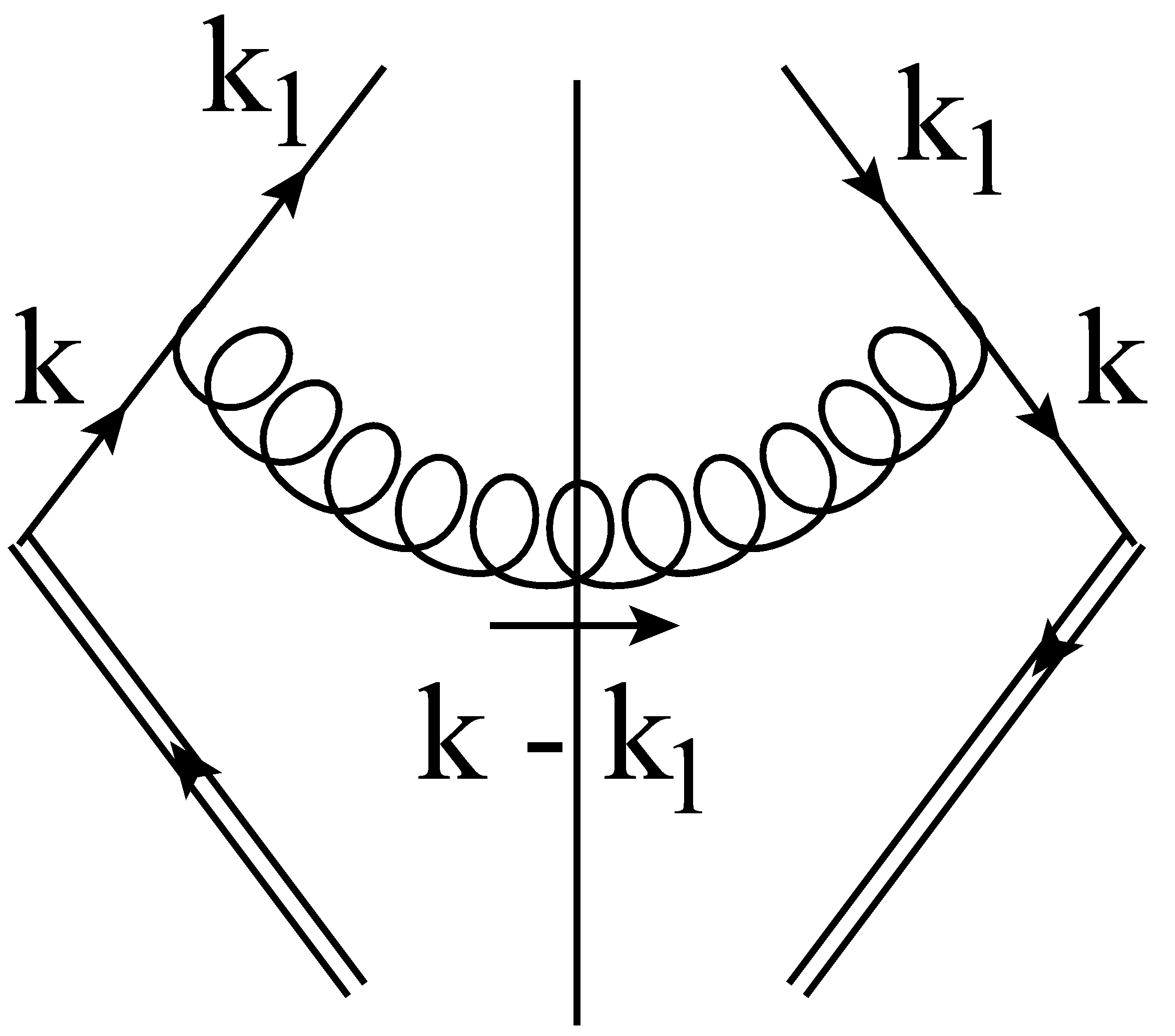}
\end{gathered} \Bigg)
\Bigg] \,
\times \notag \\
&\quad \times \,
\theta(1-z)\,
\delta\left(\tau - \frac{z}{1-z} \,\frac{k_T^2}{Q^2}\right) = 
\notag \\
&\quad=
\frac{\alpha_S}{4 \pi} \, 
2 \, C_F \, S_\epsilon \,
\theta(1-z)\,
\left[
\frac{2}{(1-z)} + \frac{(1-\epsilon) \, (1-z)}{z}
\right] \,
\times
\notag \\
&\quad \times \,
\frac{\Gamma(1-\epsilon)}{\pi^{1-\epsilon}} \,
\mu^{2 \epsilon} \,
\int d^{2-2\epsilon} \vec{k}_T 
\frac{1}{k_T^2} \,
\delta\left(\tau - \frac{z}{1-z} \,\frac{k_T^2}{Q^2}\right) = 
\notag \\
&\quad=
\frac{\alpha_S}{4 \pi} \, 4 \, C_F \, S_\epsilon \, 
\left( \frac{\mu}{Q} \right)^{2 \epsilon} \, 
\theta(1-z)\, z^{\epsilon} \,
\left[
(1-z)^{-1-\epsilon} + \frac{1-\epsilon}{2}\, \frac{(1-z)^{1-\epsilon}}{z}
\right]
\tau^{-1-\epsilon}.
\label{eq:jetferm_thrust}
\end{align}
Notice that in this case, the $\epsilon$-expansion of the above expression follows from Eq.~\eqref{eq:thrust_distr_exp} and from the analogous expansion for $z$:
\begin{align}
(1-z)^{-1-\epsilon} = 
-\frac{1}{\epsilon} \, \delta(1-z) +
\left(\frac{1}{1-z}\right)_+ - 
\epsilon \, 
\left(\frac{\log{(1-z)}}{1-z}\right)_+ +
\mathcal{O}\left(\epsilon^2\right), 
\quad \mbox{Re~}{\epsilon} < 0.
\label{eq:z_distr_exp}
\end{align}
%

%
%
%
%

\providecommand{\href}[2]{#2}\begingroup\raggedright\endgroup

\end{document}